\documentclass[12pt]{article}

\usepackage{times}
\usepackage{scicite}
\usepackage{epsfig}
\usepackage{aas_macros}

\usepackage{epstopdf, graphicx}
\usepackage[dvipsnames]{xcolor}
\usepackage{amssymb,amsmath}
\usepackage{scicite}
\usepackage{ulem}

\usepackage[font={small}]{caption}
\usepackage{fullpage}
\usepackage{amsfonts}
\usepackage{mathrsfs}
\usepackage{psfrag}

\usepackage{tabu}
\usepackage{longtable}
\usepackage{pdflscape}

\usepackage{amssymb}

\newcommand{\mlya}{${\rm Ly\alpha}$}

\newcommand{\rperp}{r_\perp}

\newcommand{\npair}{25}  
\newcommand{\nmodel}{400\ }  

\def \kms            {{\rm km~s}^{-1}}

\newcommand{\lya}{{\rm Ly}\alpha}
\newcommand{\lyb}{{\rm Ly}\beta}
\newcommand{\ion}[2]{#1 {\sc #2}}

\def \arcsec {''}
\def \mkms  {{\rm km~s^{-1}}}

\def \nesi  {10}

\def \nmagell  {6}
\def \nxshooter {2}

\newenvironment{sciabstract}{%
\begin{quote} \bf}
{\end{quote}}

\newcounter{lastnote}

\title{Measurement of the small-scale structure of the intergalactic
medium using close quasar pairs} 

\author
{Alberto Rorai$^{1,2,\ast}$, Joseph F. Hennawi$^{2,3}$, Jose O\~norbe$^{2}$, Martin White$^{4,5}$,\\ J. Xavier Prochaska$^{6}$, Girish Kulkarni$^{1,2}$, Michael Walther$^{2}$, Zarija Luki\'c$^{5}$,\\  Khee-Gan Lee$^{4}$ 
\\
\normalsize{$^{1}$ Institute of Astronomy, Cambridge, UK}\\
\normalsize{$^{2}$ Max-Planck-Institut f\"ur Astronomie, K\"onigstuhl, Germany}\\
\normalsize{$^{3}$ Department of Physics, University of California, Santa Barbara}\\
\normalsize{$^{4}$Department of Astronomy, University of California at Berkeley, CA, USA}\\
\normalsize{$^{5}$ Lawrence Berkeley National Laboratory, Berkeley, CA, USA}\\
\normalsize{$^{6}$University of California Observatories-Lick Observatory, UC Santa Cruz, CA}\\
\normalsize{$^\ast$ To whom correspondence should be addressed: arorai@ast.cam.ac.uk.}
}

\date{}

\begin{document}

\maketitle

\begin{sciabstract}
The distribution of diffuse gas in the intergalactic medium (IGM)
imprints a series of hydrogen absorption lines on the spectra of
distant background quasars known as the Lyman-$\alpha$ forest.
Cosmological hydrodynamical simulations predict that IGM density
fluctuations are suppressed below a characteristic scale where thermal
pressure balances gravity.  We measure this pressure smoothing scale
by quantifying absorption correlations in a sample of close quasar
pairs.  We compare our measurements to hydrodynamical simulations, where 
pressure smoothing is determined by the integrated thermal history
of the IGM. Our findings are consistent with standard models for photoionization heating by
the ultraviolet radiation backgrounds that reionized the universe.
\end{sciabstract}

As the dominant reservoir of baryons in the universe, the
intergalactic medium (IGM) plays a crucial role in the history and
evolution of cosmic structure.  About half a million years after the
Big Bang, the plasma of primordial baryons recombined to form the
first neutral atoms, releasing the cosmic microwave background (CMB)
and initiating the cosmic `dark ages'.  During this period primordial
neutral hydrogen and helium expanded and cooled to very low
temperatures $T\sim 20\,{\rm K}$, while dark matter driven structure
formation eventually gave rise to the first galaxies. Ultimately,
stars and supermassive black holes 
in these galaxies emitted enough ionizing
photons to reionize and reheat the universe: it is believed that soft
photons from primeval galaxies ionized hydrogen and singly ionized
helium at $z\sim 7-8$, whereas it took until $z\sim 3$ for hard
radiation emitted by quasars to doubly ionize helium \cite{Meiksin09}.
During these
reionization phase transitions, ionization fronts propagate
supersonically through the IGM, impulsively heating gas resulting in
temperature changes $\Delta T \sim 10^4~{\rm K}$\cite{Meiksin09}.
Afterwards the IGM
cools via adiabatic expansion and inverse Compton scattering off the
CMB ,but because both the cooling and dynamical times in the rarefied
IGM are long, comparable to the age of the Universe, 
memory of these thermal
events is retained \cite{HuiGnedin97,GnedHui98,HH03,Kulkarni2015}.
Thus, an empirical characterization of the IGM's thermal state across
cosmic time can constrain the nature and timing of these reionization
events.

At currently observable redshifts ($z\lesssim 7$), hydrogen 
in the IGM is mostly ionised. However, the small residual 
fraction of neutral hydrogen gives rise
to Lyman-$\alpha$ (Ly$\alpha$) absorption which is observed to be 
ubiquitous toward distant background quasars.  
This so-called Ly$\alpha$ forest is an 
established probe of the IGM and cosmic structure at
high redshifts $z \sim 2-6$. Since Ly$\alpha$ forest observations are
sensitive to gas in regions devoid of galaxies, complex and poorly understood
physical processes related to galaxy formation are not expected to
play a substantial role \cite{Kollmeier06,Desjacques2006}.  Thus the structure of
the IGM can be predicted ab initio with cosmological
hydrodynamic simulations, which have been used
to infer cosmological parameters from the Ly$\alpha$ forest
observations \cite{Viel2004,Seljak2006}. However this requires
assumptions regarding how and when
reionization injected heat into the IGM.  By comparing
simulations to observational constraints on the IGMs thermal state,
our understanding of structure formation can be leveraged to
make progress on understanding how reionization occurred.

There are two known ways to constrain the thermal state of the
IGM.  The first is the traditional approach using one-dimensional
Ly$\alpha$ forest sightlines provided by individual quasars. 
Semi-analytical models and hydrodynamical simulations 
show that IGM gas obeys a power-law relation
between the temperature $T$ and the density, which can be
written as $T = T_0(\Delta)^{\gamma-1}$ \cite{HuiGnedin97,Mcquinn2015},
where $\Delta$ is the overdensity relative to the mean, 
$T_0$ is the temperature at mean density ($\Delta=1$), and $\gamma$ is the slope
of this relation. 
Microscopic thermal motions of IGM gas Doppler-broadens \mlya\ forest lines, and any
statistic sensitive to
the smoothness of the spectra can be used to constrain the amplitude
and slope parameters $(T_0,\gamma)$
\cite{McDonald2000,Zald01,Croft2002}.
The primary drawback of this technique is the
challenge of disentangling the intrinsic small-scale
structure ($\lesssim 100\,{\rm kpc}$; all distances are in comoving units,
$1\,{\rm kpc} = 3.1\times 10^{21}\,{\rm cm}$) 
of the IGM from the thermal Doppler broadening
\cite{Peeples09a,Peeples09b,Rorai13,Puchwein15}. 

We have developed a second technique to
characterize the thermal state of the IGM \cite{Rorai13,Kulkarni2015,Onorbe2016}
which is used in this paper. The technique directly measures the intrinsic
small-scale structure by comparing
close pairs of quasars, measuring the transverse Ly$\alpha$ forest
correlations across the line-of-sight.  Although baryons in the IGM
trace dark matter fluctuations on Mpc ($1\,{\rm Mpc}=10^3\,{\rm kpc}$) scales, on smaller scales
$\lesssim 100~{\rm kpc}$ 
the gas is pressure
supported against gravitational collapse by its finite temperature
($T\sim 10^4\,{\rm K}$)
\cite{GnedHui98,schaye2001,Kulkarni2015,Onorbe2016}.
Baryonic fluctuations are suppressed relative to the
pressureless dark matter (which can collapse), and the IGM is thus
pressure smoothed on small scales. A naive guess for the pressure
smoothing scale $\lambda_P$ follows from classic Jeans argument
$\lambda_P=c_s/\sqrt{G\rho}$, where $c_s$ is the sound speed
of the gas, $G$ is the gravitational constant and $\rho$ the density
of the gas. 
However at a redshift $z$,
the actual level of pressure smoothing depends not on the prevailing
pressure/temperature at that epoch, but rather on the
temperature of the IGM in the past\cite{GnedHui98}
and must be determined from hydrodynamical simulations
\cite{Kulkarni2015,Onorbe2016}. 
The pressure smoothing scale
$\lambda_P$ thus provides an integrated record of the thermal
history of the IGM, and is sensitive to the timing and
magnitude of  heat
injection by reionization events \cite{Kulkarni2015}.
Measuring $\lambda_P$
would break the degeneracy between the small-scale structure
of the IGM and thermal Doppler broadening  \cite{Peeples09a,Peeples09b,Rorai13,Puchwein15}.

\begin{figure}[h!]
  \vskip -0.2in
     \centering{\epsfig{file=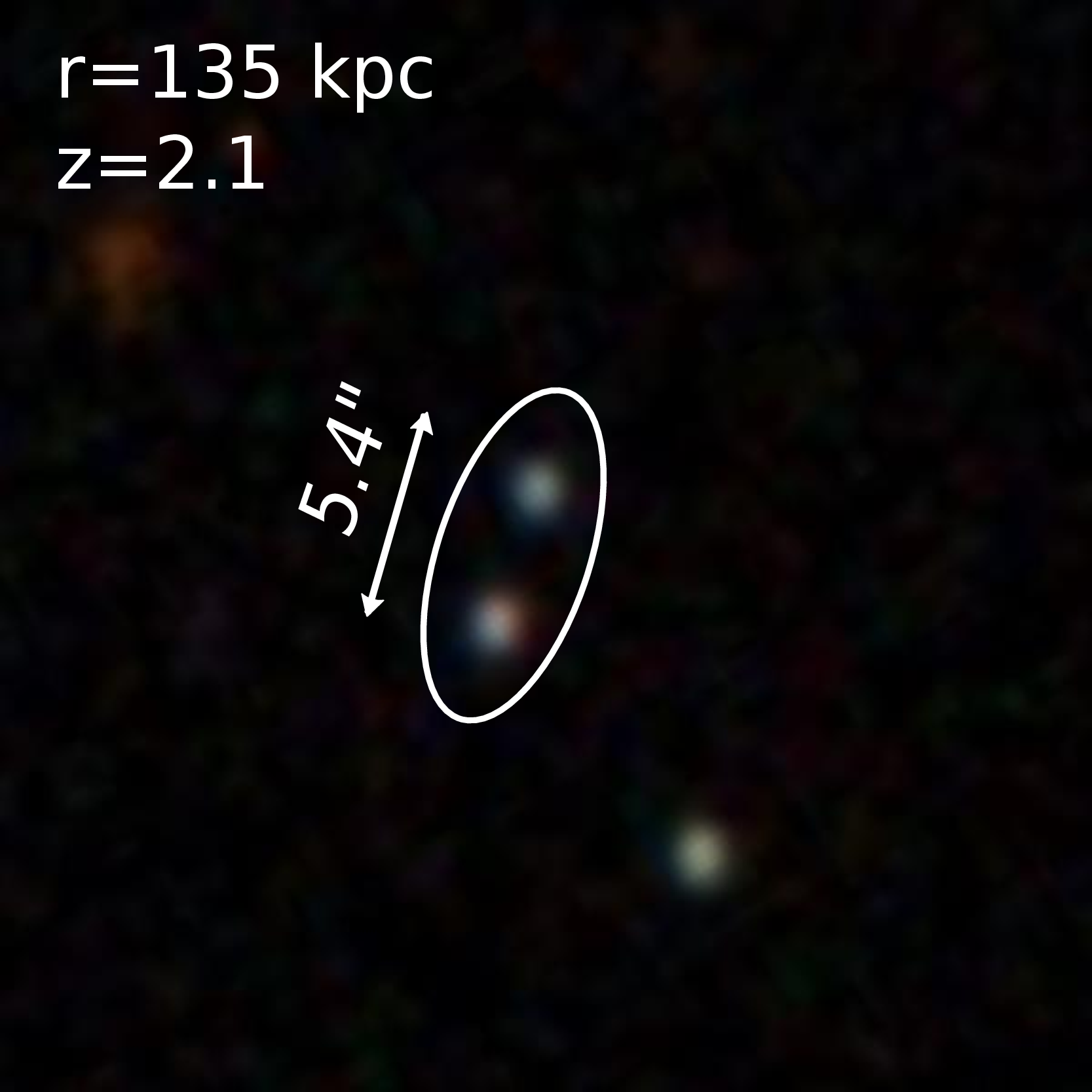,bb=10 -110 450 500 
       , width=0.25\textwidth,clip}
     \epsfig{file=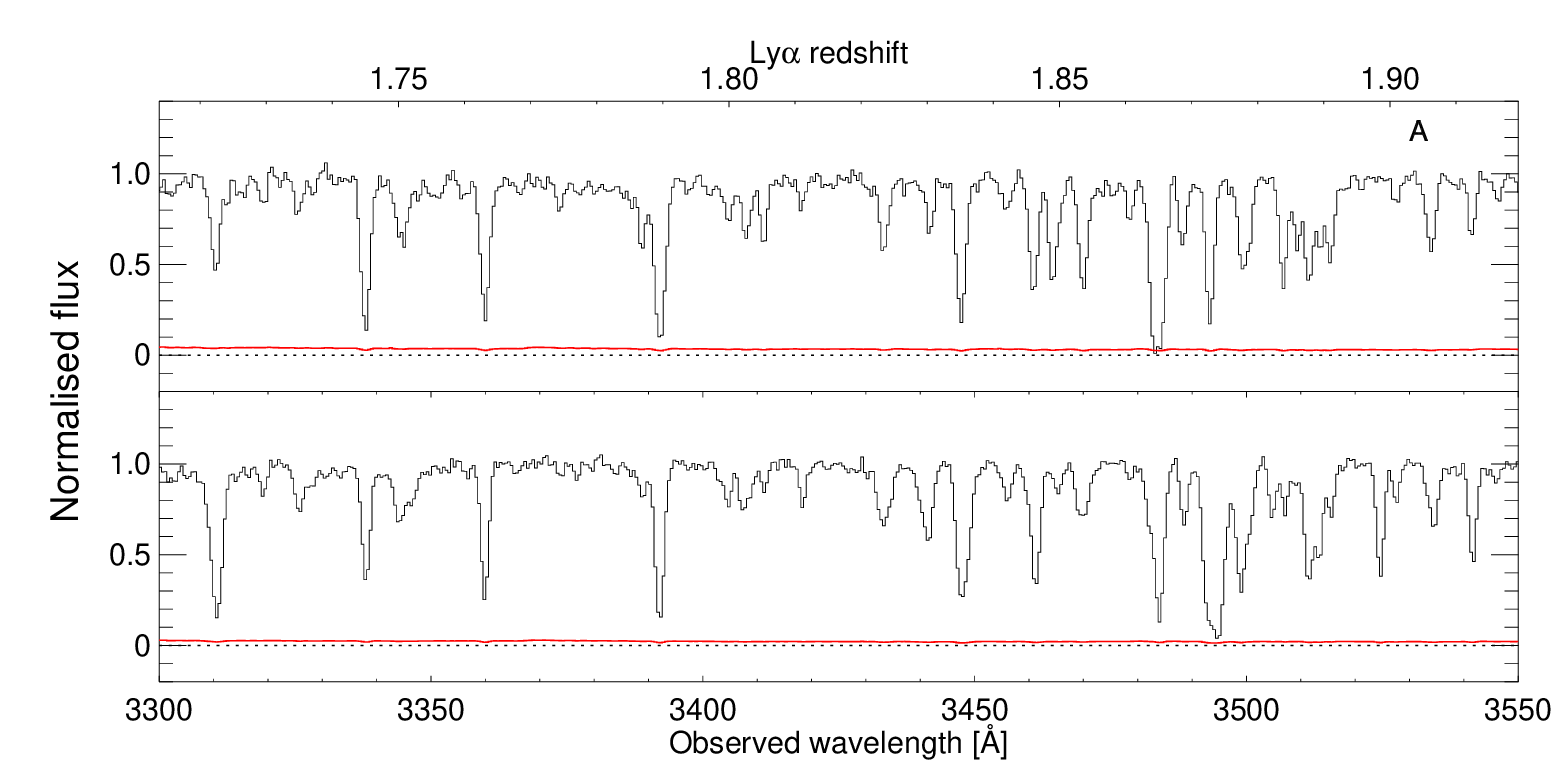,bb=80 50 720 400
       , width=0.7\textwidth,clip}
       \epsfig{file=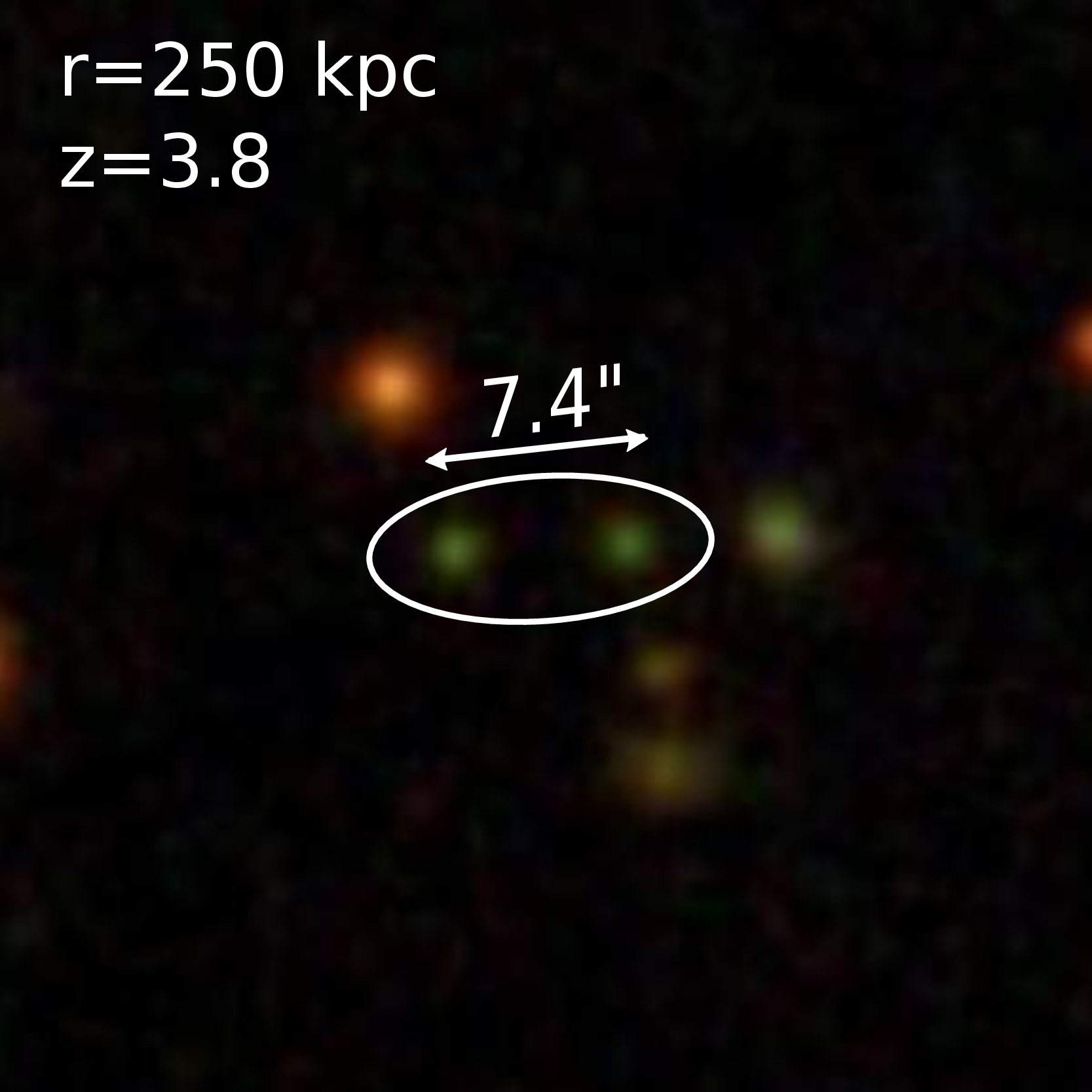,bb=10 -110 450 500 
       , width=0.25\textwidth,clip}
     \epsfig{file=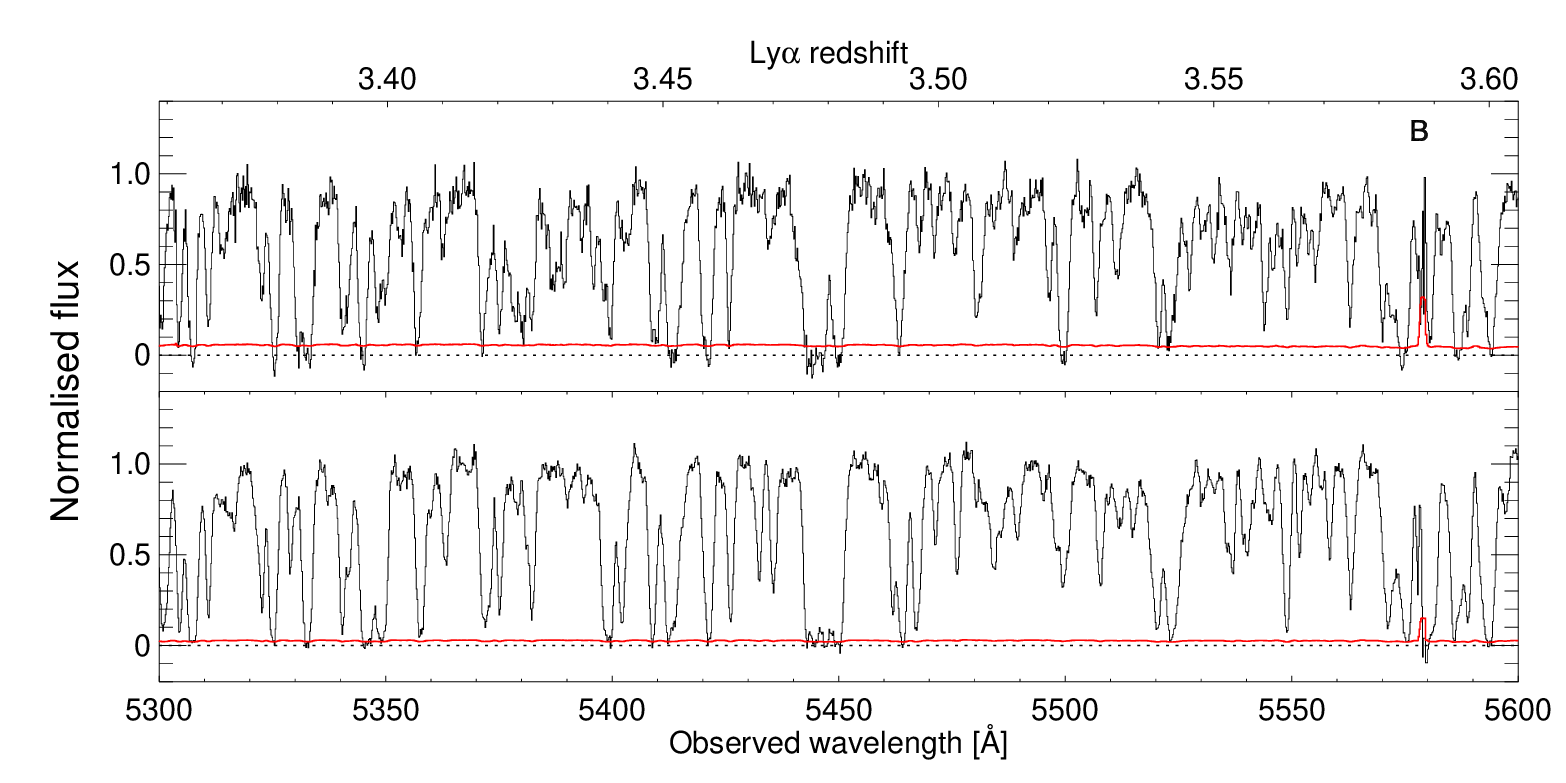,bb=80 50 720 400
       , width=0.7\textwidth,clip}}
       
	\centering{}

   \vskip -0.25in
   \caption{\label{fig:pair}\footnotesize {\bf Spectra of the quasar 
   pairs SDSS~J073522.43+295710.1, SDSS~J073522.55+295705.0 (panel A)  and SDSS~J102116.98+111227.6, 
   SDSS~J102116.47+111227.8 (panel B)}. 
   The plot shows the continuum normalized flux of the \mlya\ forest
   as a function of observed wavelength (with the corresponding 
   redshift on the upper axis). The dotted
   line marks the zero-flux level, while the red solid line shows the 
   amplitude of the $1\sigma$ noise level of each pixel.
   The two quasar pairs were observed with the Low Resolution Imaging
   Spectrograph (panel A) and the the echelle 
   spectrograph and imager (panel B) 
   at the Keck telescope.
   The angular separation between the companions 
   is $5.4''$ (panel A) and $7.4''$
   (panel B), which correspond to an impact parameter 
   of approximately 135 and 250  comoving kpc, respectively.    
   In each pair, the correlation of the absorption features is obvious,
   implying that the two spectra are probing the same 
   density structures at small scales. Analogous figures for the 
   whole sample are available \cite{supp}.}     
\end{figure}

We search for increasing coherence in the \mlya\ forest at 
progressively smaller quasar pair separations $\rperp \lesssim
300~{\rm kpc}$ (angular separation on the sky of 
$\psi\lesssim 10^{\prime\prime}$) which resolve
$\lambda_P$ \cite{Rorai13}. Previously only a handful of high-$z$ quasar
pairs with sufficiently small separations were known \cite{Foltz1984,Smette1992}.
We have conducted an observational program to identify
close quasar pairs \cite{BINARY,HIZBIN} which makes this measurement
possible over the redshift range $1.8 < z < 3.9$ \cite{supp}.
We used several telescopes to obtain spectra of \npair\
quasar pairs \cite{supp}, with transverse separations ranging from
$r_{\perp} = 100-500~{\rm kpc}$.
Fig.~\ref{fig:pair} shows the overlapping Ly$\alpha$ forests of two
quasar pairs in our sample illustrating coherent absorption
which results because their separations are comparable to the pressure smoothing
scale.

\begin{figure}[h!]
  \vskip -0.2in
   \centering{\epsfig{file=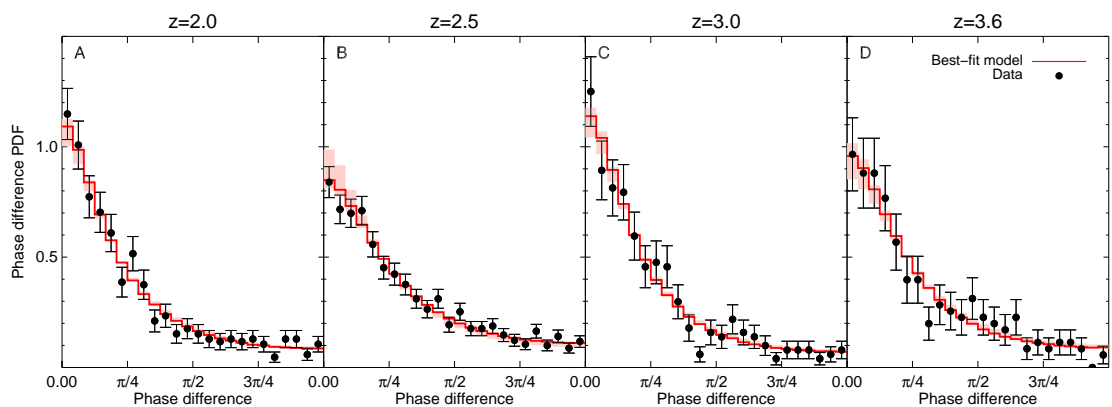,bb=50 50 1000 450 
      , width=\textwidth}}
	\centering{}

   \vskip -0.25in
   \caption{\label{fig:phases}\footnotesize {\bf Stacked phase angle
       PDFs in the four redshift bins.} The stacked PDFs are obtained by
     calculating the phase differences of \emph{all} pairs in each
     of the four redshift bins, without separating phases by
     wavenumber $k$ or transverse pair separation $r_{\perp}$.
     In this way we condense the entire
     statistical power of the phase angle data into a single PDF, at
     the price of mixing together phase angles at different $k$ and
     $r_{\perp}$, resulting in a loss of spectral and spatial
     information.   
     This is done to facilitate the visualization of the phase
     dataset, whereas our actual statistical analysis takes the 
     $k$ and  $r_{\perp}$ dependencies into account \cite{supp}.
     The black circles with (Poisson) errorbars are obtained
     from our data sample, while the red curve represents the thermal
     model which maximizes the likelihood in our phase analysis, where
     the models are obtained by post processing a dark-matter only
     simulation \cite{supp}. The red shaded regions delimit the family
     of models within the 68\% confidence level.}
\end{figure}

We apply a statistical measure of the \mlya\ forest correlations to
this quasar pair data \cite{Rorai13,supp}.
This technique, based on the phase difference
between homologous line-of-sight Fourier modes $k$
of the
spectra in the pairs, is maximally sensitive to the
pressure smoothing scale $\lambda_P$, and minimally sensitive to
the temperature-density relation parameters $(T_0,\gamma)$
\cite{Rorai13}.  The coherence of the Ly$\alpha$ forest is revealed in the
statistical distribution of these phase differences
$\theta(k,r_\perp)$, which will tend to be aligned
($\theta \sim 0$) in highly correlated spectra. We split the
sample into four redshift bins and measured phase
differences for all modes in a resolution-dependent range \cite{supp}.
Fig.~\ref{fig:phases} shows the
probability distribution function (PDF) of the phase differences (phase
angle PDF) for differing values of $k$ and $r_\perp$.
The fact that the distributions are peaked toward $|\theta|=0$
quantifies the strong correlations which are visible by eye
in Fig.~\ref{fig:pair}.

Following \cite{Rorai13}, we apply a likelihood formalism for estimating
$\lambda_P$ from our ensemble of observed
phase differences $\Delta\theta(k,r_\perp)$.
For our measurements we adopt a fast flexible semi-numerical model of
the IGM based on collisionless dark-matter only N-body simulations,
that parametrizes the IGMs thermal state
with a temperature-density relation ($T_0 ,\gamma$), 
and pressure smoothing scale $\lambda_P$. 
These thermal parameters are assigned to the
dark matter particle distribution in post-processing, and Ly$\alpha$
forest skewers are generated using the fluctuating Gunn-Peterson
approximation \cite{gp65,supp}.
This fast simulation method is employed to make our 
measurements and provides a reference for $\lambda_P$ as the smoothing length of the
dark matter density field. The $\lambda_P$ defined in this way
is directly related to the corresponding
$\lambda_P$ inferred from full hydrodynamical simulations,
which we later use to consistently interpret our results
\cite{supp}.

\begin{figure}[h!]
  \vskip -0.31in
  \centering{\epsfig{file=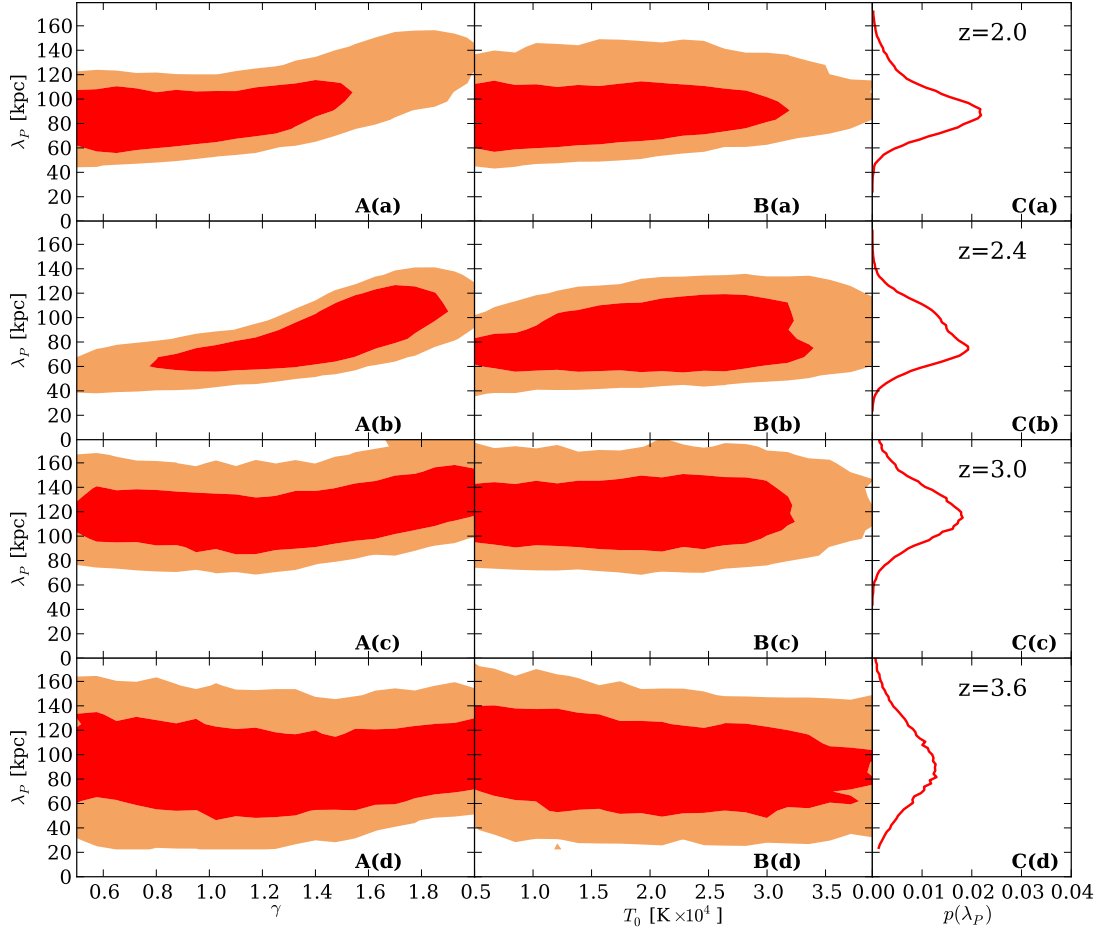, width=\textwidth}}
   \vskip -0.1in
   \caption{\label{fig:contours}\footnotesize {\bf Constraints on the IGM thermal parameters from the 
   phase angle PDF analysis.}  Panel A: 68\% (red) and 95\% (light 
   red) confidence levels of the posterior distribution in the $\gamma-\lambda_P$ plane. 
   Panel B: same as panel A, but for the $T_0-\lambda_P$ plane.
   Panel C: Posterior probability distribution for $\lambda_P$ after
   marginalizing over the other two thermal parameters $T_0$ and $\gamma$. 
   The vertical subpanels refer to $z=2,2.4,3,3.6$ 
   (a,b,c and d, respectively). The 
   standard deviation of the distribution of 
   $\lambda_P$ for the four redshift bins is $22\%,24\%,19\%$ and
   $33\%$, respectively. The uncertainty illustrated here do not include the additional
   $\sim 6\%$ increase in the uncertainties due to systematics \cite{supp}.}
\end{figure}

Imperfections in the data will cause the observed phase angle PDF to
depart from the ideal case represented by our model. 
Specifically, the combination of spectral noise and limited resolution reduces the correlations we are trying to quantify.
Metal lines and optically thick Lyman limit systems (LLSs) which are
not captured by our simulations provide a stochastic background of
absorption which may not be correlated across the two sight lines. To
account for these effects, we adopt a forward-modelling approach and
implement them into mock Ly$\alpha$ forest spectra for each of the
\nmodel models that we consider \cite{supp}. Armed with our likelihood
and forward model, we infer the posterior distribution of the thermal
parameters using a Markov chain Monte Carlo (MCMC) sampling algorithm
\cite{supp}. Fig.~\ref{fig:contours} shows contour plots of the
posterior in the thermal parameter ($T_0,\gamma,\lambda_P$)
space resulting from this analysis. The horizontal orientation of the
contours shows that, as expected from \cite{Rorai13}, the phase angle
PDF is primarily sensitive to $\lambda_P$ and depends only very
weakly on $\gamma$ and $T_0$. By marginalizing out $\gamma$ and $T_0$ we
obtain a measurement of $\lambda_P$ with statistical
errors of $\sim 20-30\%$. We explored the impact of a range of possible
systematics related to continuum fitting, imperfect knowledge of the
noise and resolution of our spectra, and uncertainties in the
abundances of metal lines and LLSs \cite{supp}. We conservatively estimate that
the combined impact of all these effects increases our uncertainties 
by at most $\sim 6\%$  \cite{supp}.

\begin{figure}[h!]
  \vskip -0.35in
  \centering{\epsfig{file=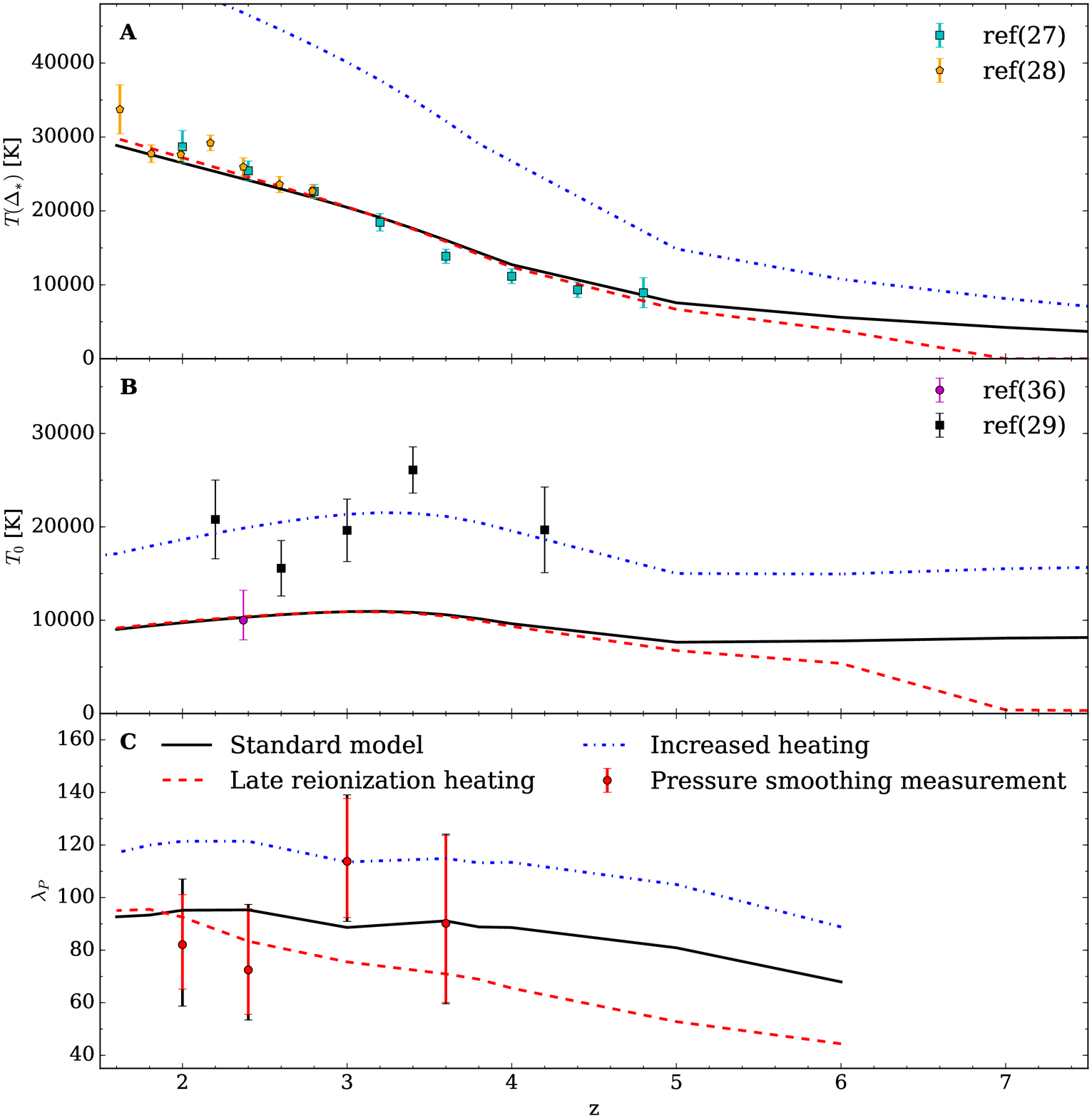, width=0.95\textwidth}}
   \vskip 0.0in
   \caption{\label{fig:model}\footnotesize {\bf Comparison of our
       pressure smoothing scale measurement to hydrodynamical
       simulations.} Panel A: The evolution of the
     temperature  $T(\Delta_{\star})$ at the typical density  $\Delta_{\star}$
     probed by the Ly$\alpha$ forest as defined in \cite{BeckerBolton2011}.
     The $T(\Delta_{\star})$ measurements and values of
     $\Delta_{\star}$ as a function of redshift are taken from
     \cite[{\rm cyan squares}]{BeckerBolton2011} 
     and Boera et al. \cite[{\rm orange pentagons}]{Boera2014}. 
     The error bars on these measurements do not include
     systematic uncertainties caused by lack of knowledge of $\lambda_P$
     scale (or equivalently the unknown thermal history), and are probably
     underestimated.
     Panel B: the temperature at the mean density $T_0$ as a function of
     redshift. The black squares shows measurements from the 
     wavelet analysis of \cite{Lidz09}, the magenta circle the results of a
     Voigt-profile fitting analysis of \mlya -forest lines from Bolton et al.
     \cite{Bolton2014}. In both cases the estimates on $T_0$ 
     are fully marginalized over
     $\gamma$, but again not $\lambda_P$, such that uncertainties
     are also  likely underestimated. 
     Panel C: The evolution of the of
     the pressure smoothing scale as a function of redshift.  The red
     points are the median of the posterior distributions of
     $\lambda_P$ in each redshift bin, while the red errorbars
     indicate the 16\% and 84\% percentiles. The
     black extensions to the errorbars show the small $\sim 6\%$
     increase in the uncertainty when systematics are included \cite{supp}.  
     The curves in all three panels show predictions from
     three hydrodynamical simulations \cite{supp}.
     The black lines use the standard 
     heating rates from \cite{Haardt12}, for which reionization
     heating starts at $z=15$.
     The blue dotted lines show a late reionization heating model
     for which reionization heating does not occur until $z_{\rm reion}=7$.
     The blue line is an Increased heating model where we
     adopt the same reionization redshift  $z_{\rm reion}=15$ of the 
     HM12 model, but increase the 
     photoheating rates by a factor of three.}
\end{figure}

To understand the implications of our $\lambda_P$ measurement we
compare it to a set of hydrodynamical simulations for
several reionization scenarios and resulting thermal histories 
\cite{supp}. This enables a comparison to recent measurements of
the IGM temperature based on the thermal Doppler broadening effect
along the line-of-sight.  The standard picture of IGM thermal
evolution is based on the \cite[{\rm hereafter\ HM12}]{Haardt12} 
synthesis model of the ultraviolet (UV)
background, which provides the photoionization and photoheating rates
of IGM gas. 
In Fig.~\ref{fig:model}A the HM12 model is
compared to recent measurements
\cite{BeckerBolton2011,Boera2014} 
of the temperature $T(\Delta_{\star}$)
of the IGM at a characteristic density $\Delta_{\star}$, where
broad agreement between data and model is observed
\cite{Puchwein15}. Fig.~\ref{fig:model}B shows that recent
measurements of $T_0$ \cite{Lidz09} require higher
temperatures at $z > 3$ than the model predicts, and are in apparent
disagreement with other measurements \cite{BeckerBolton2011}. 
Fig.~\ref{fig:model}C shows that $\lambda_P$ measurements 
are overall in good agreement with HM12 model predictions.

To study the sensitivity of $\lambda_P$ to the IGMs thermal history, we run a hydrodynamical 
simulation with an abrupt step function reionization history, for
which the IGM is cold and neutral until $z_{\rm reion}=7$, after which
it experiences the HM12 photoionization and photoheating
rates. Whereas in the HM12 model the UV background heats the IGM to
$T\sim 10^{4}~{\rm K}$ already by $z \sim 15$ \cite{Puchwein15,Onorbe2016,supp},
reionization heating is delayed until $z\sim 7$ in this late reionization heating
model, as illustrated by the red curves in
Fig.~\ref{fig:model}. While its temperature is indistinguishable from the 
HM12 scenario at all relevant redshifts, 
this alternative model yields a
smaller pressure smoothing scale, illustrating that $\lambda_P$
is indeed sensitive to reionization history. The 
smaller $\lambda_P$ values in this model are discrepant at 
1.7-$\sigma$ at $z=3$, however the overall statistical disagreement,
according to the a-squared test, is only significant at the 55\% level.
Although the precision on $\lambda_P$ achieved with our current quasar
pair dataset is not sufficient to rule out either of these models, or
set tight constraints on thermal and reionization history, this comparison
nevertheless illustrates how pressure smoothing provides additional constraints
on thermal models, which are indistinguishable from temperature measurements alone.
Finally, we consider a third increased heating model, where the HM12
photoheating rates have been increased by a factor of three
\cite{supp} at all redshifts, in an effort to better match the
higher temperatures measured by \cite{Lidz09}. Fig.~\ref{fig:model}
shows that this model disagrees with our $z < 3$  measurements of $\lambda_P$ (as well as with the Becker et al. $T(\Delta_\star)$ measurements) at the 90\% confidence level under a chi-squared test.
Our measurement of the pressure smoothing scale
appears to favor the lower IGM temperatures measured by
\cite{BeckerBolton2011,Boera2014}, and provides
independent confirmation of the standard 
picture for IGM thermal evolution.

The small-scale structure of IGM baryons could be sensitive to other
physics besides the IGMs thermal history.  For example, cold dark
matter alternatives like warm \cite{VielWDM} or fuzzy \cite{FuzzyDM}
dark matter suppress small-scale fluctuations, whereas primordial
magnetic fields have the opposite effect, increasing small-scale power
\cite{Shaw2012,Pandey2013,Chongchitnan2014}. Other astrophysics
such as radiative transfer effects during HeII reionization
\cite{MeiksinTittley12} or strong feedback from galaxy formation
\cite{Kollmeier06,Desjacques2006} could also generate additional
small-scale fluctuations. Such modifications
of either the dark matter or baryons could
modify the interpretation of our pressure smoothing scale
results. The importance of these effects relative to the
standard picture whereby the smoothness of the IGM depends on
its thermal evolution driven by cosmic reionization events, can be determined by
precisely mapping the statistics of the Ly$\alpha$ forest, from
both individual sightlines and quasar pairs, over cosmic time.

\vskip -0.12in

\clearpage

\renewcommand{\thefigure}{S\arabic{figure}}
\renewcommand{\theequation}{S\arabic{equation}}
\renewcommand{\thetable}{S\arabic{table}}
\setcounter{figure}{0}

\noindent {\Huge \bf Materials and Methods} 

\section{Overview of the Phase Angle PDF Method}\label{method_overview}

This  section  provides an overview of 
the statistical method used to estimate the pressure smoothing scale
$\lambda_P$ of the IGM. At the end of the section 
we also specify where the various aspects are described in full detail, 
so that the reader can effectively use this overview as a brief guide 
to the rest of supplementary material.

Our technique is based on a statistical analysis of the transmitted
flux of the \mlya\ forest in quasar pairs.
We define the \mlya\ flux $F$ as the ratio of the observed flux $F_o$
to the unabsorbed continuum level $F_c $. 
Following the standard practice, we analyse the 
flux contrast defined as $\delta F=F/\bar{F}-1$, where $\bar{F}$
is the observed mean flux of the \mlya\ forest at a given redshift. 
We adopt the value of $\bar{F}$ from the fitting formula of \cite{faucher08}.  

We consider the flux contrast in the coeval forest of close quasar pairs.
Two \mlya\ forest pixels in different spectra are coeval if they 
are observed at the same wavelength, implying that the \mlya\ absorption
occurred at the same redshift. If two quasars in a pair have a significant
fraction of coeval forest and if the two objects are close enough (see below),
then the pair can be used to study the transverse correlation of 
the IGM. 

We Fourier-decompose the flux contrast $\delta F(v)$ in the forest 
of the members of a pair, obtaining the Fourier coefficients 
$\delta \tilde{F}_1(k)$ and $\delta \tilde{F}_2(k)$, where the 
subscripts refer to the two members of each pair.
The components are then used to calculate
the phase differences $\theta(k)$ of 
homologous Fourier modes (i.e. with the same wavenumber $k$): 
\begin{equation}
\theta(k)=\arccos\left(\frac{\Re[\delta \tilde{F}^*_1(k)
\delta \tilde{F}_2(k)]}{\sqrt{|\delta \tilde{F}_1(k)|^2|\delta \tilde{F}_2(k)|^2}}\right).
\label{eqn:phase}
\end{equation}
The alignment of these phases quantifies the coherence of the two
spectra and are the fundamental ingredient of our statistical
analysis.  Note that the normalization factor of the flux contrast
(i.e. $\bar{F})$ is divided out in this formula, making phases
insensitive to its value.  In \cite[{\rm hereafter RHW}]{Rorai13} it is shown that the
shape of the statistical distribution of these phase differences is
sensitive to the pressure smoothing scale of the IGM. This phase angle
probability distribution function (PDF) can be characterized for an ensemble
of pairs as a function of the wavenumber $k$ and of the transverse
separation $r_{\perp}$. We find that the shape of the phase angle PDF
$P_{\rm WC}(\theta)$ can be approximated as a Wrapped-Cauchy (WC) distribution
\begin{equation}
P_{\rm WC}(\theta)= \frac{1}{2\pi}\frac{1-\zeta^2}{1+ \zeta^2 - 2\zeta \cos\theta}, 
\label{WCD}
\end{equation}
which is fully characterized  by a single parameter known as 
the concentration $\zeta$. In the limit where $\zeta \rightarrow 1$,
the distribution tends to a Dirac delta function $\delta_D(x)$, which is
the phase difference distribution of identical spectra. Conversely, 
$\zeta =0 $ results in a uniform distribution, the expected distribution
for totally uncorrelated spectra. A negative $\zeta$ gives distributions peaked at
$\theta = \pi $ which are non-physical in this context. 

In RHW they used a semi-numerical model
of the \mlya\ forest based on collisionless dark-matter 
only simulations to study the 
dependencies of the phase angle PDF on $[k,r_{\perp}]$, on the
smoothing scale $\lambda_P$, as well as temperature-density
relation of the IGM defined by $T=T_0\Delta^{\gamma-1}$.
Here $T_0$ is the temperature at mean density, $\Delta$ is
the density divided by the mean of the Universe and $\gamma$ is the index 
of the relationship. 
At a fixed
wavenumber $k$, a large separation relative to the pressure smoothing
scale results in a flatter distribution of $\theta$, 
which approaches uniformity for $r_{\perp} \gg \lambda_P$ (i.e. incoherent
spectra and thus random phases). 
Conversely, the distribution approaches the fully coherent
limit of a Dirac delta function for $r_{\perp} \ll \lambda_P$ (almost identical spectra), 
and the transitions from a strongly peaked distribution to a uniform one
occurs when $r_{\perp}$ is comparable to the smoothing scale 
$\lambda_P$.  At fixed $r_{\perp}$, lower $k$-modes
(i.e. larger scales) are more correlated (smaller $\theta$
values) than high-$k$ modes. This is expected, because sight lines separated by a distance which is small relative to
the wavelength of a $k$-mode (i.e. $kr_{\perp}\ll 1$), probe essentially the
same density fluctuations associated to the wavenumber $k$. The results of RHW suggested 
that the phase difference statistic is almost  insensitive to the parameters
governing the temperature-density relationship $T_0$ and $\gamma$, whilst
having a strong dependence on $\lambda_P$.

By comparing the observed phase differences
$\{\theta \}$ in a quasar pair to
the predicted phase angle PDFs for a large set of models, we
can set constraints on $\lambda_P$. To achieve
this goal we define a likelihood $\mathscr{L}$ of the ensemble of measured phase differences for a 
given IGM model, after RHW:
\begin{equation}
 \mathscr{L}(\{\theta \}|\lambda_P,T_0,\gamma)=
 \prod_{i} P_{\rm WC}(\theta(k_i)|\zeta(k_i,r_{\perp}|\lambda_P,T_0,\gamma)), 
\label{diaglik}
\end{equation}
which exploits the form of the phase angle PDF defined in eqn.~(\ref{WCD}).
Here the product is performed over all the modes $k_i$ for
a given quasar pair with separations $r_{\perp}$. This formalism
can be easily generalized to an ensemble of pairs at different
separations $r_{\perp}$.
 The concentration parameter $\zeta$ of the WC distribution  is a function of the wavenumber $k$, the separation and 
the IGM model parameters $\{\lambda_P,T_0,\gamma\}$.
We generate a grid of thermal models by post-processing a dark-matter only
simulation (described in \S~\ref{dm_models}), assuming that the IGM is optically thin and in ionization equilibrium,
and describing the pressure smoothing with a convolution of the 
DM particle distribution with a Gaussian kernel. 
To properly evaluate the likelihood of phases as in eqn.~(\ref{diaglik}) we  
need to take into account observational effects like noise and resolution,
which may modify the distribution of phases. In addition, we must also
consider other astrophysical sources  of absorption in the forest 
like metals and strong \ion{H}{i}  absorbers. 
We follow the approach of forward modelling these effects in the simulations, 
rather than trying to subtract off their effect from the data. 

Given a grid of thermal models of the IGM, we can then explore 
the likelihood in the parameter space to obtain constraints
via standard MCMC techniques. To optimize
the sampling in parameter space, we used an irregular parameter
grid combined with an interpolation scheme based on Gaussian 
processes. 
The results of the measurement are tested against a broad
range of possible systematic effects, including a potential bias in
our estimation of the resolution or the noise, as well as our
imprecise knowledge of the number of metal line and strong \ion{H}{i}
absorbers.

As described below, we run a set of hydrodynamic simulations of the IGM
with various assumptions on the thermal and reionization history. We
use them to generate mock samples of pairs to which we apply the phase
analysis. By comparing the outcome with the results of our measurement
we are able to gain insights into the agreement between the current
theoretical picture of the IGM and the new constraints from pressure smoothing
$\lambda_P$ measured from quasar pairs.

The supplementary text is organized as follows: 
we present the data sample and the selection criteria in  
\S~\ref{sample}, with information on instruments used,
resolution, noise level, redshift and angular separation distributions. 
We then discuss the effect of 
noise and resolution on the phase distribution, whilst particular care
of justifying our assumptions on the full-width at half maximum (FWHM) 
of the resolution kernels (\S~\ref{power_spectrum_test}). 
In  \S~\ref{phase method} 
we explain how phase differences are calculated 
from the data, in particular how we address the problem of determining
the Fourier coefficients of an irregularly-sampled function. 
The semi-numerical dark matter simulation models on which we base our phase angle
PDF measurement are described 
in \S~\ref{forward modeling}, where we also show 
how observational effects such as noise, resolution, and contaminants
are forward-modelled in the simulations. 
We then briefly present the statistical tools used in determining 
the constraints on the pressure smoothing scale (\S~\ref{MCMC_interp}),
in particular the design of the parameter grid, the Gaussian process interpolation, 
and the MCMC technique. 
To facilitate the visualization of the sensitivity of this method, we 
construct plots in \S~\ref{stack_pdf}, where we stack together phase distributions 
from different separations and wave numbers 
in order to reduce the dimensionality of the statistic. 
In \S~\ref{consistency_tests} we quantify 
the effect of different sources of systematic errors
that could impact our analysis. These include continuum fitting uncertainties 
(\S~\ref{continuum_uncertainty}), the 
forward modeling of resolution (\S~\ref{res_stability}),  
noise (\S~\ref{noise_stability}), and of the presence
of contaminants (\S~\ref{contaminant_stability}).
Finally, in \S~\ref{hydro comparison}  we explain
how we apply the phase analysis to 
a set of hydrodynamical simulations, allowing the comparison 
of the transverse coherence in real data to the prediction
of realistic IGM models. 

\section{Experimental Design and the Quasar Pair Dataset}\label{sample}

\subsection{Quasar Pair Survey}

Our goal is to characterize the correlated
HI \mlya\ forest absorption in close quasar pairs at $2
\lesssim z \lesssim 4$ with separations $< 450$ kpc (comoving)
corresponding to
angular separations $\psi \lesssim 18''$ (depending on redshift).  
As quasar pairs are rare, we have leveraged several large photometric and spectroscopic survey datasets
to find the pairs, and then performed dedicated follow-up observations on a number of
large telescopes to obtain higher quality data for the \mlya\ forest
phase correlation analysis.

The starting point of our experiment is to identify close quasar pairs
with angular separations $\psi$ corresponding to comoving
separations of $r_{\perp} < 450$\,kpc (here and in the following, the conversion between angular separation and impact parameter is done assuming the same cosmology specified in \S~\ref{dm_models}).
To find the pairs, we mine the
photometric and spectroscopic databases of the Sloan Digital Sky
Survey \cite[{\rm SDSS}]{SDSS09}, the Baryonic Oscillation
Spectroscopic Survey \cite[{\rm BOSS DR12}]{BOSS12}, and the 2dF quasar
redshift survey \cite[{\rm 2QZ}]{2QZ} surveys.
Modern spectroscopic surveys
bias their selection select against close pairs of quasars to avoid fiber
collisions.  For the SDSS and BOSS, the finite size of optical fibers
precludes discovery of pairs with separation $<55\arcsec$ and
$<62\arcsec$, corresponding to to 1.37\,Mpc and 1.55\,Mpc at $z=2$,
whereas for the 2QZ survey the fiber collision scale varies from $\sim
30-40\arcsec$ corresponding to $750-1000$\,kpc.  In the regions where
spectroscopic plates overlap, this fiber collision limit can be
circumvented. However, only $\approx 30\%$ of the SDSS
spectroscopic footprint and $\approx 40\%$ of the BOSS footprint are
in overlap regions.  Unfortunately, small separation quasar pairs are
rare enough that these overlap regions are not sufficient for building
up large samples of quasar pairs close enough for our purposes.

To circumvent this fiber collision limitation, we have conducted
a comprehensive spectroscopic survey to discover additional close
quasar pairs and to follow-up the best examples for our scientific
interests. Close quasar pair candidates are selected from the
photometric quasar catalogue of \cite{bovy11,bovy12} as well as its recent
extension by \cite{Dipompeo15}, and are confirmed via spectroscopy on 4m
class telescopes including: the 3.5m telescope at Apache Point
Observatory (APO), the Mayall 4m telescope at Kitt Peak National
Observatory (KPNO), the Multiple Mirror 6.5m Telescope, and the Calar
Alto Observatory (CAHA) 3.5m telescope. Our continuing effort to
discover quasar pairs is described in \cite{Hennawi04}, \cite{BINARY},
and \cite{HIZBIN}.

Our quasar pair survey has gathered
science-grade follow-up optical spectra on large-aperture telescopes
using spectrometers with a diverse range of capabilities. This
includes data from Keck, Gemini North and South, Magellan, the Large
Binocular Telescope (LBT), and the Very Large Telescope (VLT). This large spectroscopic
quasar pair data set constitutes the parent sample for the final data
set used in this work.

\subsection{Spectroscopic Observations}

After applying the selection criteria (see \S~\ref{sample_definition}) for the 
pressure smoothing scale 
measurement to our quasar pair spectroscopic database, only spectroscopy from Keck, Magellan, and
the VLT contribute to the final sample. We describe each of these observational setups in turn.

At the W.M.~Keck Observatory, we have exploited two optical
spectrometers to obtain spectra of quasar pairs. One of these was the
Low Resolution Imaging Spectrograph \cite[{\rm LRIS}]{LRIS}, which we
used to observe 7 of the quasar pairs studied here during a 
series of observing runs in 2004-2008. We generally used the
multi-slit mode with custom designed slit masks that enabled the
placement of slits on other known quasars or quasar candidates in the
field.  LRIS is a double spectrograph with two arms giving
simultaneous coverage of the near-UV and red.  For our current science
goals only the near-UV side (LRIS-B) is relevant, since it covers the
$\lya$ forest. We used the D460 dichroic with the $1200$ lines
mm$^{-1}$ grism blazed at $3400$\,\AA\ on the blue side, resulting in
wavelength coverage of $\approx 3300-4200$\,\AA, a dispersion of
$0.50$~\AA\ per pixel, and the $1''$ slits give a FWHM resolution of
about $170\mkms$. About half of our LRIS observations were taken after the
atmospheric dispersion corrector was installed, which reduced
slit-losses
in the UV.
More information about the observations, data reduction, and a detailed
observing log are given in \cite{QPQ6}. 

At Keck we also used the Echellette Spectrometer and Imager
\cite[{\rm ESI}]{ESI} to obtain spectra of \nesi\ quasar pairs
analysed here.  ESI provides continuous spectral coverage from
4000\,\AA\ to 10000\,\AA\ at a resolution of FWHM = $64 \mkms$. These
data have been previously analyzed for \ion{C}{iv} correlations
between neighbouring sight lines \cite{Martin2010} and for the
analysis of an intriguing triplet of strong absorption systems
\cite{Ellison2007}.  Those papers give  more
information about the data acquisition, reduction, and the
observing logs.

Observations for \nmagell\ pairs used in this work were obtained with
the Magellan Clay telescope using the Magellan Echellette Spectrograph
\cite[{\rm MagE}]{MAGE} during the nights of Universal Time (UT) 
2008 January 07-08, UT
2008 April 5-7, and UT 2009 March 22-26. These data
cover the wavelength range $\lambda \approx 3050 - 10300$ \AA\, and
have a FWHM = 62$\mkms$ or 51 $\mkms$, depending on the 
slit aperture employed. For more
more information about the Magellan observations as well the 
observing logs see \cite{QPQ6}. 

The XSHOOTER spectrograph \cite{Vernet2011} at VLT provided 
\nxshooter\ of the pairs of our sample. XSHOOTER's three arms
provide wavelength  coverage in the range between 
3000 and 25000 \AA. One pair
(SDSS~J000450.90-084452.0, SDSS~J000450.66-084449.6 ) has been observed in a program dedicated to this project.
The slit aperture used was 0.5\arcsec, yielding an estimated  
resolution of 30$\mkms$ FWHM. The reduction of the data was performed
using a custom data reduction pipeline 
kindly made available to us by George Becker. Data for one of the pairs
used in our analysis (SDSS~J091338.97-010704.6, 
SDSS~J091338.30-010708.7) was obtained from the VLT/XSHOOTER archive 
(program ID:089.A-0855, PI: Finley,H.) and
reduced. This slit aperture for these observations was 1.0\arcsec resulting in a
slightly lower resolution of 69$\mkms$ FWHM.

\subsection{Sample Definition}\label{sample_definition}

Here we present the selection criteria that we imposed on our quasar pair spectroscopic
database to arrive at the final dataset analyzed in this paper.

We first applied a broad cut to select quasar pairs suitable for the
characterizing correlated \mlya\ forest absorption.  An obvious
prerequisite is the existence of a segment of overlapping \mlya\ forest
between the two quasars in the pair, which can be expressed as $(1+
z_{\rm fg})\lambda_{\rm \lya} > (1+z_{\rm bg}) \lambda_{\rm \lyb}$.
Here $\lambda_{\rm \lya}=1215.67\,{\rm \AA}$ and $\lambda_{\rm \lyb}=1025.72$ are the rest-frame wavelengths
of $\lya$ and $\lyb$ that define the region of usable \mlya\ forest, and
$z_{\rm fg}$ and $z_{\rm bg}$ are the redshifts of the foreground (f/g) and
background (b/g) quasar, respectively. 
To avoid cases
where this segment is too small to contribute meaningfully to the statistics,
we define the overlapping fraction $f_{\rm ov}$ of the \mlya\ forest as
\begin{equation}\label{eqn:fov}
f_{\rm ov}= \frac{(1+z_{\rm fg})\lambda_{\lya} - (1+z_{\rm bg})\lambda_{\lya}}
{(1+z_{\rm fg})(\lambda_{\lya}-\lambda_{\lyb})}
\end{equation}
and we set a lower threshold at $f_{\rm ov} = 0.3$, removing in this way 
projected quasar pairs with large redshift separations. 

A second cut is applied to the transverse separation of the pair
$r_{\perp}$, evaluated at the redshift of the f/g quasar $z_{\rm fg}$.
RHW showed that the
most informative pairs are those with impact parameter comparable to
the pressure smoothing scale.
Existing small-scale measurements of the
line-of-sight power spectrum of the \mlya\ forest \cite{Zald01,Lidz09} exclude
pressure smoothing scales larger than $\gtrsim 300-400\,{\rm kpc}$, hence we restrict
our analysis to pairs with $r_{\perp} < 450\,{\rm kpc}$ (comoving).

Based on the redshift and impact parameter distribution of the data, we decided to
divide the data in a set of redshift bins:
$1.8-2.2, 2.2-2.7, 2.7-3.3$ and $3.3-3.9$.
The lower limit $z=1.8$ is set to avoid the forest close to the atmospheric
cutoff (at $\lambda \sim 3000$ \AA), and the bins are wider at higher $z$ to enclose a sufficiently 
large sample of pairs.

To avoid contamination from the
quasar proximity zone we consider only rest-frame wavelengths blueward
of $\lambda_{\rm \lya,max}=1190$ \AA . Given the particularly large redshift
uncertainty $\sim 1000\kms$ of the quasars in our dataset,
contamination from Ly$\beta$ absorption is also a concern, and we
restrict attention to rest-frame wavelengths redward of
$\lambda_{\rm \lya, min}=1040$. 
The coeval
forest in a pair on which we calculate phases is thus defined in the range 
$\lambda \in [\lambda_{\rm \lya,min}(1+z_{\rm bg}),
\lambda_{\rm \lya,max}(1+z_{\rm fg})]$, 
which is narrower than the one implied by the $f_{\rm ov}$ defined above.
The corresponding redshift range is delimited by 
$z_{\rm min}=\lambda_{\rm \lya,min}/\lambda_{\rm \lya }$ and 
$z_{\rm max}=\lambda_{\rm \lya,max}/\lambda_{\rm \lya }$.

The set of pairs is then visually
inspected in order to find contaminants. Specifically, some quasars
exhibit  strong associated absorption lines known as Broad Absorption Lines (BAL),
which are thought to be produced in the vicinity of the black hole and 
may reach velocities up to $v\sim 10,000$ km/s. For this reason they
could be blueshifted into the \mlya\ forest, causing blending with IGM 
absorption. Since we are not able to model this blending, we remove
from the sample all the pairs in which one of the two spectra is
contaminated by BAL.

Strong absorption lines with neutral hydrogen column density 
$N_{\rm HI} \gtrsim 10^{17}\,{\rm cm^2}$,
are believed to be associated with galaxies or their circumgalactic
media. These so called Lyman Limit systems (LLSs) are a significant
source of contamination, since their correlation properties across
quasar pair sight lines are determined by the details of galaxy formation,
rather than by the simpler low-density hydrodynamics governing the
$\lya$ forest. Standard practice in studies of $\lya$ forest
statistics is to mask out these LLS, forward model them, or some
combination of these approaches \cite{KG2015}. The strategy we
adopt here is to identify and mask all objects for which we can
clearly see damping wings in the spectrum, which corresponds
approximately to a column density threshold of $N_{\rm HI}\gtrsim 10^{19}
{\rm cm}^{-2}$, which encompasses Damped Lyman-$\alpha$ systems (DLAs)
($N_{\rm HI} > 10^{20.3} {\rm cm}^{-2}$) as well as so called super-LLSs
($10^{19}\,{\rm cm}^{-2} < N_{\rm HI} <10^{20.3} {\rm cm}^{-2}$). As our
algorithm for measuring phase angles requires contiguous spectral
coverage, when strong absorbers are present, we simply split the
unmasked data into multiple segments specified by
redefining  $z_{\rm min}$ and $z_{\rm max}$ for each of them. 
Column densities in the range
$10^{17}\,{\rm cm}^{-2} < N_{\rm HI} < 10^{19} {\rm cm}^{-2}$ cannot be
reliably identified in spectra of our wavelength coverage, ${\rm
  S\slash N}$ ratio, and resolution, so rather than masking, we
directly forward model such absorbers by adding them to our models.
In \S~\ref{contaminants} we discuss the details of this procedure, and the
impact of these strong absorbers on our results in detail.

At the end of this process we obtain a list of paired
wavelength regions of coeval 
\mlya\ forest, delimited by $z_{\rm min}$ and 
$z_{\rm max}$. Each quasar pair may have more than a single 
segment. 
For each of these segments, we evaluate the signal-to-noise ratio  
per observed frame Angstrom ${\rm S}\slash {\rm N}_{\rm
1\AA}$ in the two spectra,  
between $z_{\rm min}$ and $z_{\rm max}$.
We only include in the final sample those pairs for 
which both members of the pair have
${\rm S}\slash {\rm N}_{\rm 1\AA} > 10$. To determine ${\rm S}\slash
{\rm N}_{\rm 1\AA}$, we first compute the median signal-to-noise ratio 
 per pixel ${\rm S}\slash {\rm
N}_{\rm px}$ of the spectrum over the interval $[(1 + z_{\rm
min}),(1 + z_{\rm max})]\lambda_{\rm \lya}$.
We then multiply ${\rm S}\slash {\rm N}_{\rm px}$ by the factor
$\sqrt{1\,{\rm \AA} \slash \Delta\lambda_{\rm px}}$ to put the different
spectral resolution data utilized in this study onto a common scale,
where $\Delta\lambda_{\rm px}$ is the median pixel width in each
spectrum.

The vast majority of the quasar pairs identified by our survey 
are binaries with small redshift separations, or
projected quasar pairs at different redshifts.  At the angular
separations we consider $\sim 4-10\arcsec$, only a small fraction of
quasar pairs at the same redshift are expected to be gravitational
lenses, i.e. a double image of the same source, owing to the relative
paucity of such wide-separation lenses \cite{Hennawi2007}.  We
nevertheless conducted a literature search for all quasar pairs
selected by the criteria above and found two which were known to be
gravitational lenses. Of the remaining quasar pairs, none have spectra
of sufficient similarity to warrant the lensing hypothesis, nor did
any show evidence for a lens galaxy in the SDSS imaging (for such wide
separation lensed quasars, the lens is often visible as a group of
galaxies even in the relatively shallow SDSS imaging).  In principle,
lenses can also be used to study the coherence of the IGM if the
lens redshift is precisely known, in which case the dependence of the
impact parameter with redshift can be easily modelled. However, lenses
typically probe small impact parameters at \mlya\ forest
redshifts of $100\,{\rm pc}-10\,{\rm kpc}$ \cite{Rauch01}.
The smallest separations are likely insensitive to the pressure smoothing
scale, whereas even for larger $\sim 10\,{\rm kpc}$ separations, the
analysis is likely much more sensitive to incoherence caused by LLSs
and/or metal line absorbers (see \S~\ref{contaminants}) which would require
a much more careful treatment and modeling procedure. For this reason,
we discard the pairs identified as gravitational lenses.

The final sample obtained from our selection criteria  is
illustrated in Fig.~\ref{rz_dist}. 
The lines trace the coeval \mlya\ forest in the pairs,
following the evolution of the impact parameter as a function of
redshift.  The extent of the overlapping segments depends on the
respective redshifts of the quasar in the pair (see
eqn.~(\ref{eqn:fov})), as well as on our masking of strong absorbers,
which appear as gaps in the lines.  A
complete list of the coeval \mlya -forest segments is provided in
Table
\ref{tab:sample}, together with all the relevant parameters for each pair.
The redshift binning determines a further split of the sample, 
as a result of intersection of each of the segments with the four $z$ bins. 
All intersections that span a redshift interval smaller than $\Delta z=0.08$  are discarded.

\subsection{Continuum Fitting and Data Preparation}

We fitted the continuum manually using a fitting algorithm that
performs a cubic spline interpolation between manually inserted break
points, resulting in a continuum tracing the undulations and
emission features of the quasars which are obvious to the reader. 
These features, of course, are more
easily discerned in the higher ${\rm S\slash N}$ spectra.
This led, in part, to
the imposed {\rm S\slash N} criterion of our sample.  At the typical spectral
resolution of our data, one generally expects the normalized flux to
lie below unity (in the absence of noise) owing to integrated
\mlya\ opacity from the IGM.  We took this into full consideration when
generating the spline continuum and also allowed for the expected
increase in \mlya\ opacity with increasing redshift. We emphasize that
phase correlation statistic that we employ to measure the pressure smoothing scale
is not particularly sensitive to errors in the continuum-placement, as we explicitly
demonstrate in \S~\ref{consistency_tests}. 

Since our analysis involves a statistic computed in 
velocity space, we transform the wavelengths into velocities
according to
\begin{equation}
\Delta v=c\log(\lambda/\lambda_0),
\end{equation}
where $c$ denotes the speed of light, 
$\Delta v$ is the relative velocity between two points responsible
for resonant \mlya\ absorption at observed wavelengths $\lambda$ and
$\lambda_0$. Here $\lambda_0$ is an arbitrary reference wavelength, typically the
lowest in a segment, taken as the origin of the velocity space. 
For matter moving with the Hubble flow, this velocity $\Delta v$ corresponds
to a comoving distance of
\begin{equation}
\Delta x= \frac{(1+z)\Delta v}{H(z)}.
\end{equation}
where the $H(z)$ is the Hubble parameter at the observed redshift,
which is calculated assuming  the same cosmology specified in 
\S~\ref{dm_models}.

\section{Noise and Resolution}\label{power_spectrum_test}
  
The distribution of phases may deviate from its intrinsic shape determined
by IGM physics, depending on the level of noise and the spectral resolution
of the observations.  Specifically, noise results in  
scatter of the Fourier components into which we decompose the spectra, randomly
changing both their moduli and their phases. Since phases are scattered 
independently in the two spectra of a quasar pair,  
noise will broaden the phase difference PDF. The broadening will be 
more evident in the high-$k$ modes, which are damped by the resolution 
cut-off and whose signal is therefore dominated by the noise. These considerations
call attention to the estimates of the noise and resolution of our spectra,
which are necessary inputs to our  forward modelling procedure, which will be 
described in \S~\ref{forward modeling}.

The data reduction pipelines used to reduce the data deliver accurate
estimates for the noise, which takes into account the distinct
contributions from photon counting noise from both the sky background
and the object, as well as detector readout noise. The accuracy of the
algorithms used to estimate the spectral noise in our data reduction
pipelines were recently characterized in \cite{QPQ4}, where it was
found that the pipeline delivers noise estimates that are accurate to
$\lesssim 10\%$.

The resolution is a more delicate issue. For slit spectroscopy, the
width of calibration spectral lines provided by an arc lamp,  
taken during daytime, can be used to
determine the resolving power of the spectrometer for a source
(i.e. the arc lamps) which uniformly illuminates the spectrograph
slit. We refer to this resolution measured from arc lines as the slit
resolution of the spectrometer. For Keck/LRIS we directly measure this
resolution from the FWHM of arc-lines, whereas for echelle
spectrometer Keck/ESI, Magellan/MagE, and VLT/XSHOOTER we adopt the
slit resolution reported in the instrument manuals (also based on
arc-line widths) for the order in question and the slit used. These values
of the slit resolution are listed as the FWHM in Table~\ref{tab:sample}. 

However, the true resolution of a spectrum of a point source will
depend on the seeing during the observation, and could be
significantly smaller than the slit resolution if the data is obtained
in good seeing conditions. For these reason we expect the value of the
FWHM listed in Table~\ref{tab:sample} to typically
overestimate the true resolution of the spectra.

To assess the accuracy of these resolution estimates we compute the
line-of-sight power spectrum of the flux contrast $\delta F$ and
compare it with a measurement from a distinct high-resolution quasar sample. 
The idea behind this test
is that smearing of the spectra due to the resolution of the
spectrograph modifies the flux power spectrum in a known way. Given an
estimate for the resolution, we can undo the effect of this smearing
in our data, and compare our resulting resolution-corrected power
spectrum to the power spectrum of the \mlya\ forest from an independent
fully resolved dataset. If, as we suspect, the slit resolution FWHM is larger than
that of our actual data, this will be manifest as a mismatch between
these two power spectra, and we can correspondingly adjust our resolution
estimate to the value that produces the best match. In addition to calibrating
our spectral resolution, this comparison provides a basic
sanity check on the various steps of the procedure (data reduction, noise
estimates, continuum normalization, masking strong absorbers) required to generate
the $\delta F$ fields from our spectra.
 
We consider a sample of quasars composed of 38 spectra observed with the Ultraviolet and Visual Echelle Spectrograph  
\cite[{\rm UVES}]{Dallaglio2008} and 37 spectra from the
Keck Observatory Database of Ionized Absorption 
\cite[{\rm KODIAQ}]{Omeara2015} project observed with the High Resolution Echelle
Spectrometer \cite[{\rm HIRES}]{HIRES}. We 
select objects with S/N $>20$ per 6 km/s interval inside the usable
\mlya\ interval. The resolution of the sample is 6$\mkms$ FWHM, except
some of the HIRES spectra which have FWHM of 3$\mkms$. 
The sample covers the redshift range 
between $z=1.7$ and $z=3.5$.

Similar to our approach, DLAs are identified in the high-resolution
data through their damping wings and masked. Metals absorbers are
identified redwards of the \mlya\ line by looking for doublets lines
for common transitions (\ion{Si}{iv}, \ion{C}{iv}, 
\ion{Mg}{ii}, \ion{Al}{iii}, \ion{Fe}{ii}). We then mask
spectral regions of the \mlya\ forest where we expect other lines
originating from these absorbers for different possible metal
transitions (removing 60$\mkms$ around the central redshift).

In order to calculate the flux power spectrum of both the quasar pair data and the
high-resolution data, we follow 
the approach of \cite{BOSSpower}, which we summarize here. 
The flux contrast $\delta=F/\bar{F}-1$  is considered to be the sum of 
the contribution of the forest and noise: $\delta=\delta_{F}+
\delta_n$. Under the assumption that the Fourier modes of the two
components are uncorrelated, we can decompose the total power as 
the sum of the noise power and of the forest power. The latter can
therefore be calculated as
\begin{equation}
P_{\rm 1D}(k)=\frac{P^{\rm raw}(k)-P^{\rm noise}(k)}{W^2(k,\sigma_{R},\Delta v)}
\end{equation}
where $P^{\rm raw}(k)$ is the power spectrum of the observed spectrum and
$P^{\rm noise}(k)$ is the power spectrum of the noise estimated from
the pipeline. $W$ is a filtering function which models the response
of the spectrograph, and depends on the wave number $k$, on the resolution
width $\sigma_R$ and on the pixel scale $\Delta v$ in the following way
\begin{equation}
W(k)=\exp\left(-\frac{(k\sigma_R)^2}{2} \right)\times \frac{\sin(k\Delta v/2)}
{k\Delta v/2} .
\label{eq:power_spectrum_correction}
\end{equation}

The only relevant difference between our approach and
the method of \cite{BOSSpower} is that our spectra do not
have a regular binning in velocity space, therefore we need to use 
the Lomb-Scargle periodogram to compute the power, along the lines of
the technique described in \S~\ref{sec:LSFA}.

When calculating the flux power spectrum of the quasar pair sample 
we divide the data into the three lower redshift bins used in the phase
analysis, i.e.
$\Delta z= 1.8-2.2, 2.2-2.7, 2.7-3.3$ (called ``coarse'' bins
in the rest of this section). 
We ignore the highest-redshift bin 
for this study, because of the small number of quasar pairs in this bin and
the fact that our high-resolution dataset does not extend beyond $z = 3.5$,
precluding a meaningful comparison. We will thus assume that the
results we obtain about the resolution can nevertheless be extended beyond $z=3.3$.  The
power spectrum is first calculated separately on the segments of
\mlya\ forest we defined in \S~\ref{sample_definition}, and
subsequently averaged in logarithmic bins in $k$ and over all
segments. Uncertainties are calculated by bootstrapping over these
segments. Since the forest of quasar pairs is correlated, we are
probably underestimating the errors in this way.  However, the aim of
this section is not to do a rigorous statistical analysis but only to
verify the approximate agreement of the power spectrum of the pair
sample with high resolution data.

When calculating the power spectrum of the UVES and HIRES sample we
initially adopt redshift bins spaced by $\Delta z=0.2$ between $z=1.7$
and $z=3.5$. We will refer to these small intervals as 
$\Delta_{j,\rm HR}z $, identified by the index $j$, to 
distinguish them from the redshift bins relative to the 
pair analysis.
The power and the relative errors are then obtained in
the same way as for the quasar pair data.  However, the redshift bins
in which the quasar pair data are subdivided are coarser, and the pairs path length distribution is not uniform across these bins.
Since the power level evolves with redshift, we need to take into
account the redshift distribution of the forest path length in
coarse bins when comparing the quasar pair power spectrum to the high-resolution data.
To do this, we start from
the high-resolution power in the $\Delta_{j,\rm HR} z=0.2$ bin, and we
then use a weighted average to calculate the corresponding power in the
coarser redshift intervals in which the pair sample is binned. The weights depend on the path length
of the forest in pairs within each $\Delta_{j,\rm HR}z $ bin.  
More precisely we define
\begin{equation}
w_{j}=\frac{\sum_i |\Delta_i z \cap \Delta_{j,\rm HR} z|}
{\sum_i |\Delta_i z|} ,
\end{equation}
where $\Delta_i z$ is the redshift interval spanned by the $i$th segment
of the pair sample in the considered coarse redshift bin and the sum is performed over all
the segments in the coarse bin. Note that the denominator is equal to the 
total path length of the \mlya\ forest in the quasar pairs
sample for the said bin.
The power from the high resolution sample is therefore computed as 
\begin{equation}
P_{\rm coarse}(k)=\sum_j P_{j,\rm HR}(k)w_j
\end{equation}
where $P_{j,\rm HR}$ is the power spectrum of the high-resolution
data in the  $\Delta_{j,\rm HR}z $  redshift interval. 
This is repeated for each coarse bin.

The results are shown in Fig.~\ref{fig:power_spectrum}, 
together with literature values of the power calculated from the BOSS
sample \cite{BOSSpower}.
The BOSS power spectrum was measured in the same
 $\Delta_{j,\rm HR} z $ bins as our high-resolution data, so we 
applied  the same weighted average described above to produce 
this figure. The BOSS dataset only samples redshift $z>2.1$,
for which reason we could not calculate the relative power in the
lowest-redshift coarse bin.  
The power spectrum
calculated from our pair sample is in broad agreement
with the power from high-resolution data,
but has a clear 
excess at the high-$k$ end. We attribute this discrepancy to an underestimation
of the resolution, following the argument presented above. In fact, if the 
resolution is underestimated, the filtering correction to the power in
eqn.~\ref{eq:power_spectrum_correction} would overcompensate for 
the resolution cut at high-$k$, adding spurious power.  

We choose to apply a correction factor to all of our resolution estimates
in order to improve the match of the power spectra. By visual inspection, 
we find that if we decrease the FWHM of all our pairs by 20\%, the agreement between the quasar
pair and high-resolution power spectra is better
(Fig.~\ref{fig:power_spectrum}). This correction
defines our default value for the resolution assumed throughout
the text.
The choice of the correction factor is not made via a rigorous 
statistical analysis, and is somewhat arbitrary. Furthermore, we are 
adopting a fixed factor for all data, while each pair has been observed
under particular conditions and using a specific slit width. These simplifications
are ultimately justified by the weak sensitivity of phase differences to 
resolution, as we show in a consistency test a posteriori (see \S~\ref{res_stability}).

\section{Phase Calculation on Pair Spectra}\label{phase method}

Our method relies on the calculation of phase differences of Fourier
components. However, Fourier transforming the 
the \mlya\ forest of real observed spectra requires further elaboration.
While in simulations we generate 
mock spectra on perfectly regular grids in velocity space, the pixels
of observed spectra are in most of the cases unevenly spaced. 
Since the discrete Fourier transformation is defined for evenly-sampled
functions, we have to either interpolate or rebin the data onto a
regular grid, or to use approximate methods without modifying
the sampling. The two methodologies have opposite advantages
and disadvantages, so we decide to implement both and to check that
they lead to consistent results (\S~\ref{test_LSSA}). 
This test will also demonstrate that our results are not affected by 
pixel rebinning at the data-reduction level. 

\subsection{Method 1: Least-Square Spectral Analysis}
\label{sec:LSFA}
A widely-used approach to generalize Fourier transformation 
to irregularly-sampled series is the so called least-square
spectral analysis (LSSA). It consists of 
fitting a function $f(x_i)$ with a linear combinations of 
trigonometric functions $\cos(k_jx_i)$ and $\sin(k_jx_i)$, 
where $\{x_i\}$ is the set of points where $f$ is sampled and
$\{k_j\}$ are the wave numbers of the modes  that we want 
to fit. 
This leads for example to the Lomb-Scargle periodogram \cite{Lomb76}, 
a method often employed to calculate the power spectrum of 
a signal. It is also possible to follow this strategy to 
recover the phase information, which is what we want to
calculate in quasar spectra. We follow
for this purpose the method described in \cite{Lomb_phases}
which we  briefly summarize here. 

The Fourier decomposition is given by the minimization,
for each different $k_j$ of 
\begin{equation}\label{LSFA}
|| f({\bf x}) - c_j \cos(k_{j}{\bf x}) + s_j\sin(k_{j}{\bf x}) ||
\end{equation}
where in our case $\bf{x}$ is the array of the velocity-space pixels in a spectrum,
 $f$ is the transmitted flux of the \mlya\ forest, $|| g({\bf x}) ||=\sum_i g^2(x_i)$ 
denotes the squared norm and $(c_j s_j)$ are the coefficients that we need
to determine. Put another way,
we want to find the projection of $f$ on the functional subspace defined
by the linear combinations of $\cos(k_j{\bf x})$ and $\sin(k_j{\bf x})$. In the case
where $x_i$ are evenly spaced and $k_j=2\pi j/L$, with $L$ being the
total length of the spectrum, this is equivalent to the standard  Fourier decomposition. For generic
$\{x_i\}$ and $\{k_j\}$ the linear subspaces relative to different $k$ may
not be orthogonal and may not form a complete functional basis, so this 
fitting procedure cannot be properly regarded as a decomposition.

The minimization of eqn.~\ref{LSFA}
is obtained via the Moore-Penrose
pseudo-inverse matrix \cite{Penrose55} applied to the linear
system
\begin{equation}
f({\bf x}) = (c_j\ s_j) \Omega_j
\end{equation}
where $\Omega_j$ is defined as 
\begin{equation}
\Omega_j =\left(\begin{array}{ccc}
\cos(k_jx_1) & ... & \cos(k_jx_n)\\
\sin(k_jx_1) & ... & \sin(k_jx_n)\\
\end{array}\right)  .
\end{equation}
The pseudo inverse is then $\Omega_j^+=\Omega_j^T(\Omega_j\Omega_j^T)^{-1}$
and the coefficients are estimated by 
\begin{equation}\label{pseudo_invertion}
(c_j\ s_j)= f({\bf x})\Omega_j^+ .
\end{equation}
According to the pseudo-inverse properties, these coefficients are exactly the 
ones that minimizes $|| f({\bf x}) - (c_j\ s_j) \Omega_j ||$, i.e. eqn.~\ref{LSFA}. 
When the system has a solution  this norm is zero, but for our problem
this is never the case. This definition of the pseudo-inverse
requires that $\Omega_j\Omega_j^T$ is invertible, which is however always satisfied
for reasonable pixel distributions. 

By explicitly writing
eqn.~\ref{pseudo_invertion} we obtain
\begin{equation}
(c_j\  s_j) =\left(\begin{array}{c}
\sum_i f(x_i)\cos(k_jx_i)\\
\sum_i f(x_i)\sin(k_jx_i)\\
\end{array}\right)^T
\left( \begin{array}{cc}
\sum_i\cos^2(k_jx_i) & \sum_i\cos(k_jx_i)\sin(k_jx_i)\\
\sum_i\cos(k_jx_i)\sin(k_jx_i) & \sum_i\sin^2(k_jx_i)
\end{array}
\right)^{-1}
\end{equation}
where the diagonal terms are non-zero because $\sin(k_j\bf{x})$ and 
$\cos(k_j\bf{x})$ are not orthogonal in general. 
Nevertheless it is possible to apply a phase shift to the coordinates such that, for 
a given $k_j$, the non diagonal terms vanish \cite{Lomb76}. 
It can be shown that the shift is equal to 
\begin{equation}
T_j=\frac{1}{2k}\arctan \frac{\sum_i\sin(k_jx_i)}{\sum_i\cos(k_jx_i)}.
\end{equation}
After diagonalization, the equation above simplifies to
\begin{equation}
(c_j\ s_j)= \left( \frac{\sum_if(x_i)\cos(k_j(x_i-T_j))}{\sum_i\cos^2(k_j(x_i-T_j))}\ \  
\frac{\sum_if(x_i)\sin(k_j(x_i-T_j))}{\sum_i\sin^2(k_j(x_i-T_j))}\right), 
\end{equation}
which is the expression we are looking for. 
The power spectrum immediately follows from this result as $P(k_j) = c_j^2+s_j^2$.

If we need to recover phase information we must
consider that phases are changed by the Lomb shift, therefore we 
have to apply the inverse translation at each $k$. 
This is easily done 
by defining the Fourier coefficients in the complex representation as
\begin{equation}
F(k_j)=(c_j +i s_j) e^{ik_jT_j} .
\end{equation}

We are now ready to calculate phase differences in the usual way
\begin{equation}
\theta(k) = \arccos\left(\frac{\Re[\tilde{F}_1^*(k)\tilde{F}_2(k)]}{|\tilde{F}_1(k)||\tilde{F}_2(k)|}\right)
\end{equation}
where $F_1$ and $F_2$ are the transmitted fluxes of the \mlya\ forest
in the two spectra of the pair.

A final caveat concerns non-orthogonality: the Fourier components
do not form an orthogonal basis if the pixel spacing is irregular. 
For this reason, the cross products $C_{l,m}=\sum_i 
\exp(-i(k_m-k_l)x_i)$ could be different from zero, therefore inducing a 
systematic correlation between the Fourier coefficients
and potentially a correlation between phase differences. 
We anticipate that this effect should be important only at scales 
comparable to the pixel separation, however we force the 
orthogonality of the components in the following way: 
after calculating the coefficient $\tilde{F}(k_0)$
we subtract the corresponding component from the original function
by defining
\begin{equation}
F'({\bf x})=F({\bf x})-\tilde{F}(k_0)e^{-ik_0{\bf x}}.
\end{equation}
Then we calculate the next coefficient $\tilde{F}(k_1)$  on the residual
function $F'({\bf x})$. We iterate this procedure until all the coefficients are calculated.
This algorithm is a standard orthogonalization process, and requires 
specifying the order on which the components are subtracted. The most natural
choice for us is starting with the large scale modes, i.e. with the lowest wavenumber,
which are the least affected by noise and resolution. This procedure cannot
obviously remove intrinsic correlations between different Fourier modes
originated from cosmological or physical processes. However it was
shown in RHW that these are negligible in the relevant range of wavenumbers.

\subsection{Method 2: Rebinning on a Regular Grid}\label{interpolation}
A second possibility is to rebin the observed flux pixels onto a
regular grid, to allow the standard calculation of the Fourier
coefficients. The advantage of this method is that we avoid
approximations deriving from the least-square evaluation of the
phases, but on the other hand, we do not have a clear picture of how
the rebinning modifies the Fourier phases. The pros and cons of this
approach are complementary to the LSSA procedure described in the
previous section, therefore we decide to adopt both of them and check
that the results are consistent, assuring in this way that our
approximate calculation of phases is not a source of bias
(\S~\ref{test_LSSA}).

In order to consistently calculate phase differences, not only is it
necessary to bin the pixels of each spectrum onto a regular grid in velocity
space, but also to use the same regular grid for both spectra of
the pair.  We define the common regular grid from the original arrays
via a simple procedure. For a single spectrum with $N$ irregular
pixels located at $\{v^0_i \}$, the step of the regularized array
is $\Delta v=(v^0_{N}-v^0_1)/(N-1)$ and the full vector
 $\{v_i=v^0_1+i\Delta v\, \forall i=0..N\}$.  When considering two spectra with
different pixels arrays $\{u^0_i\}$ and $\{w^0_i\}$, having
respectively $N$ and $M$ points, we define the grid in the common
velocity interval $I=[\max(u^0_1,w^0_1),\min(u^0_N,w^0_M)]$. We then
count the number of pixels encompassed within this interval for each
of the two spectra, and we take the smallest of the two numbers to be the
cardinality of the common grid grid $N_g$. In this way we avoid
oversampling in the rare cases where one spectra is observed with a
finer pixel scale than the other.  The spacing is then simply
$|I|/(N_g-1)$, where $|I|$ is the total length of the interval.  We
finally rebin the transmitted fluxes onto the newly-defined pixel
vector and we are set to compute the phase differences by standard
Fourier analysis.

\subsection{Methods Comparison}\label{test_LSSA}

We have presented two possible ways of 
calculating phases of irregularly sampled functions: one employs
least-square spectral analysis (LSSA), the other rebins the function onto a regular grid
and applies the standard discrete Fourier transform. 
Since the two methods imply complementary approximations, checking that
they lead to consistent phase distributions is a good test of the accuracy
of these calculations.
In Fig.~\ref{fig:method_comparison} we show the distributions of the 
observed  phases binned in $k$ and $r_{\perp}$, adopting both methods.
This figure is constructed by calculating the phases of all the segments in the 
redshift bin $z\in 1.8-2.2$, without restriction on  $k$ and 
$r_{\perp}$. Subsequently we group the phases according to the wavenumber
and separation, by subdividing the $k$-space in three bins ($0.008-0.01$
km$^{-1}$ s, $0.01-0.04$ km$^{-1}$ s and $0.04-0.07$ km$^{-1}$ s) 
and the $r_{\perp}$-space in 
the intervals $100-200$ kpc, $200-300$ kpc, and $300-400$ kpc. 
The phase PDF of each of the nine subgroups 
is shown in the nine panels in Fig.~\ref{fig:method_comparison}. 
We choose to do this test at low redshift because the data resolution
is typically lower and the pixel sampling coarser, therefore it should
be easier to highlight problems in the phase calculation on an
irregularly-sampled velocity grid.

In all cases the two methods agree well, and most importantly
the statistical estimator that we use in the likelihood, i.e. the wrapped-Cauchy 
concentration parameters, are practically identical. This can be seen by fitting
the wrapped-Cauchy function to the two distribution,
which are indistinguishable at all $r_{\perp}$ and $k$
(Fig.~\ref{fig:method_comparison}). 
We also emphasize that 
the approximate Fourier transformation is also part of the forward-modelling 
of simulations (see \S~\ref{FM_summary}), so even in the case of a significant
effect on phase distributions, it would be taken into 
account in our calibration. 
From now on the standard method adopted to calculate phase
differences, both in data and simulations, is the LSSA technique.

\section{Forward Modeling the Phase Angle PDF}\label{forward modeling}
To connect the observed phase differences in quasar pairs with
the quantity we want to measure, the pressure smoothing scale, we run 
a grid of semi-numerical
models based on a dark-matter simulation. 
We adopt an approximate scheme that enables a detailed exploration of
IGM thermal parameter space, defined by the pressure smoothing scale
$\lambda_P$,
the temperature at the mean density $T_0$, and the slope of the 
temperature-density relationship $\gamma$.  

A proper comparison of data to models requires that we
account for 
aspects of the data which are not present in our idealized models, such
as noise, resolution, and the presence of
contaminants such as metal lines and strong HI absorption systems. 
Since it is not straightforward to subtract these effects from
distribution of phase angles, we adopt a forward-modeling
approach, which involves producing models with the same properties
as our data. 
In this way the analysis of phase angles is calibrated against the simulations in a
consistent way.  Briefly speaking, the calibration is obtained by
creating, for each observed pair, an entire ensemble of simulated
copies, with the same transverse separation, the same noise amplitude
and the same resolution, but varying the underlying IGM properties
that we want to study.
 
In this section we summarize our IGM model and we describe each step of
the forward-modeling procedure.

\subsection{Dark-Matter Simulations and Parameter Grid}\label{dm_models}

Ideally, we would like to explore a large set of thermal 
models of the IGM. However, because the pressure smoothing scale $\lambda_P$
is sensitive to the entire thermal history, the dynamic
range we would need to span is not easily achievable 
with current hydrodynamical simulations of
cosmological volumes. Therefore we opt for a fast, approximate
method based on N-body simulations, which enables to easily 
cover the relevant space of IGM parameters. In \S~\ref{hydro comparison}
we will use a set of  hydrodynamical simulations 
as a reference to compare our results with the prediction
of more accurate models. 

We base our model of the Ly$\alpha$ forest on a N-body 
dark matter (DM) only simulation.  In this scheme, the
simulation provides the dark matter density and velocity fields
\cite{Croft98,MeiksinWhite2001}, and the gas density and temperature are
computed using simple scaling relations motivated by the results of
full hydrodynamical simulations
\cite{HuiGnedin97,GnedHui98,GnedinBaker2003}. We do not 
consider the effect of
uncertainties on the cosmological parameters, as they are constrained
by various large-scale structure and CMB measurements to much higher
precision than the thermal parameters governing the IGM.

We used an updated version version of the TreePM code
\cite{TreePM} to evolve $2048^3$ equal mass ($2.5\times
10^{5}\,h^{-1}M_\odot$) particles in a periodic cube of side length
$L_{\rm box}=30\,h^{-1}$Mpc with a Plummer equivalent smoothing of
$1.2\,h^{-1}$kpc.  The initial conditions were generated by displacing
particles from a regular grid using second order Lagrangian
perturbation theory at $z=150$.  This TreePM code has been compared to
a number of other codes and has been shown to perform well for such simulations
\cite{Hei08}.  Recently the code has been modified to use a hybrid
Message Passage Interface (MPI) + Open Multi-Processing (OpenMP)
approach which is particularly efficient for current
supercomputers.
We also adopt the cosmological parameter from \cite{Planck2014}, i.e. 
density parameters for the cosmological constant of $\Omega_{\Lambda}=0.68$, for the matter of $\Omega_m=0.32$ and 
reduced Hubble constant $h=0.67$. We focus on four 
snapshot at $z=2,2.4,3,3.6$, approximately at the center of 
the redshift intervals in which we bin the data.

The baryon density field is obtained by smoothing the dark matter
distribution; this mimics the effect of the pressure
smoothing (see \cite{MeiksinWhite2001} for a discussion about this kind of 
approximations). For any given thermal model, we adopt a constant pressure
smoothing scale $\lambda_P$, rather than computing it as a function of the
temperature, and this value is allowed to vary as a free parameter
(see below). 
The dark matter distribution is convolved
with a window function $W_{\rm IGM}$ in real space. By the convolution 
theorem, this operation is equivalent to multiplying the Fourier 
components of the density field by the Fourier-transform of the 
window function 
\begin{equation}
 \delta_{\rm IGM}(\vec{k})=W_{\rm IGM}(\vec{k},\lambda_P)\delta_{\rm DM}(\vec{k}) .
\end{equation}
For example, for a Gaussian kernel with width $\sigma=\lambda_P$ the Fourier-
transformed is
$W_{\rm IGM}(k)=\exp (-k^2\lambda_P^2/2)$, which would truncate the 3D power spectrum at $k \sim 1/\lambda_P$.

For computational reason,  it is convenient to adopt a 
function with a  finite-support 
\begin{equation}
\delta_{\rm IGM}(x) \propto \sum_i m_i K(|x-x_i|,R_P)
\end{equation}
where $m_i$ and $x_i$ are the mass and position of the particle $i$, $K(r)$ is the kernel, 
and $R_P$ the smoothing parameter which sets the pressure smoothing scale.
 We adopt the following cubic spline kernel
\begin{equation}
K(r,R_P)=\frac{8}{\pi R_P^3}
  \begin{cases}
      1-6\left(\frac{r}{R_P}\right)^2+6\left(\frac{r}{R_P}\right)^3 & \frac{r}{R_P} \leq \frac{1}{2} \\ 
      2\left(1-\frac{r}{R_P}\right)^3 & \frac{1}{2}<\frac{r}{R_P}\leq 1 \\ 
      0 & \frac{r}{R_P} >1
  \end{cases}.\label{eqn:kernel}
\end{equation}
In the central regions the shape of $K(r)$ very closely resembles a Gaussian
with $\sigma \sim R_P/3.25 $, and we will henceforth take this $R_P/3.25$ to be our definition
of $\lambda_P$, which we refer to as the pressure smoothing scale.
An analogous smoothing procedure is also 
applied to the particle velocities. Following
\cite{Rorai13}, the mean inter-particle separation of our 
simulation cube $\delta l= L_{\rm box}\slash N_{\rm p}^{1\slash 3}$ 
sets the minimum pressure smoothing scale
that we can resolve with our dark matter simulation, hence we can safely model values
of $\lambda_P > 22 {\rm kpc}$.

Following the standard approach,  we assume a tight relation between
temperature and density which is well approximated by a power law
\cite{HuiGnedin97},
\begin{equation}
T(\delta)=T_0 (1+\delta)^{\gamma-1} \label{eqn:rhoT} .
\end{equation}
Typical values for $T_0$ are on the order of $10^4$ K, while $\gamma$ is expected to be around
unity and to asymptotically approach the value of $\gamma_{\infty}=1.6$, if there is no other
heat injection besides photoionization heating \cite{HuiGnedin97}.

The optical depth for Ly$\alpha$ absorption is proportional to the density of
neutral hydrogen $n_{HI}$, which, if the gas is highly ionized (neutral fraction $x_{HI}\ll 1$)
and in photoionization equilibrium, can be calculated as in \cite{gp65} :
\begin{equation}
\label{eqn:nHI}
n_{HI} = \alpha(T) n_{H}^2/ \Gamma 
\end{equation}
where $\Gamma$ is the photoionization rate due to a uniform intergalactic ultraviolet background (UVB), 
and $\alpha(T)$ is the recombination coefficient which scales as $ T^{-0.7}$ at typical IGM temperatures. 
These approximations result in a power law relation between Ly$\alpha$ optical depth $\tau$ and
overdensity often referred to as the fluctuating Gunn-Petersonn approximation (FGPA):
$\tau\propto (1+\delta)^{2-0.7(\gamma-1)}$. 
We compute the observed optical depth in redshift-space via the following
convolution of the real-space optical depth 
\begin{equation}
 \tau(v)=\int_{-\infty}^{\infty} \tau(x) \Phi(Hax+v_{p,\parallel}(x)-v, b(x))dx \label{eqn:tau},
\end{equation}
where $Hax$ is the real-space position in velocity units,
$v_{p,\parallel}(x)$ is the longitudinal component of the peculiar
velocity of the IGM at location $x$, and $\Phi$ is the normalized
Voigt profile (which we approximate with a Gaussian) characterized by
the thermal width $b=\sqrt{2K_BT/m_H}$, where $K_B$ is the 
Boltzmann constant and $m_H$ the mass of the hydrogen atom. 
We derive the temperature from the baryon density 
via the temperature-density
relation (see eqn.~\ref{eqn:rhoT}).  The observed flux transmission is 
then given by $F(v)=e^{-\tau(v)}$. 
We follow the standard approach, and treat the metagalactic
photoionization rate $\Gamma$ as a free parameter, whose value is
fixed a posteriori by requiring the mean flux of our Ly$\alpha$
skewers $\langle \exp(-\tau)\rangle$ to match the measured values from
\cite{faucher08}.  This amounts to a simple constant re-scaling of the
optical depth.  The value of the mean flux is taken to be
fixed, and thus assumed to be known with infinite precision. This is
justified, because in practice, the relative measurement errors on the
mean flux are very small in comparison to uncertainties of the
thermal parameters we wish to study.

To summarize, our models of the Ly$\alpha$ forest are uniquely described by the three parameters
($T_0, \gamma$, $\lambda_P$), and these three parameters
are considered to be independent. In particular the pressure smoothing
scale is not tied to the instantaneous temperature at mean density $T_0$,
due to its non-trivial dependence on the full past thermal history 
\cite{GnedHui98}, and both this
dependence and the thermal history are not well understood. 
For each of the three redshift bins we generate 400 models 
spanning the parameter space in the following 
intervals $\lambda_P \in 22-200$ kpc, $T_0 \in 5000-30000$ K and
$\gamma \in 0.5-2.0$.

In each model, we extract synthetic spectra parallel to the line of sight,
following the recipe described above. The sight lines are 30 Mpc$/h$ long
and have 1024 pixels each, giving a pixel scale of  about 29 kpc$/h$ or 
3.3 km s$^{-1}$. The positions of these spectra 
is dictated by the separation of the pairs in the observed sample, as described in the next section.

\subsection{Transverse Separation}\label{rperp_evolution}
\begin{figure}   
\psfrag{r [kpc]}[c][][1.]{$r_{\perp}$ [kpc]}
\centering \centerline{\epsfig{file=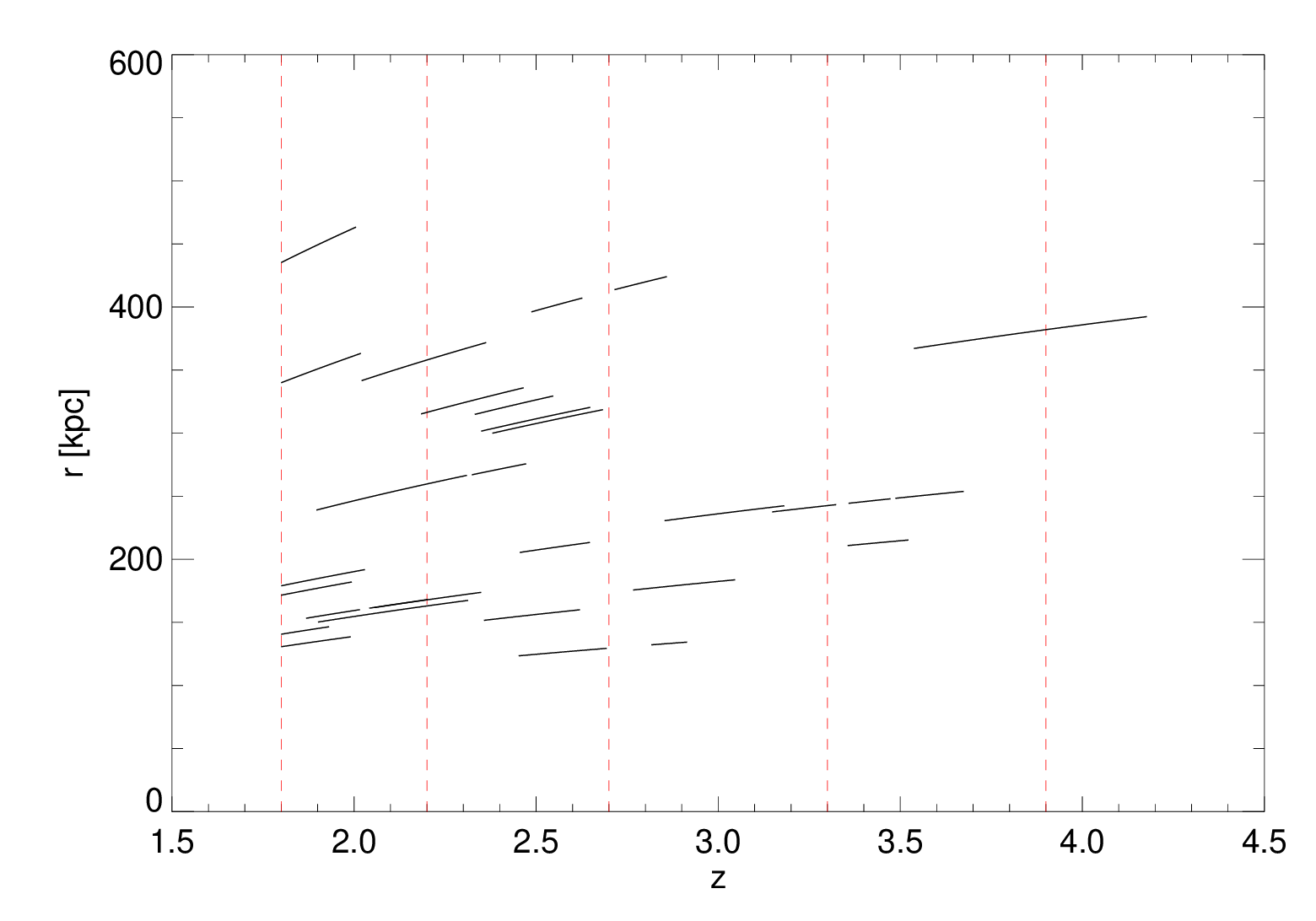,
      width=\textwidth}}
  \vskip -0.1in
\caption{\label{rz_dist} {\bf The distribution of our sample in redshift $z$ and in
transverse separation $r_{\perp}$.} Each black line represents a segment
of overlapping \mlya\ forest in a pair. The length of a segment depends on 
the redshift difference between the two quasars and on 
the presence of DLAs or other
contaminants that require excluding part of the forest. The atmospheric
cutoff sets a lower limit for all pairs at $z\approx 1.7$. 
The lines are curved because the impact parameter evolves with redshift 
converging toward us.
The vertical red dotted lines delimit the four redshift intervals 
to which we split the sample.}
\end{figure}

\begin{figure}   
\centering \centerline{\epsfig{file=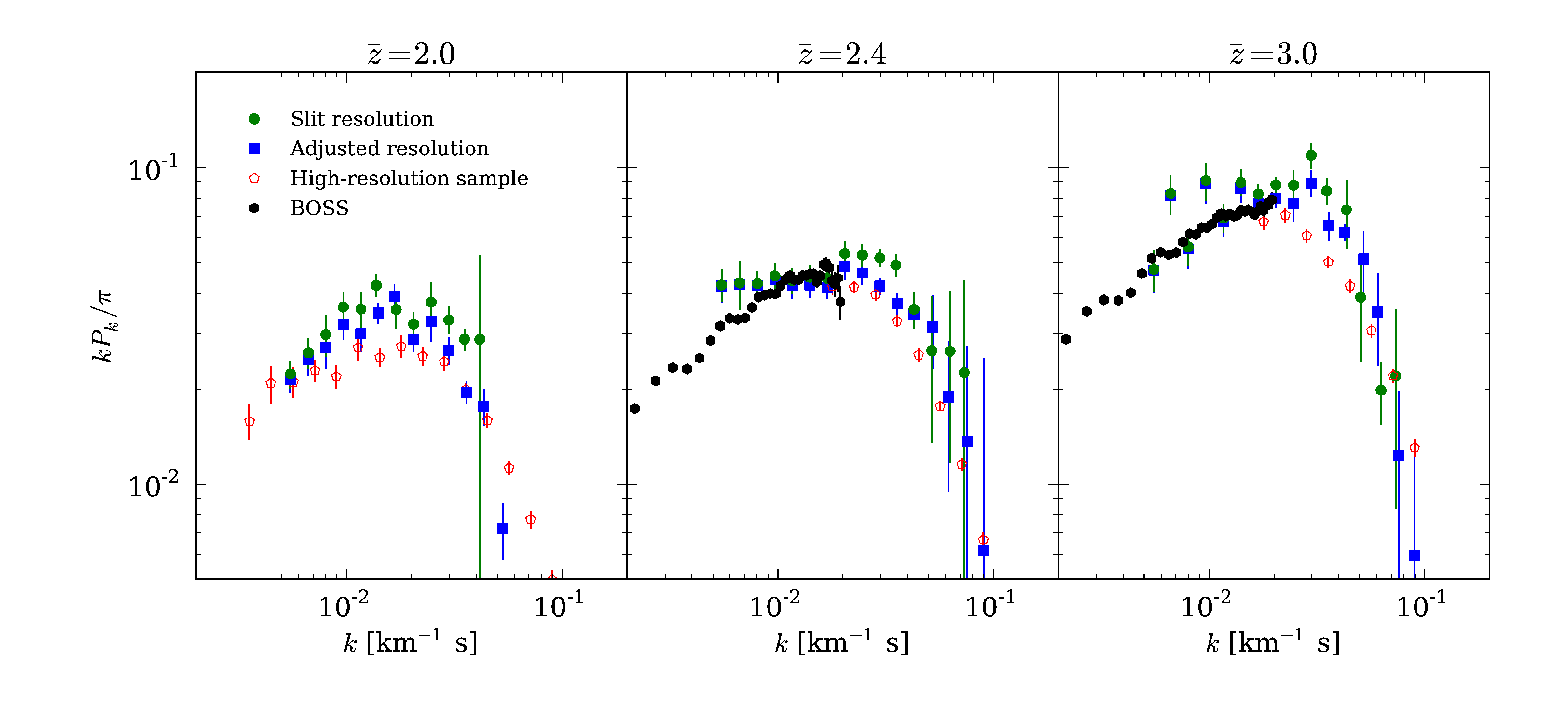,
      width=\textwidth}}
  \vskip -0.1in
\caption{\label{fig:power_spectrum} {\bf Test of the resolution
modeling} using the power spectrum of the \mlya\ flux contrast
in three redshift bins. The noise power has been subtracted and a 
Gaussian resolution correction has been applied, as described in the text. 
The red open pentagons are calculated from a sample of high-resolution 
spectra observed with UVES and HIRES, and are used as a 
benchmark for the comparison with 
our data. The green circles mark the power spectrum of the \mlya\ forest in 
our pair sample, assuming the nominal resolution of the instrument, whereas
for the blue squares we increased our assumed  resolution by 20\% . 
Uncertainties are estimated via a bootstrap technique, but they are likely to be 
underestimated because the spectra are not all statistically 
independent, as there is correlation 
of the \mlya\ forest power between the members of each pair. 
Where possible, we use 
published measurements of the power spectrum \cite{BOSSpower} 
to verify the consistency of our calculation at low-$k$
(black circles). The power 
spectrum calculated from the pair sample is overall noisier, as expected,
than the high-resolution one. We find that assuming the nominal resolution
of the instrument overestimates the signal, indicating that
the resolution is typically underestimated. A correction of 
$20\%$ to the resolution FWHM estimates accommodates the 
discrepancy between the power at an acceptable level, and is 
taken as our standard assumption in the
forward-modelling procedure. 
}
\end{figure}

\begin{figure}   

    \psfrag{r=150.0 kpc}[c][][1.]{$r_{\perp}=150$ kpc}
    \psfrag{r=250.0 kpc}[c][][1.]{$r_{\perp}=250$ kpc}
    \psfrag{r=350.0 kpc}[c][][1.]{$r_{\perp}=350$ kpc}
\centering \centerline{\epsfig{file=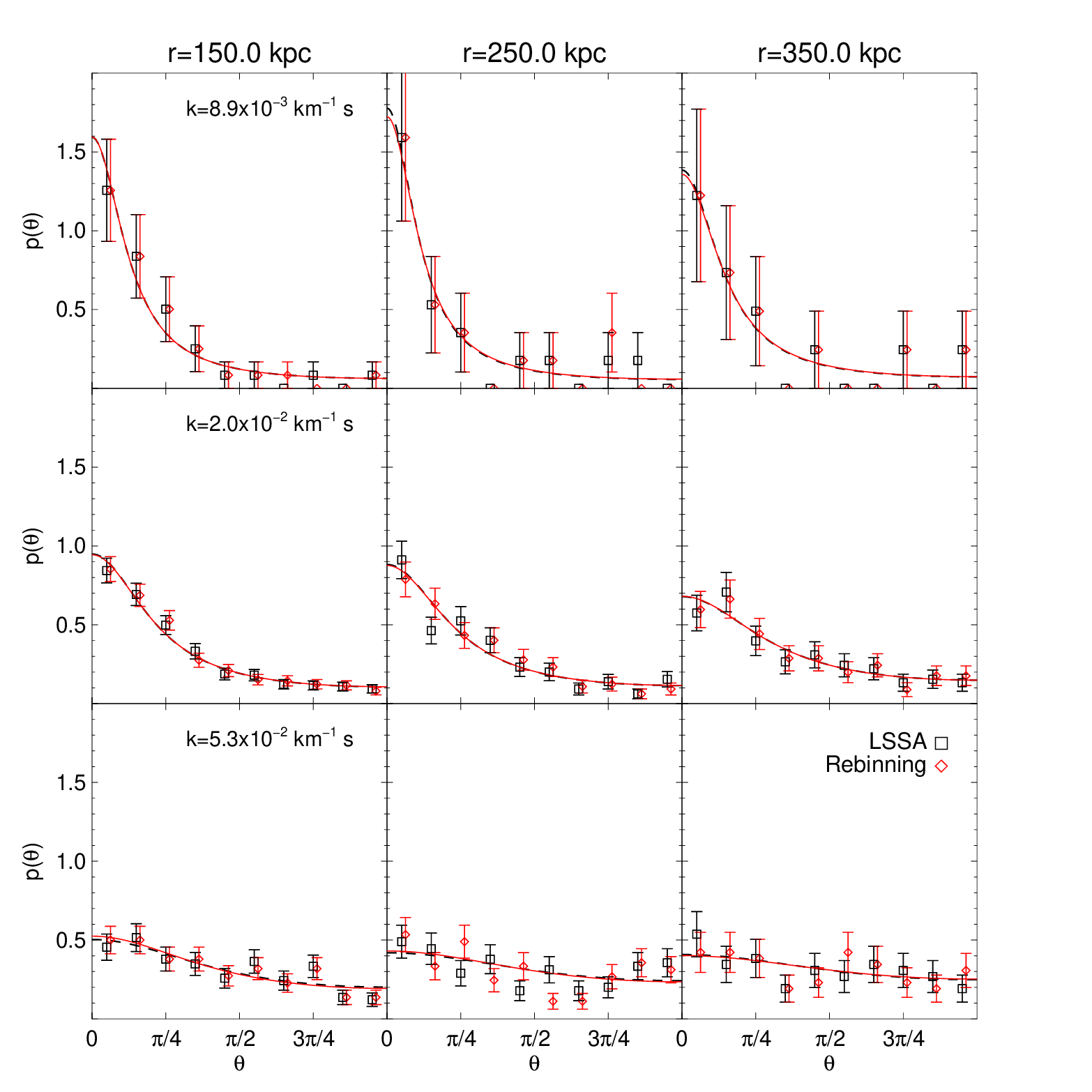,
      width=\textwidth}}
  \vskip -0.1in
\caption{\label{fig:method_comparison} {\bf Comparison of the two
phase calculation methods} applied to data at 
$z=2$: least square spectral analysis 
(black squares) and  rebinning on a regular grid to apply Fourier
transformation (red diamonds). The best-fitting wrapped-Cauchy 
functions are shown as solid lines, with the same colors. The distributions
are derived by first calculating all the phase differences in the sample
of pairs within the redshift range $z \in 1.8-2.2$, then subdividing
them in nine sets according to the transverse separation $r_{\perp}$ 
and the wavenumber $k$, as described in the text. 
The two methods agree well in all cases, and the wrapped-Cauchy
fits are close to each other, implying that the phase distributions
are statistically equivalent. This demonstrates the robustness of  
the approximated method that we use to calculate phases.
}
\end{figure}

Two quasars separated on the sky by an observed angle $\psi$ have a
transverse distance dependent on their redshift.
If we are studying 
\mlya\ absorption, the transverse separation between the coeval forest
in the two spectra is an evolving function of the wavelength, since 
the sight lines are convergent toward us. This transverse separation
can be written as
\begin{equation}
r_{\perp}(z_{\text{abs}})=D_{A}(z_{\text{abs}})\psi (1+z_{\text{abs}})
\end{equation}
where $z_{\text{abs}}=\lambda/\lambda_{\alpha}-1$ is the \mlya\ absorption 
redshift and $D_{A}$ is the angular diameter distance, which depends on
the adopted cosmological parameters.  The variation
of $r_{\perp}$ across our redshift bins is not negligible, especially for
the longer segments of forest, as Fig.~\ref{rz_dist} suggests. 
Since we know that phases are dependent on $r_{\perp}$,
we should take this fact into account. 
Extracting the \mlya\ forest along convergent skewers in the simulation would 
be complicated to implement, and furthermore our simulation cube 30 Mpc$/h$ constitutes
only a small fraction of the path length in a typical segment.  Instead we account
for the variation of $r_{\perp}$ with redshift with the following strategy.
We extract skewers parallel to the coordinate axis of the simulation cube, but
for each observed pair we compute a full ensemble of synthetic pairs with 
separations uniformly distributed over the range covered by $r_{\perp}(z_{\text{abs}})$
within the redshift limits of the segment. 
In practice, if the coeval \mlya\ forest of the pair lies between
$z_{\text{min}}$ and $z_{\text{max}}$, we simulate 400 pairs randomly located in 
the box and with separation $\{r=r_{\perp}(z_i)\}$, where the 400 redshifts $z_i$ are
logarithmically spaced between $z_{\text{min}}$ and $z_{\text{max}}$. 
The logarithmic spacing is chosen to achieve linear spacing in $v(z)$, 
which is the coordinate on which Fourier coefficients are calculated. 

\subsection{Contaminants: Lyman-Limit Systems and Metal Lines}\label{contaminants}

Our approximate semi-numerical model of the \mlya\ forest based on a
smoothed dark-matter only simulation cannot reliably model the
stronger absorption lines resulting from LLSs, or the metal lines
which also contaminate the forest.  LLSs, as well as a significant
fraction of metal lines, are believed to be predominantly associated
with dense gas in the circumgalactic medium of galaxies, which is not
captured by our simple approach which only models low-density hydrogen
gas in the IGM.  Furthermore our procedure for calculating $n_{\rm
  HI}$ from a uniform UV background in
eqn.~\ref{eqn:nHI} assumes the
gas is highly ionized and optically thin to Lyman limit absorption,
thus ignoring self-shielding effects relevant in LLS and 
low-ionization metal species.  We thus take LLS and metals into
account by adding them to our simulated spectra, according to their
measured abundances, rather than by directly identifying and masking
them. The strongest LLS absorbers with $\log
N_{HI}>19.0$ are easily identified in our spectra via their damping
wings, and these systems are directly masked.  We add these contaminants
to our simulated spectra following the same procedure described in \cite{KG2015}. 

We adopt the estimate of \cite{Ribaudo2011} for the LLS total abundance:
\begin{equation}
l_{\rm LLS}(z)=l_{z0}(1+z)^{\gamma {\rm LLS}} 
\label{eq:lls_evolution}
\end{equation} 
where $l_{\rm LLS}(z)$ is the number of LLS per unit  
redshift, and the parameters take the values $l_{z0}=0.1157$ 
and $\gamma_{\rm LLS}=1.83$. 
Every time we want to include LLS in our \mlya\ forest models at 
redshift $z$, we multiply $l_{\rm LSS}(z)$ by the total path 
length of the synthetic spectra, expressed in redshift, to obtain
the total number of absorbers. 
We then assume that the column density distribution $f(N_{HI})$
follows a power law in the column density range $10^{16.5}-10^{19}$
cm$^{-2}$\cite{Prochaska2010}.
Following \cite{KG2015}, we adopt the steepest slope for the power law,
and we also add partial Lyman Limit Systems (pLLS) in the density range
$10^{16.5}-10^{17.2}$ cm$^{-2}$. The slope of the pLLS column density
distribution has been inferred by \cite{Prochaska2010} from the total 
mean free path of ionizing photons, and it is $\beta_{\rm pLLS}=-2$.
These choices are motivated in \cite{KG2015}, as they improve the fit of hydrodynamical
simulations of the \mlya\ forest flux PDF as measured from BOSS. We will test a posteriori
the sensitivity of our measurement to variation of the LLS abundance
(\S~\ref{contaminant_stability}).
In summary, the distribution from which we add HI strong absorbers
can be written as  
\begin{equation}
f(N_{HI})=  \begin{cases}
      k_1 N^{-2}_{HI} & 10^{16.5} < N_{HI} \leq 10^{17.2} \\ 
      k_2 N^{-1.2} & 10^{17.2}< N_{HI} \leq 10^{19.0} 
  \end{cases}.\label{eqn:lls_dist}
\end{equation}
where the column densities are expressed in cm$^{-2}$, and the 
coefficients $k_1$ and $k_2$ are determined by imposing
continuity at $N_{HI}=17.2$ cm$^{-2}$ and by requiring the abundance ratio 
of pLLS to LLS to be $l_{\rm pLLS}/l_{\rm LLS}=1.8$ (see
the red line in figure 15 of \cite{KG2015}). Note that 
unlike \cite{KG2015}, we do not add super-LLSs with 
$N_{HI}>10^{19}$ cm$^{-2}$, because we have sufficient
resolution and signal-to-noise to identify and mask them directly.

Again following \cite{KG2015}, we add metal lines to our simulated 
\mlya\ forests based  on lower-redshift quasar spectra from BOSS. 
We randomly pick segments of quasar spectra in the same observed 
wavelengths of our simulated forest, but in the rest-frame region
1260-1390 \AA, such that they are redder then the \mlya\ line, but
bluer than all the relevant metal transitions. All the absorption lines
in such segments will be due to metals in the IGM, so they effectively
represent a realization of the metal lines distribution in the path length
of the analyzed \mlya\ forest. These realizations constitute our
model for metal contamination in the forest, which is included in 
our forward-modelling procedure. 
We use a metal catalogue \cite{Lundgren2009},
which lists
absorbers in SDSS \cite{Schneider2010} and BOSS quasar
spectra \cite{Paris2012}. The SDSS spectra were included
in order to increase the number of $z_{\rm qso} \approx 1.9$-$2.0$ quasars
needed to introduce metals into the $z \geq 2.3$ \mlya\ forest
mock spectra, which are not well sampled by the BOSS quasar target
selection \cite{Ross2012}. We emphasize that we work with
the “raw” absorber catalogue, i.e., the individual absorption lines
have not been identified in terms of metal species or redshift. For
each quasar, the catalogue provides a line list with the observed
wavelength, observed frame equivalent width $W_r$, FWHM, and detection S/N, $W_r/\sigma_{W_r}$ . To
ensure a clean catalogue, we use only $W_r/\sigma_{W_r} \geq 3.5 $ absorbers
in the catalogue that were identified from quasar spectra with
S/N $>$ 15 per \AA\ redward of \mlya .
The latter criterion
ensures that even relatively weak lines (with EW $\gtrsim 0.5$ \AA) are
accounted for in our catalogue. 

In order to add noiseless lines to our model spectra, we assume that
they all lie on the flat part of the curve of growth, motivated by the
fact that most metal lines detected at BOSS/SDSS resolution are
saturated. In this regime the EW depends mostly on the Doppler parameter
$b$ of the lines, and only weakly on the central opacity $\tau_0$. We make
the assumption that $\tau_0 = 3$ for all the lines, and derive $b$
from the relation
\begin{equation}
b=c \frac{W_r}{2\sqrt{\ln (\tau_0 /\ln 2)}}
\end{equation}
where $c$ is the speed of light. The line optical depth profile is than assumed to be 
Gaussian with normalization $\tau_0$ and width $b$. 
The key point is that in the saturated regime the results are very weakly
dependent on $\tau_0$, which justifies our arbitrary choice. 

\subsection{Resolution}\label{resolution}

RHW pointed out that phases have the mathematical
property of being invariant under convolution with symmetric
kernels. This mathematical fact however applies only to noiseless
data, analogous to the situation of a general deconvolution problem.
In fact, in the presence of noise phase scattering is enhanced for
high-$k$ modes where the signal from the forest is suppressed due to
resolution. Correlated phases for a given mode are de-correlated 
by noise and their intrinsic
probability distributions is flattened depending on the noise level
and the resolution kernel.
It follows that our forward modeling needs to reproduce the
combined effect of resolution and noise, unless the data have
very high signal-to-noise ratio. We also deduce that the Fourier modes
suppressed by the resolution cutoff (see
eqn.~\ref{eq:power_spectrum_correction}) are unreliable, because they
are dominated by noise. For this reason we set an upper limit on the
usable $k$-range for each quasar pair spectrum depending on the
spectral resolution, $k_R=1/\sigma_R\approx 2.355/\text{FWHM}$, where
FWHM is the full-width at half-maximum defining the spectral
resolution, and $\sigma_R\approx {\rm FWHM}\slash 2.355$ is the
standard deviation.  We conservatively assume the FWHM to correspond
to the nominal resolution of the instrument determined by the slit
width, which we know to be a lower limit on the actual resolution,
i.e.  an upper limit on the FWHM (see \S~\ref{power_spectrum_test} for
further discussion).

In our forward-modeling we convolve our simulated 
spectra with a Gaussian kernel with FWHM defined by the resolution
 of the spectrograph. Although the resolution 
is wavelength-dependent, we use a constant width
for each \mlya\ forest segment which is specified in
Table~\ref{tab:sample}. This width corresponds to the FWHM at the average 
wavelength of each segment, where the average is defined as the midpoint  in
velocity space, which can be shown to be
\begin{equation}
\bar{\lambda}=\frac{\lambda_1\lambda_2\ln (\lambda_2/\lambda_1)}{\lambda_2-\lambda_1} ,
\end{equation}
where $\lambda_1$ and $\lambda_2$ are the minimum and the maximum
observed wavelength of the segment, respectively. For our default value of the
resolution we adopt a correction factor to the slit resolution based on a comparison with
the \mlya\ power-spectrum measured from high-resolution spectra, as illustrated 
in \S~\ref{power_spectrum_test}.

The quality of the data varies significantly in our sample,
with the ${\rm S\slash N}$ per \AA ngstr\"om varying between approximately
10 and 120.
Noise randomizes phases and hence makes the shape of the
phase angle PDF flatter.  Phases calculated from pairs with different
${\rm S\slash N}$ are affected to different degrees, which is why each quasar in
our sample demands its own specific calibration.

\begin{figure}   
\centering \centerline{\epsfig{file=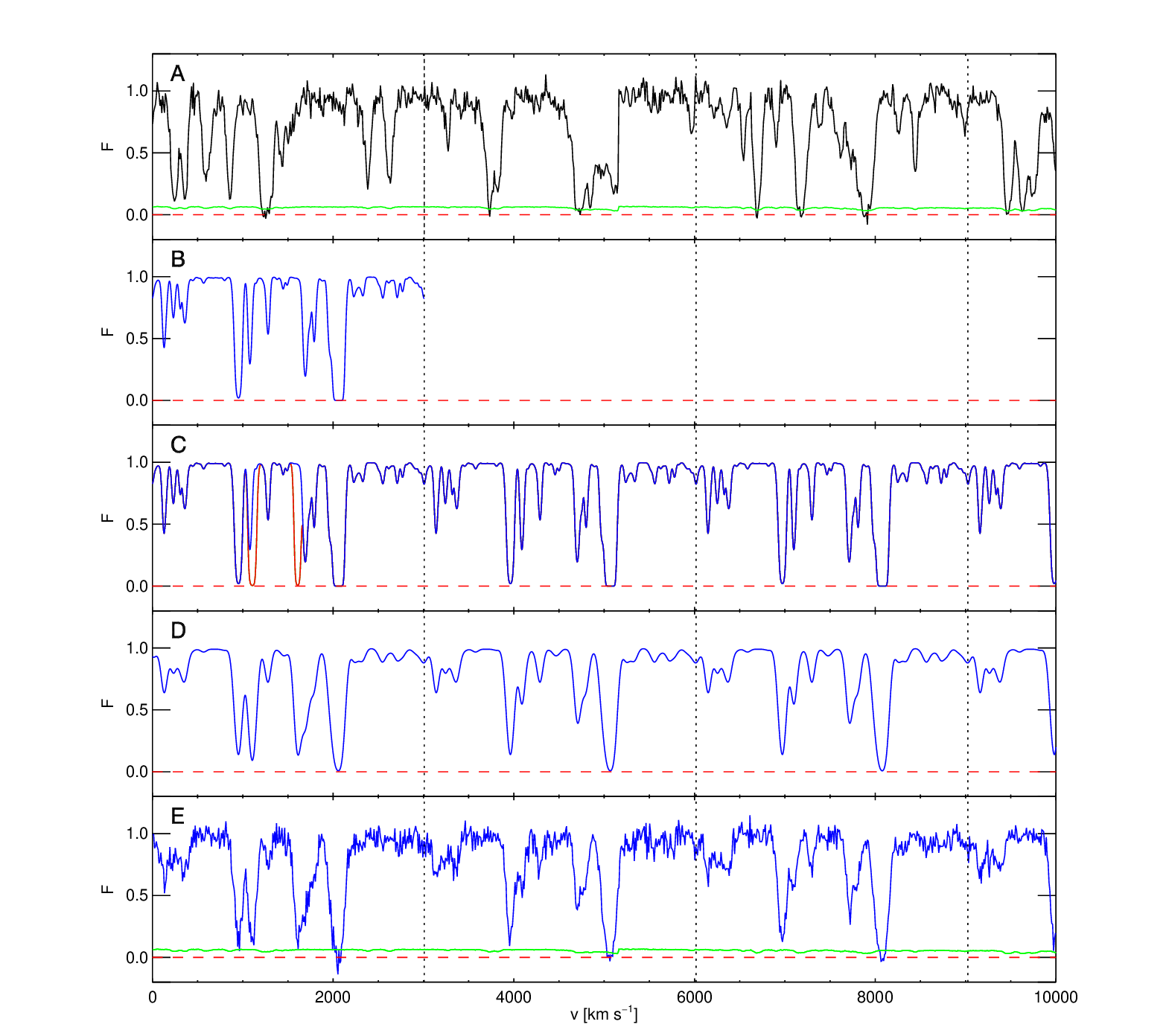,
      width=\textwidth}}
  \vskip -0.1in
\caption{\label{fig:forward_modeling} {\bf The main
steps in our forward-modelling procedure.} The upper panel shows the 
spectrum of the quasar SDSS~J111610.69+411814.4, observed with ESI. The zero level and the 
estimated noise are marked in dashed red and green, respectively. 
Vertical dotted lines divide the spectrum in regions of 30 Mpc$/h$,
which is the length of the box size of our N-body simulation on which our
semi-numerical models are based. The second panel shows in blue a simulated 
spectrum from one of these models, with $\lambda_P=85$ kpc, $T_0=20000$ K
and $\gamma=1.6$. In the third panel the synthetic
spectrum is periodically replicated and the optical depth of metal lines (red)
and LLS (green) is added (we have chosen a particular sight line where both contaminants
are present for illustration purposes). In the fourth panel we show the extended
spectrum convolved with a Gaussian kernel with FWHM matched to the resolution
of the spectrograph, which in this case amount to 64 km s$^{-1}$. In the last
panel we add noise pixel by pixel according to the pipeline estimate at each
wavelength. The zero point of the velocity scale is arbitrarily set to
the minimum redshift of the segment (for the observed spectra) 
or to the edge of the box (for the simulations).} 
\end{figure}

A further complication stems from the wavelength dependence of the
noise, which is typically higher at smaller $\lambda$, 
because of the lower spectrograph
sensitivity in the near-UV.  We model this wavelength dependence by
adding Gaussian  noise to our model spectra with a
standard-deviation given by the wavelength dependent 1$\sigma$ noise
vector of each spectrum, produced by the data reduction pipeline.  In
order to add wavelength dependent noise in this way, we need to somehow extend the simulated
sight lines, which are only $30$ Mpc$/h$ long ($3008$ km/s at $z=2$), 
so that they are comparable
to the length of the observed spectral segments ($\sim 15000$ km/s for
$\Delta z=0.2$ at $z=3$). This is done by periodically
replicating each simulated spectrum until its size matches that of the forest segment on which it
is calibrated (see Fig.~\ref{fig:forward_modeling}).
This procedure is allowed because the periodic replication 
does not affect the phase distribution of the modes, but it 
is effectively used as a convenient resampling of the same 
sight lines under different noise conditions, in order to 
take into account the wavelength-dependent sensitivity of 
the instruments.

The flux of the spectrum obtained after the periodic replication is
finally rebinned on to the same pixel grid as the observed spectrum,
which is always coarser than the one used in our simulation.  Once
this is done, we are able to generate Gaussian noise, matched pixel by
pixel, to the estimated wavelength-dependent noise of the data. As our
quasar spectra have been continuum normalized, the noise vectors are
also divided by the same continuum such that the appropriate level of
noise is added.

\subsection{Forward-Modeling of the Simulation}\label{FM_summary}

The methodology described in the three previous paragraphs 
constitutes our forward-modelling procedure, which alters our
simulated spectra to have the same properties as real spectra
observed through a telescope with finite resolution and integration
time, and containing contaminant metal lines and LLSs.
Our forward-modelled simulations can be directly compared
to observations, enabling our statistical phase angle PDF analysis and thermal
parameter study. As we have explained above, the forward-modelling procedure is
tailored to individually reproduce the properties of each spectrum in each quasar
pair, and must be applied to all the IGM models that we want to test.
The general procedure that we follow to perform
the phase angle PDF analysis on our dataset is summarised below.

Our goal is to evaluate the likelihood in eqn.~(\ref{diaglik}) for any
thermal model $ \left\lbrace T_0,\gamma,\lambda_P\right\rbrace$ 
using phase differences from our pair sample.
Given a quasar pair separated on the sky by an angle $\psi$, with
overlapping \mlya\ forest intersecting one of our redshift bins $Z=[z_1,z_2]$,
we have to forward-model the simulated spectra and to 
determine the phase angle PDF
for an IGM model with $\left\lbrace T_0,\gamma,\lambda_P\right\rbrace$ at each
$k$ for the appropriate impact parameters. This operation is structured as follows: 
\begin{itemize}
\item  We determine the overlapping portion of the \mlya\ forest of the two QSOs
  which intersects the redshift bin $Z$. This segment will have a comoving separation 
  varying with redshift as $r_{\perp}(z)=D_{A}(z) \psi (1+z)$ (see section
  \S~\ref{rperp_evolution}).

\item  We generate 400 pairs from the 
  simulated box distributed in transverse separations $r_{\perp}$
  depending on $r_{\perp}(z)$ as described in \S~\ref{rperp_evolution}.
		
\item The optical depth of the total sample of 400 sight lines 
	is renormalized in order to match the literature value of the 
	mean \mlya\ flux of the central redshift of the bin.

\item All 400 pairs are forward-modeled according to the properties of 
  the two observed spectra in the pair, which have in general 
  different resolutions and S/N. This is done through the following four steps:
  
  \begin{enumerate}
  \item Simulated spectra are periodically replicated until they match the length of
    the observed segments of forest.
    
  \item The optical depth of metal lines and LLSs is added to the sight lines.
    
  \item The spectra are convolved with a Gaussian kernel with
    FWHM set by the spectral resolution.
    
  \item Simulated spectra are then rebinned onto the same pixel grid as the data.
    
  \item Gaussian uncorrelated noise is added to the simulated flux
    with a standard-deviation determined by the $1\sigma$ noise vector
    of the observed spectrum.
  \end{enumerate}
  
\item We calculate phase angle differences from the simulated pairs and estimate the
  wrapped-Cauchy concentration parameters $\zeta$ (see eqn.~(\ref{WCD}))
  at each bin in $k$. We thus predict the phase angle PDF $P_{WC}(\theta(k))$  as a 
  function of $k$ for the considered pair. Having rebinned the sight
  lines  at step 3, we have to calculate phases with the LSSA method 
  as we do with data.

  This procedure is then repeated for each observed quasar pair, for
  each IGM model, and in each redshift bin.  Given the forward-modeled
  phase angle PDFs, we can then compute the likelihood in
  eqn.~(\ref{diaglik}) by taking the product over all quasar pairs and
  over all modes sampled by a given quasar pair.  The product over
  modes is evaluated only for modes below the limiting wavenumber
  $k_R$,
  which is set by resolution. This provides values of
  the likelihood function for all pairs over the entire model
  parameter space, allowing us to infer the thermal properties of the
  IGM at each redshift.

\end{itemize}

\section{MCMC Exploration of the Phase Angle PDF Likelihood}\label{MCMC_interp}

Our parameter grid consists of 400 models in the space defined by
$T_0,\gamma,\lambda_P$. Based on previous measurement and on a
preliminary analysis, we adopt the following flat prior on the
parameters: $T_0 \in 5000-30000$ K, $\gamma \in 0.5-2.0$,
$\lambda_P \in 22-200$ kpc. The lower limit in $\lambda_P$ is
dictated by the resolution of our dark-matter simulation: according to
a previous test (see RHW) we can only study smoothing scales larger
than the mean interparticle separation.  We first calculate 50 sets of
smoothed density and velocity skewers from our dark-matter simulation,
as described in \S~\ref{dm_models}, corresponding to 50 different
choices of $\lambda_P$, logarithmically spaced between 20 kpc and 200
kpc.

We build our grid of models to efficiently sample not only the 3D
thermal parameter space, but also the projected 2D and 1D subspaces of
the parameters. Regular Cartesian grids do not perform well because
when projecting a cubic grid onto a
plane all points in a row are projected to a single point.  
The algorithm adopted to create the 
parameter grid is the following:
\begin{itemize}
	\item The parameter space is broadly subdivided in $8 \times 4 \times 4 = 128$
	bins (in $\lambda_P,T_0,\gamma $). 
	
	\item The first point is chosen randomly in the center of one of the 128 bins.
	
	\item Each of the next points is then added to one of the bins $\tilde{B}$ with the smallest number of points in it (chosen randomly). 
	
	\item Once the bin $\tilde{B}$ is chosen, we pick one of the discrete 
	values of $\lambda_P$ 
	within that bin. We always choose the value of $\lambda_P$ with the smallest
	occurrence within the set of the previous points. 
	
	\item We  then define a $4 \times 4$ subgrid in the $T_0$-$\gamma$ plane, limited to the area defined by the projection of $\tilde{B}$.  
	
	\item Analogously to what we have done with $\lambda_P$, we pick one of the 
	16 subcells and assign to the current point the value of $T_0 $ and $\gamma$ at its
	center. The subcell is chosen to be the least represented among the previous 
	points. 
	
	\item This process is repeated until we have reached the desired number of 
	points in the parameter space.

\end{itemize}

In this way we define a grid of 400 points in $T_0,\gamma$ and
$\lambda_P$ for all four redshift bins.  We then calculate the phase
angle PDF likelihood of all the models as described in the previous section,
then we use Gaussian process interpolation to interpolate this likelihood
to any arbitrary point in the parameter space.  Gaussian processes are a statistical
method to interpolate irregular grids, which assume that the quantity
to interpolate is the result of a stochastic process characterized by
a Gaussian distribution. Detailed explanations and its application to
the matter power spectrum can be found in
\cite{Heitmann2010,Heitmann09,Lawrence09}.

In our case, the only input for the Gaussian process interpolation
is the choice of the 'smoothing lengths', which quantify
the extent of the correlation of the likelihood along each dimension
of the parameter space. We choose these smoothing
lengths to be a multiple of the average spacings of our parameter grid. The
choice of these smoothing lengths is somewhat arbitrary, but we
checked that the inferred posterior distributions of thermal parameters
do not change in response to reasonable
variations of these smoothing lengths.

We employ a publicly available MCMC package {\sc emcee} \cite{MCpackage}. The results of
the MCMC runs are shown in
Fig.~3 of the main text, in the
form of 65\% and 95\% confidence levels in the $\gamma$-
$\lambda_P$ and $T_0$-$\lambda_P$ planes, and in the form
of posterior probability distribution of $\lambda_P$.

\section{Visual Comparison of the Phase Angle PDF to Models}\label{stack_pdf}

Our parameter estimation is based on MCMC sampling of a multi-dimensional
parameter space, which delivers a posterior distribution for
$\lambda_P$, $T_0$, and $\gamma$. Nevertheless, it is instructive to
construct visual data for model comparisons. This is challenging
because the information in the phase angle PDF is spread out across
many $k$ and $r_{\perp}$. So we adopt a procedure of binning modes and
separations together, which also allows us to overplot similarly binned phase angle
PDFs from our models calibrated with our forward-modelling procedure. 
(\S~\ref{dm_models}).

\begin{figure}   

    \psfrag{r=150.0 kpc}[c][][1.]{$r_{\perp}=150$ kpc}
    \psfrag{r=350.0 kpc}[c][][1.]{$r_{\perp}=350$ kpc}
\centering \centerline{\epsfig{file=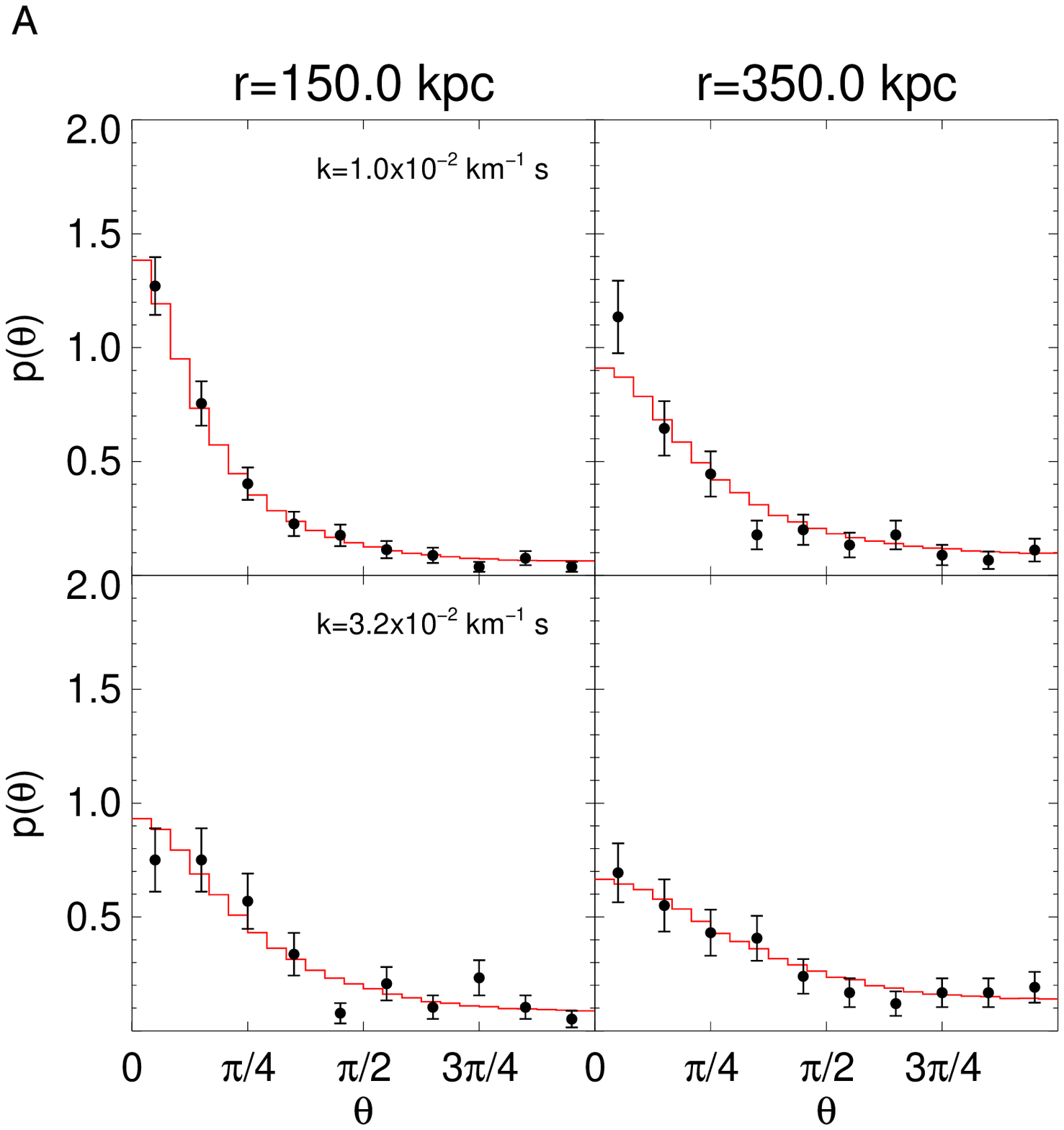,
      width=0.5\textwidth}
	  \epsfig{file=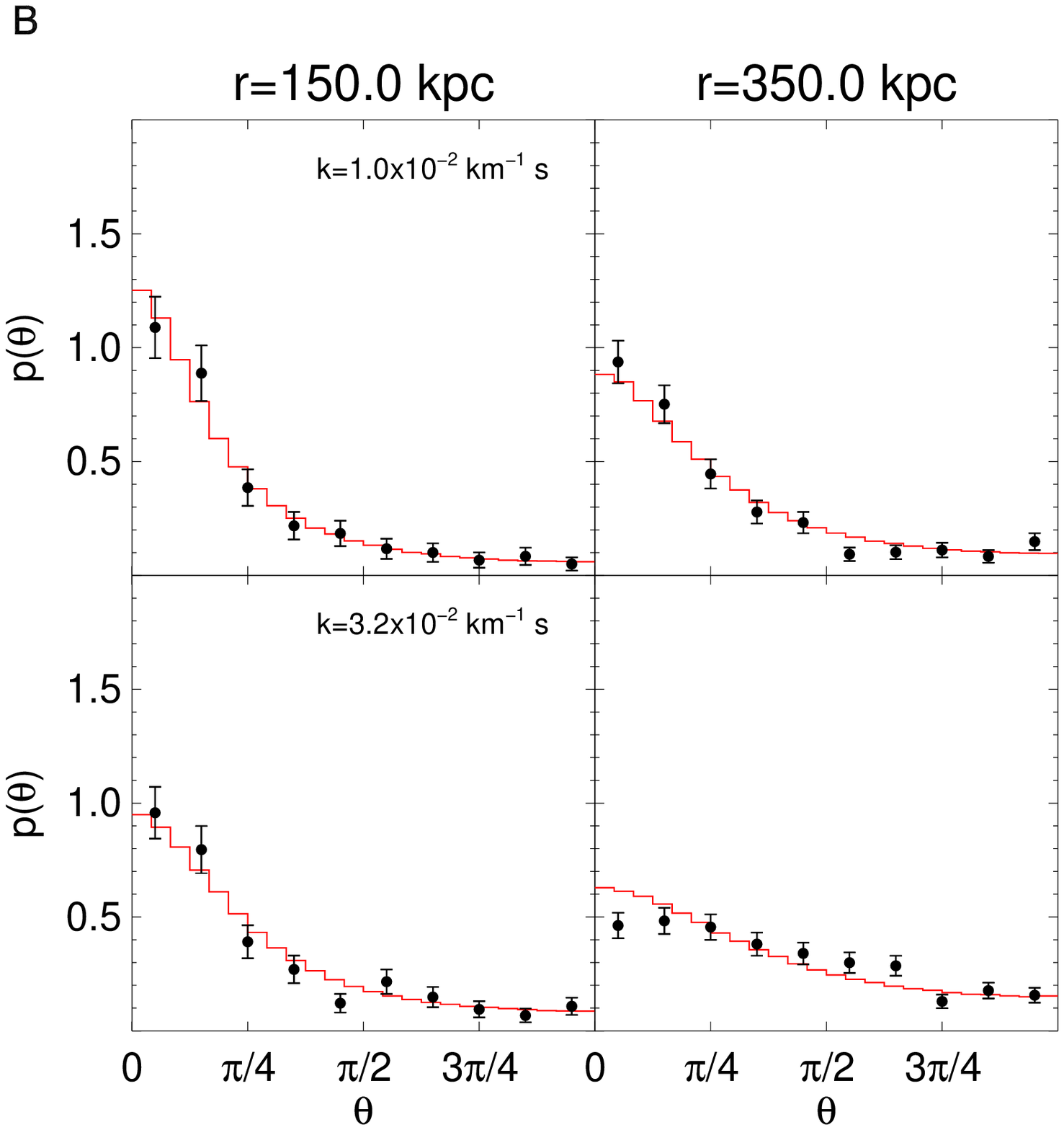,
      width=0.5\textwidth}}
  \vskip -0.1in
  \centering \centerline{\epsfig{file=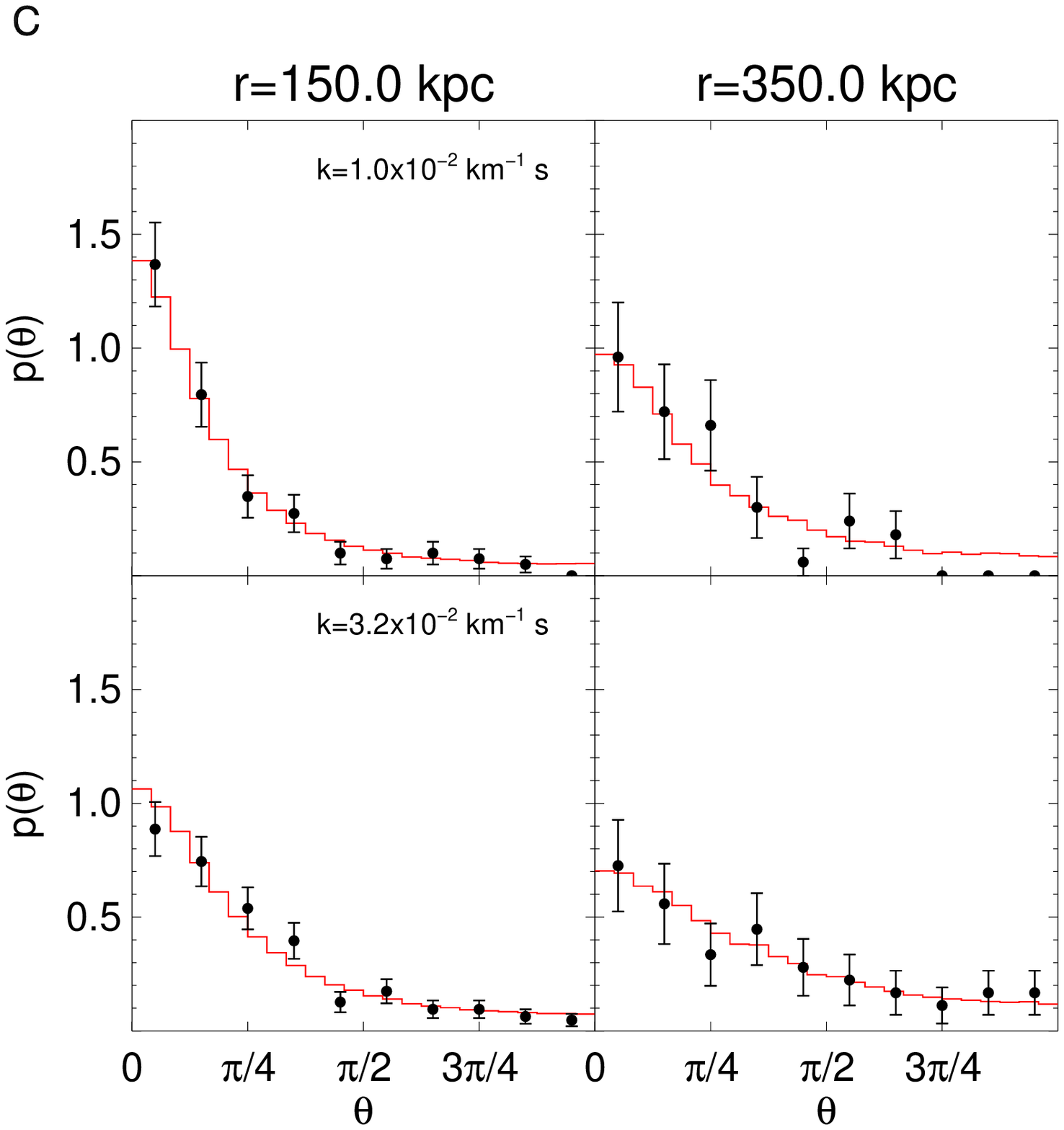,
      width=0.5\textwidth}
	  \epsfig{file=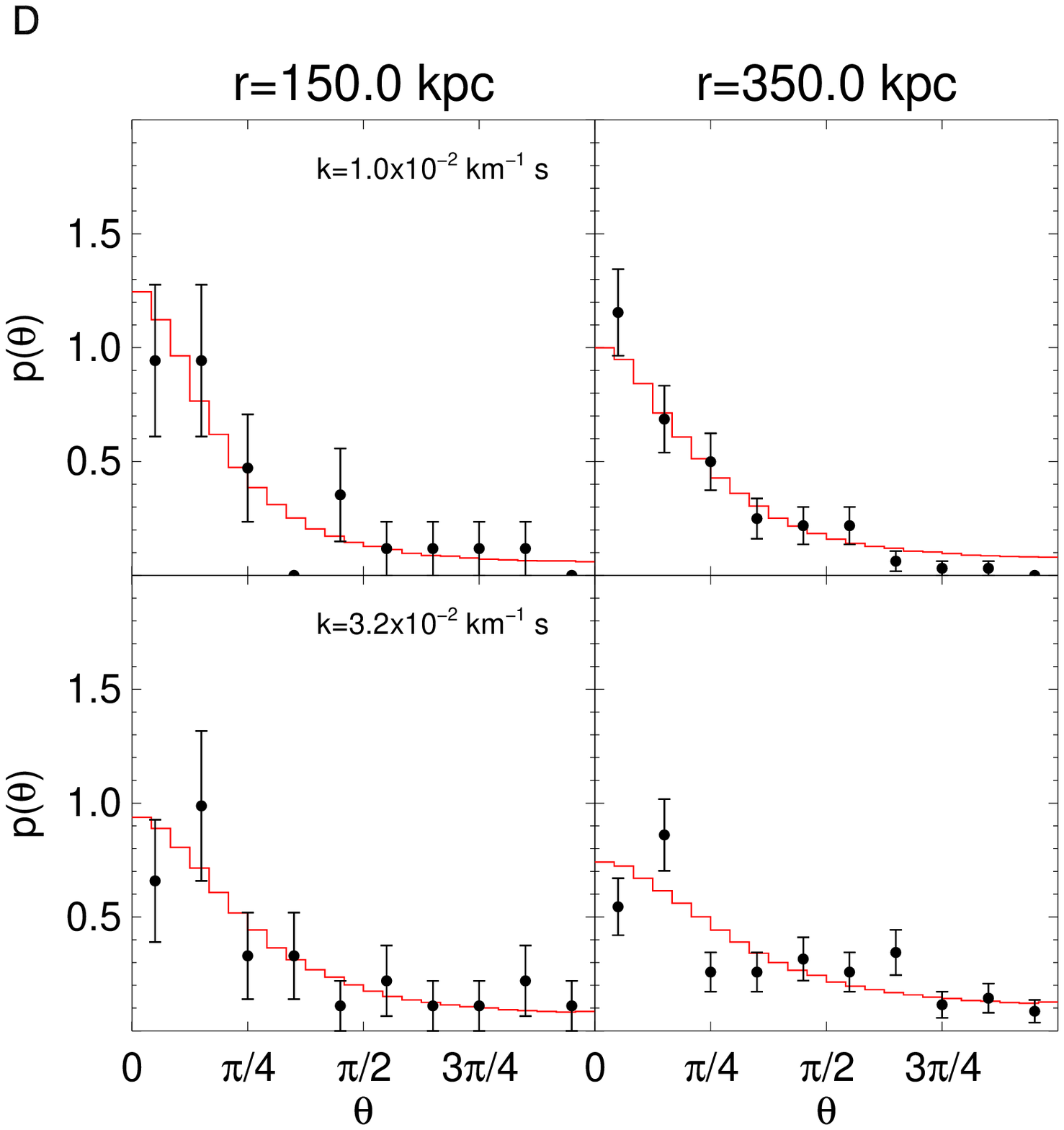,
      width=0.5\textwidth}}
  \vskip -0.1in
\caption{\label{all_phases_pdf} {\bf Observed phase distributions.} 
Black points: Phase 
difference PDF calculated  from real data calculated 
at redshift $z=2,2.4,3,3.6$ (from left to right and from 
top to bottom). In each of the four plots the phase difference are 
binned in $k$ and $r_{\perp}$ (see text
for the exact definition of the bins). The central values of the 
$r_{\perp}$ bins are reported at the top of each column, and the 
central values of the $k$ bins are reported in the legend of the 
left columns. Error bars are calculated assuming a Poisson distribution.
Red histogram: the phase difference PDF of the best-likelihood
model, calculated after fully modelling noise, resolution and contaminants,
as described in the text.
}
\end{figure}

The measured phases are binned and plotted in
Fig.~\ref{all_phases_pdf} in the same way described in
\S~\ref{test_LSSA}, which we used to produce Fig.
\ref{fig:method_comparison}. In this case, we only use two bins for
$k$ and two bins for $r_{\perp}$, for all four redshift bins in our
analysis.  The $k$ bins are defined by the intervals
$0.005-0.02, 0.02-0.05$ km$^{-1}$ s (from top to bottom in each of the
four subplots), and the $r_{\perp}$ bins by $50-250, 250-450$ kpc
(left to right).  The values quoted by the labels in the figure are
the (rounded) central values of the bins. In constructing this plot
from the data, we are grouping together phases from different pairs, with different
noise and resolution properties. In order to sensibly compare these
observed distributions to our model, we employ our forward-modelling
procedure. The forward-modelling approach, summarized in
\S~\ref{FM_summary} creates a sample of mock spectra which exactly
reproduces the distribution of separation, wavenumber, noise and
resolution of the data, but exploiting the much larger path length available
in the simulations, resulting in a much smaller statistical error.  For any model, this mock data
sample can therefore be used to facilitate 
a meaningful visual comparison between models and data. After forward-modelling
a simulation, we calculate the phase differences in the synthetic
pairs and group them in the same exact way as has been done for the
data. The phase angle PDFs of the similarly binned mock data
are shown as the red histograms in
Fig.~\ref{all_phases_pdf}. For this comparison we
have chosen the model in the parameter grid
that provides the maximum likelihood (at each redshift) evaluated as in
eqn. \ref{diaglik}.

Fig.~2 is built in an analogous way, except that we define
only one single bin which encompasses all the phases in each redshift bin, both
for the data points and for the simulated models. 
The shaded regions in that figure are calculated by taking as the boundaries the two 
semi-numerical models with the closest $\lambda_P$ to the 
16\% and 84\% percentiles of the $\lambda_P$ posterior distribution. 
Given the weak dependence on $T_0$ and $\gamma$, we choose the values of 
these parameters to be as close as possible to $T_0=15000$ K $\gamma=1 $,
i.e. around the middle of the allowed range.  

\section{Quantifying Systematic Errors}\label{consistency_tests}
The phase angle PDF method has not been previously analysed for
the impact of systematics on it have not been previously considered. 
In this section we
assess the robustness of our results with respect to possible 
sources of systematic error. 
First we consider the sensitivity of our results to the fit of the unabsorbed continuum
level  (\S~\ref{continuum_uncertainty}), then we determine how
sensitive we are to assumptions made about the the spectral resolution
(\S~\ref{res_stability}) and on our knowledge of 
the noise level (\S~\ref{noise_stability}). In \S~\ref{contaminant_stability}
investigate the impact of metal lines and LLS in decreasing the transverse
coherence of the \mlya\ forest in quasar pairs, and determine how
how much our results would change if the assumed abundance of these contaminants
is varied. 

\subsection{Continuum Fitting}\label{continuum_uncertainty}

\begin{figure}   

    \psfrag{r=150.0 kpc}[c][][1.]{$r_{\perp}=150$ kpc}
    \psfrag{r=250.0 kpc}[c][][1.]{$r_{\perp}=250$ kpc}
    \psfrag{r=350.0 kpc}[c][][1.]{$r_{\perp}=350$ kpc}
\centering \centerline{\epsfig{file=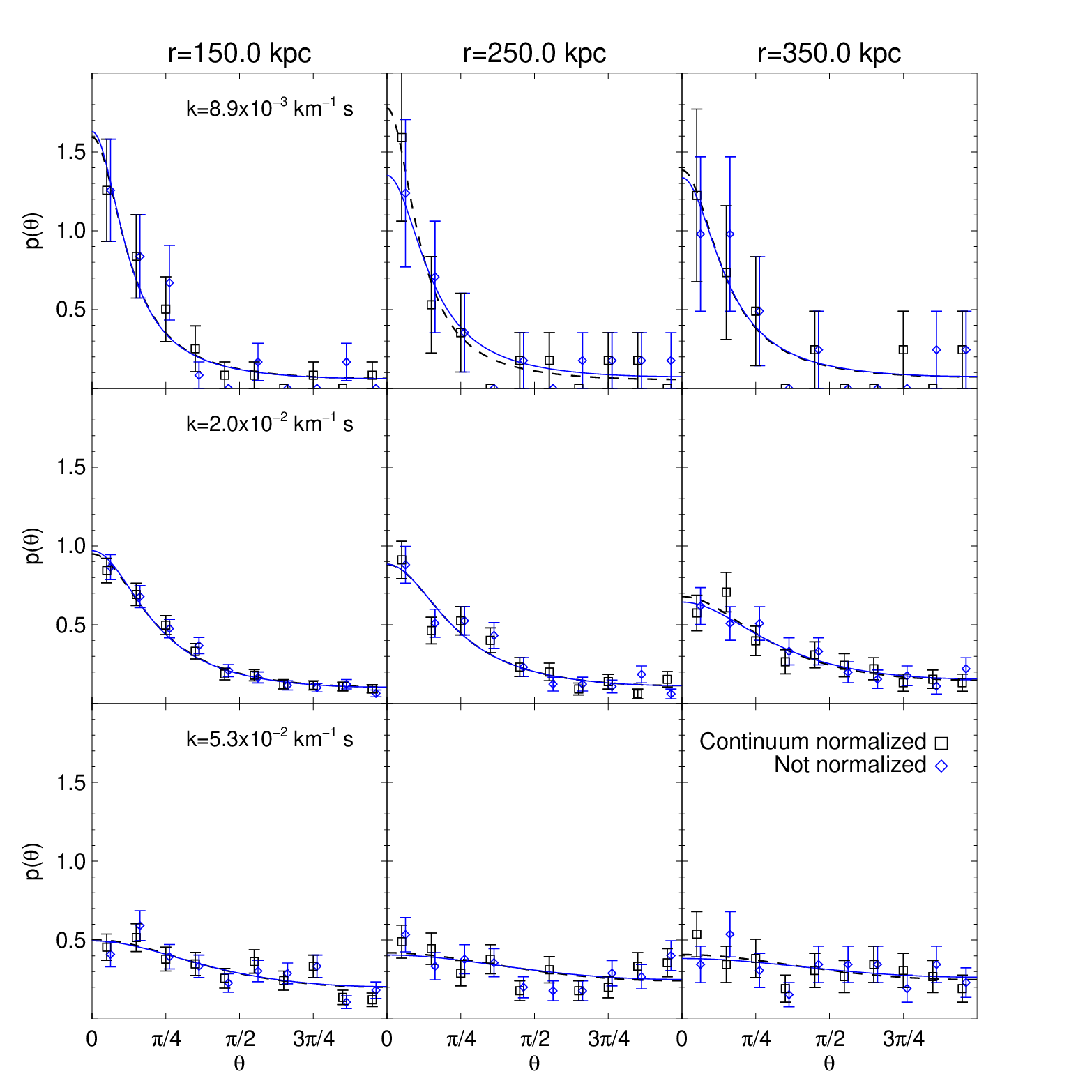,
      width=\textwidth}}
  \vskip -0.1in
\caption{\label{no_continuum} {\bf Effect of the uncertainty on 
continuum placement.}
The plot shows the phases of real data calculated at redshift
$z=2$ with continuum normalization (black squares),
or without continuum renormalization. (blue diamonds). The corresponding
wrapped-Cauchy best fits are shown as solid lines, matched by color. We show the 
comparison for the same $r_{\perp}$ and $k$ bins as in Fig.~\ref{fig:method_comparison}. 
The distributions in the two cases agree well, and the wrapped-Cauchy
fits are in most cases close to each other, 
implying that the phase distributions
in the three cases are statistically equivalent. This demonstrates that  
the phase statistic is insensitive to uncertainties in the continuum 
placement.
}
\end{figure}

Phase differences are insensitive to the normalization of the flux,
and this fact led us to argue that they are robust against
uncertainties in our continuum fitting. This statement would be
mathematically exact if continuum error could be described as simply a
multiplicative uncertainty in the flux normalization. However, in
realistic spectra the true underlying continuum may have emission
lines or a different slope than the one estimated by continuum
fitting, but as long as these features affect scales larger  
than the largest modes we analyse ($v \gtrsim 1500$ km/s)
we expect little effect on the phase angle PDF. To prove this
explicitly we adopt a conservative approach: we calculate the
phase angle PDF directly from the observed flux without
  fitting the continuum of the spectra at all. We then compare the results
with the standard case of continuum-normalized spectra.  The results
are shown in Fig.~\ref{no_continuum} for our redshift bin at $z=2$. The agreement between
the non-normalized flux and the standard continuum-normalized case 
is clear at all $r_{\perp}$ and $k$ (binned as in
Fig.~\ref{fig:method_comparison}).  The differences
between the best-fit wrapped-Cauchy distributions, plotted as
the solid lines are even smaller.  
The only significant discrepancy occurs in a single bin
($r_{perp}=250$ kpc, $k=8.9 \times 10^{-3}$ s/km, top center),
which may be caused by a strong feature at large scales in the
non-normalized continuum of a spectrum.  This difference is however smaller
than the statistical error indicated by the errorbars.

\subsection{Resolution}\label{res_stability}

In \S~\ref{power_spectrum_test} we described how we used the 
line-of-sight flux power spectrum from a distinct sample of high-resolution quasar spectra
to calibrate our resolution estimate. In this section we present a test
to show that our results are not significantly impacted by 
the exact value of the resolution that we assume.
This test is performed by repeating the measurement after varying 
our assumptions on the spectral resolution, exploring the cases 
where it is chosen to be 10\% higher or lower then the default value we adopted.

\begin{figure}   
\centering \centerline{\epsfig{file=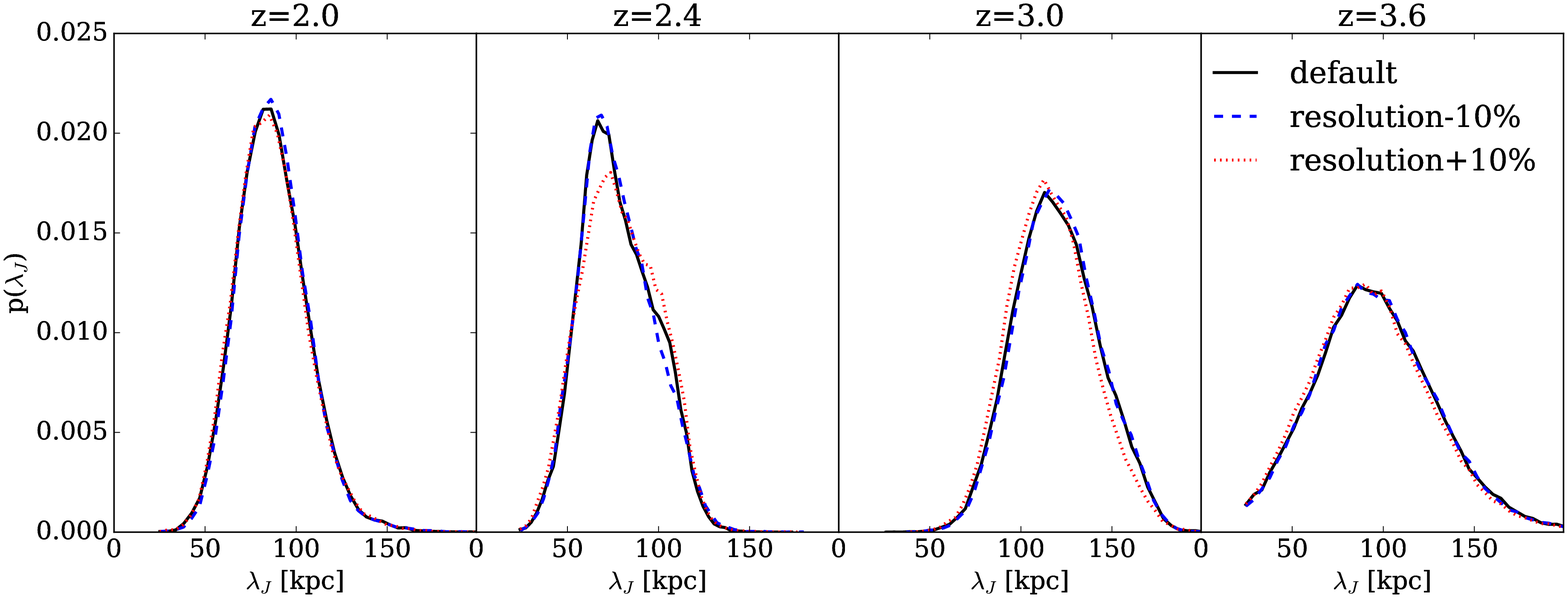,
      width=\textwidth}}
  \vskip -0.1in
\caption{\label{fig:res_stability} {\bf Impact of the uncertainty
	on resolution modeling.}
	Posterior probability distribution 
	for $\lambda_P$ under different assumptions regarding the 
	resolution. The 
	default run is shown in blue, where the FWHM is rescaled to achieve
	a good match of the flux power spectrum, as explained in the text. 
	The green and the red curves show cases where the resolution is assumed
	to be 10\% lower or higher, respectively. 
	At all redshifts the variation of the median value in the posterior
	distribution of the pressure smoothing scale is
	less than  1\%, more than an order of magnitude smaller than the statistical error 
	on $\lambda_P$. The exact values are reported in Table~\ref{tab:systematics}.
	} 
\end{figure}

The results of this resolution test are illustrated in
Fig.~\ref{fig:res_stability}. The bias in $\lambda_{P}$ resulting
from the resolution correction are in all cases a tiny fraction of our
statistical errors. We calculate the differences between the median of
the posterior distribution of $\lambda_P$ in the default run and in
these two test runs.  We choose the the larger of these two
differences as our estimate for the error associated with our
uncertain knowledge of the resolution, which amounts to less than 1\%
for all redshift bins (Table~\ref{tab:systematics}).  
These errors are an order of magnitude smaller than the
purely statistical uncertainties, which we quantify as the difference
$\Delta^{+1\sigma}\lambda_P=\lambda_{P,84}-\lambda_{P,50}$ between the
84th percentile and the median of the posterior distribution of
$\lambda_P$, and the difference
$\Delta_{-1\sigma}\lambda_P=\lambda_{P,50}-\lambda_{P,16}$ between the
median and the 16th percentile (the relevant percentile values are
listed for all redshift bins in Table~\ref{tab:systematics}).  We
therefore conclude that our measurement is not substantially affected
by our lack of knowledge of the exact value of the spectral
resolution. This partly follows from the mathematical properties of
phases that they are invariant under convolution of the function with
symmetric kernels (see RHW for more details). Furthermore, we are
excluding from the analysis the high-$k$ Fourier modes beyond the
resolution limit (see \S~\ref{resolution}). The possibility of
narrowing the dynamic range in order to exclude unresolved modes is
one of the main motivations to adopt a Fourier-space statistic.

\subsection{Noise}\label{noise_stability}
We adopt the same approach described in the previous section to
quantify the impact of uncertainties in our noise model on our
results.  In particular we want to understand the bias on $\lambda_P$
we would obtain  if the noise model estimated by our data
reduction pipeline is incorrect, which would result in a systematic error
in our forward-modelling procedure. If our estimates of the noise
are too low, the coherence of the quasar pairs in our models would
be higher than they actually are in reality.  As a consequence, the
lower coherence of the data relative to the (incorrectly modelled) more coherent
models would be interpreted as a smaller pressure smoothing scale.  To quantitatively assess the
magnitude of this effect, we repeat the phase analysis under the assumption that the
noise level is 10\% higher or lower than the value estimated by our
data reduction pipeline.

\begin{figure}   
\centering \centerline{\epsfig{file=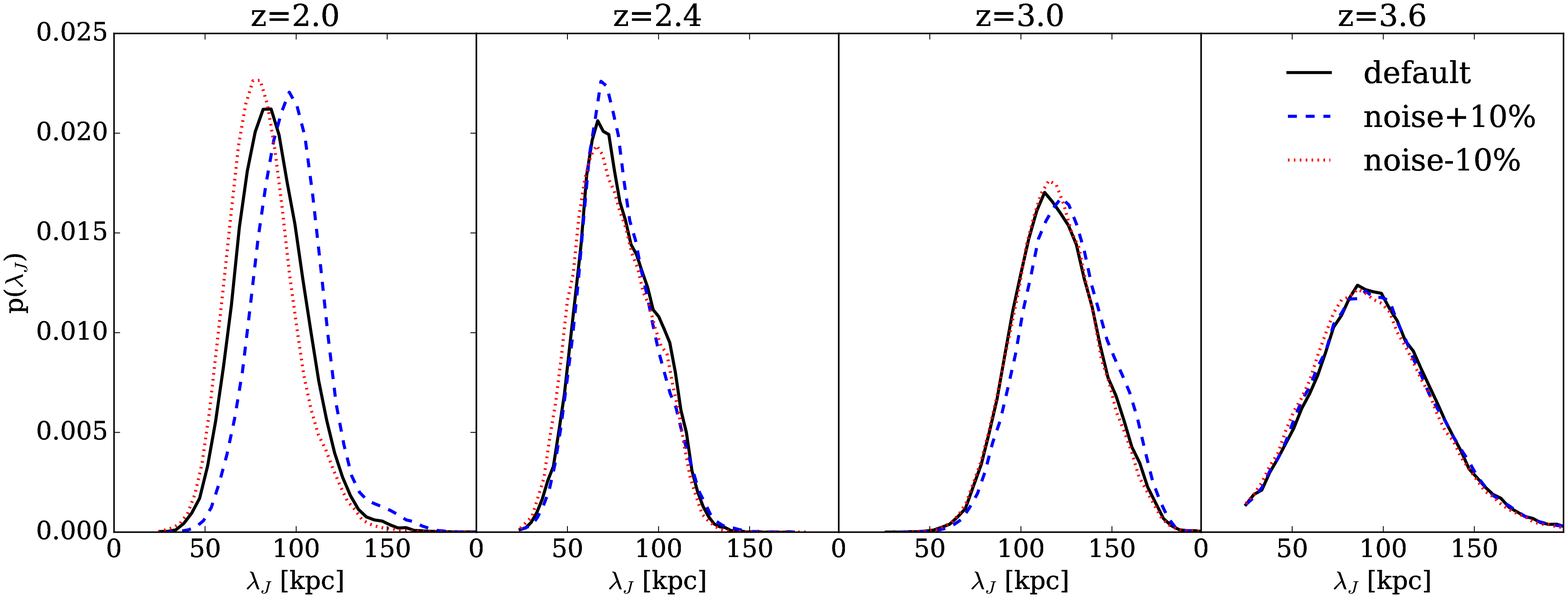,
      width=\textwidth}}
  \vskip -0.1in
\caption{\label{fig:noise_stability} {\bf Impact of the uncertainty
	on noise modeling.}
	Same as Fig.~\ref{fig:res_stability}, but this time adjusting 
	the noise level 	in the simulated spectra by 10\%. 
	The variation of the median $\lambda_P$ is different depending on
	the redshift and on the data quality. The precise values are reported
	in Table \ref{tab:systematics}. } 
\end{figure}

The results are shown in Fig.~\ref{fig:noise_stability}. The discrepancy
caused by varying the noise by 10\% from the pipeline estimates
differs substantially in the four redshift bins. This is caused by the 
variation in the signal-to-noise ratio within the data sample. Although
the lower redshift quasar pairs in our sample tend to be brighter on
average, the spectrograph throughputs degrade rapidly toward the UV and 
the signal from the \mlya\ forest is weaker at lower redshift. There is therefore
no reason to expect a clear redshift trend. 
Similar to the previous section, we estimate the associated uncertainties as
the highest of the two differences between the median values
of the posterior distribution of $\lambda_P$ in the $\pm 10\%$ test
runs and the default case, at each redshift. The values are reported
in the fifth column of Table~\ref{tab:systematics}. The highest
variations (9.9\% and 4.7\%) occur at $z=2$ and $z=3$, respectively, 
while at $z=2.4$ and $z=3.6$ the estimated uncertainty is below the 
percentage level.

\subsection{Metal Lines and Lyman Limit Systems}
\label{contaminant_stability}

\begin{figure}   
\centering \centerline{\epsfig{file=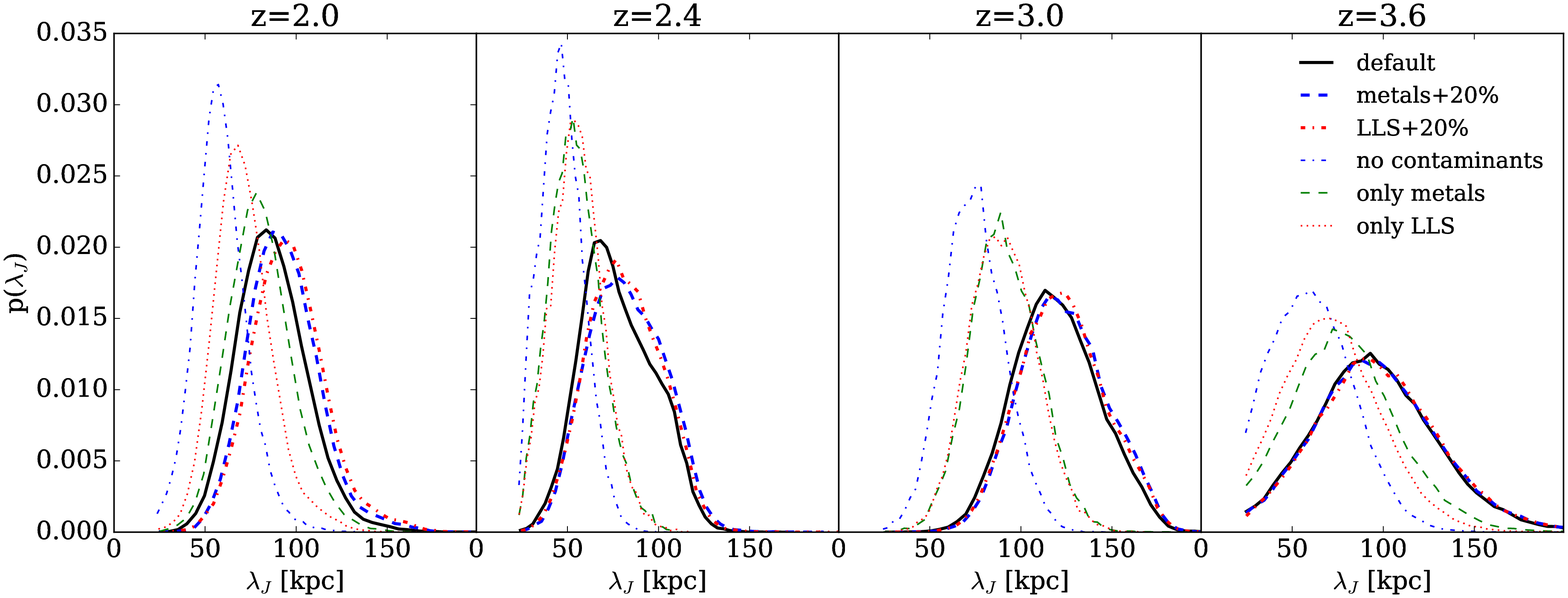,
      width=\textwidth}}
  \vskip -0.1in
\caption{\label{contaminants_tests} {Effect of contaminants modeling.} Posterior probability distribution for $\lambda_P$ under different assumptions regarding the 
properties of the contaminants. Our default 
run is represented by the blue line. The cyan dot-dashed line 
represent the case where
both metals and LLS are completely neglected. The effect of metals alone is demonstrated by the violet dashed curve, where LLS are not included. Conversely, the orange dotted lines show models containing only LLS 
contamination. Contamination in the models can lead to underestimation
of the smoothing scale if not corrected. 
We tested the stability of our results with respect
to the modelling of contaminants, by varying by 20\% the abundance of 
metal lines (green curves) or LLS (red). The consequent increase 
in the median value of the $\lambda_P$ posterior distribution is at most 9\% (see text for a complete
discussion).}
\end{figure}

The last systematic we need to address is the contamination due to
metal lines and LLSs. In our default forward model
both of these components are added to the synthetic spectra as 
described in \S~\ref{contaminants}. In Fig.~\ref{contaminants_tests}
we present the resulting posterior distributions for $\lambda_P$ when
one or both of these contaminants are neglected, compared to the 
default case where they are fully taken into account. 
It demonstrates that the pressure smoothing
scale is significantly underestimated if this effect is not modelled,
and the precision is overestimated. The effect is similar but
smaller if only one of the two type of contaminants is considered, as shown
by the posterior distribution of $\lambda_P$ when only LLS  
or metal lines are included.

We reiterate that we add 
these contaminants to simulated pairs in a completely random uncorrelated 
fashion, whereas in reality, given the small impact parameters probed
by our quasar pair sight lines, metals and LLSs could be significantly auto-correlated \cite{Ellison2007,Martin2010,Rubin2015}, and LLSs may also cross-correlate with the Ly$\alpha$ forest absorption \cite{Fumagalli2014}. 
If such correlations exist in reality and are strong enough to
influence the phase coherence, it would make the real quasar pair data
more coherent compared to our models, where we have assumed a
maximally incoherent distribution of metals and LLSs. In that case the
mismatch between metal and LLS coherence in the data relative to the model,
would be interpreted as a higher pressure smoothing scale in 
the data, i.e. we would overestimate $\lambda_P$.

To quantify the impact of our treatment of metals and LLSs on our
results we modify our forward modelling procedure, increasing the
respective abundance of metals and LLS by 20\%. The posterior
probability distribution for $\lambda_P$ obtained from our MCMC is
then compared to our default forward model, analogous to our treatment
of resolution and noise in the the previous sections. 

To increase the LLS abundance, we simply scale the 
factor $l_{z0}$ in eqn.~\ref{eq:lls_evolution}
up by 20\%. In this way we increase the occurrence of 
LLS at all column densities by the same amount.
Metal line contamination is artificially enhanced  
as follows. As described in \S~\ref{contaminants}, metals
are injected into our spectra by sampling co-eval spectral 
segments of lower-$z$ quasars, for
which the metal lines are not blended with the \mlya\ forest. In practice, only
a fraction of these spectral segments contain metal lines. To increase
the metal line abundance in our mock spectra, we simply artificially increase
the probability of picking a segment with metal lines by 20\%. 
On average this procedure increases the abundance of metal lines in the models
by the desired amount.
The results are shown in Fig.~\ref{contaminants_tests} and 
summarized in the last two columns of Table~\ref{tab:systematics}.

The variation on the estimated $\lambda_P$ (i.e. the median of the 
posterior distribution) caused by a change
in metal or LLS abundance by 20\% is in all case less than 10\%, with a decreasing trend towards high redshift. These discrepancies
are taken as an estimate of the error associated with the 
uncertainty on metal contamination and LLS. When added in
quadrature to the purely statistical uncertainty, they only 
 contribute to the total error by a small amount.

\subsection{Systematic Uncertainties: Summary}
In this section we have tested the potential biases associated with
uncertainties on spectral resolution, noise, metal contamination and
LLS. In Table~\ref{tab:systematics} we report all the corresponding
errors we estimated. These errors are always calculated as the
difference between the median of the posterior distribution of
$\lambda_P$ in the default case and the median in a test case where
these effects are varied. When we have two test cases where we vary
the systematic in opposite directions (as done for the noise and the
resolution), we conservatively take the maximum of the two
differences. We report these uncertainties, organized by type and by
redshift, in Table~\ref{tab:systematics}.  In calculating the total
error (i.e. statistical+systematic, last column in Table~\ref{tab:systematics}), we add the systematic errors in quadrature to
the purely statistical uncertainties from the MCMC posterior for $\lambda_P$ for our
default case. The latter are quantified as the difference
$\Delta^{+1\sigma}\lambda_P=\lambda_{P,84}-\lambda_{P,50}$ between the
84th percentile and the median of the posterior distribution of
$\lambda_P$, and the difference
$\Delta_{-1\sigma}\lambda_P=\lambda_{P,50}-\lambda_{P,16}$ between the
median and the 16th percentile.  The impact of systematic errors
is illustrated by the black extensions to the statistical errorbars (red) shown in
Fig.~4 of the main text, which have been calculated in this way.

\section{The Smoothing Scale in Hydrodynamical Simulation}\label{hydro comparison}

As we discussed in \S~\ref{dm_models}, we opt for a more flexible
semi-numerical model of the thermal state of the IGM based on
collisionless dark-matter only simulations to conduct our data
analysis. The reason for this is that it is a great computational
challenge to simulate the full set of thermal histories that could
give rise to a large range of pressure smoothing scale that the data
analysis requires.  In this section we discuss an ensemble of full
hydrodynamical simulations that we run.  Our approach is to
treat the hydro simulation as a mock dataset, and then run our
phase angle PDF method on the hydro simulations to infer the smoothing
scale in the context of the DM models.
We show that there is a tight
relation between the semi-numerical pressure smoothing scale in the DM
models, and the analogous scale in the hydro simulation. This forms
the basis for Fig.~4 in the main text,
where we applied the phase angle PDF method
to a set of hydro simulations, and we provide the details of how
that Figure was constructed.

\subsection{Hydrodynamical Simulations}\label{sec:nyx_sims}

For the hydrodynamics simulations, we use Nyx, an N-body/gas dynamics
code for large-scale cosmological simulations \cite{Nyx,Lukic15}.  It
follows the temporal evolution of dark matter gravitationally coupled
to an inviscid ideal fluid in an expanding universe. The gas dynamics
are evolved using a finite volume approach on a three-dimensional
Eulerian grid.  Dark matter is represented as discrete particles
moving under the influence of the gravitational field. The same mesh
structure that is used to update fluid quantities is also used to
compute the gravitational field and to evolve the particles via a
particle-mesh method. Adaptive mesh refinement (AMR) can be used for
the hydrodynamics and self-gravity solver.  However, to model
Lyman-alpha absorption, it is most important to correctly describe the
thermodynamic state of the IGM at relatively low densities, from below
the cosmic mean, to about ten times the mean density.  Because these
are vast regions of the universe, employing AMR to resolve them does
not provide a significant advantage.  In addition, these regions are
devoid of galaxies, thus physical processes related to galaxy
formation are not expected to play a significant role in determining
thermal state of the IGM \cite{Kollmeier06}.  For these reasons, we do
not use the AMR capability in these runs.
The simulations we run are periodic-boundary cubes, 20 Mpc$/h$ on a side,
with $1024^3 $ cells, resulting in approximately 20 kpc/$h$ cell size. This box
size and resolution are chosen in order to achieve convergence in the low
density IGM \cite{Lukic15}.   In addition, we have verified that the phase angle PDF
determined from pairs of spectra is converged in simulations of this
resolution, and that it is insensitive to the box size. In these
simulations we use the following cosmological
parameters, consistent with
the latest cosmological constraints from the CMB  
\cite{Planck2015}: $\Omega_m=0.3192,\Omega_b=0.04964,
\Omega_{\Lambda}=0.6808,h=0.67038$ and $\sigma_8=0.826$.
Note that these parameters are slightly different than those used in the 
dark-matter simulations described in \S~\ref{dm_models}. 
In any case, changes in the dark-matter models only affect the 
definition of $\lambda_P$, not its physical interpretation and its
connection to the thermal history which is instead calibrated on
hydrodynamical simulation, as it will become clear in section 
\S~\ref{phase_analysis_hydro}.
In any case, the sensitivity of our results to the cosmological parameters
is weak, as shown in  \S~\ref{sec:convergence_test}.

\begin{figure}   
\centering \centerline{\epsfig{file=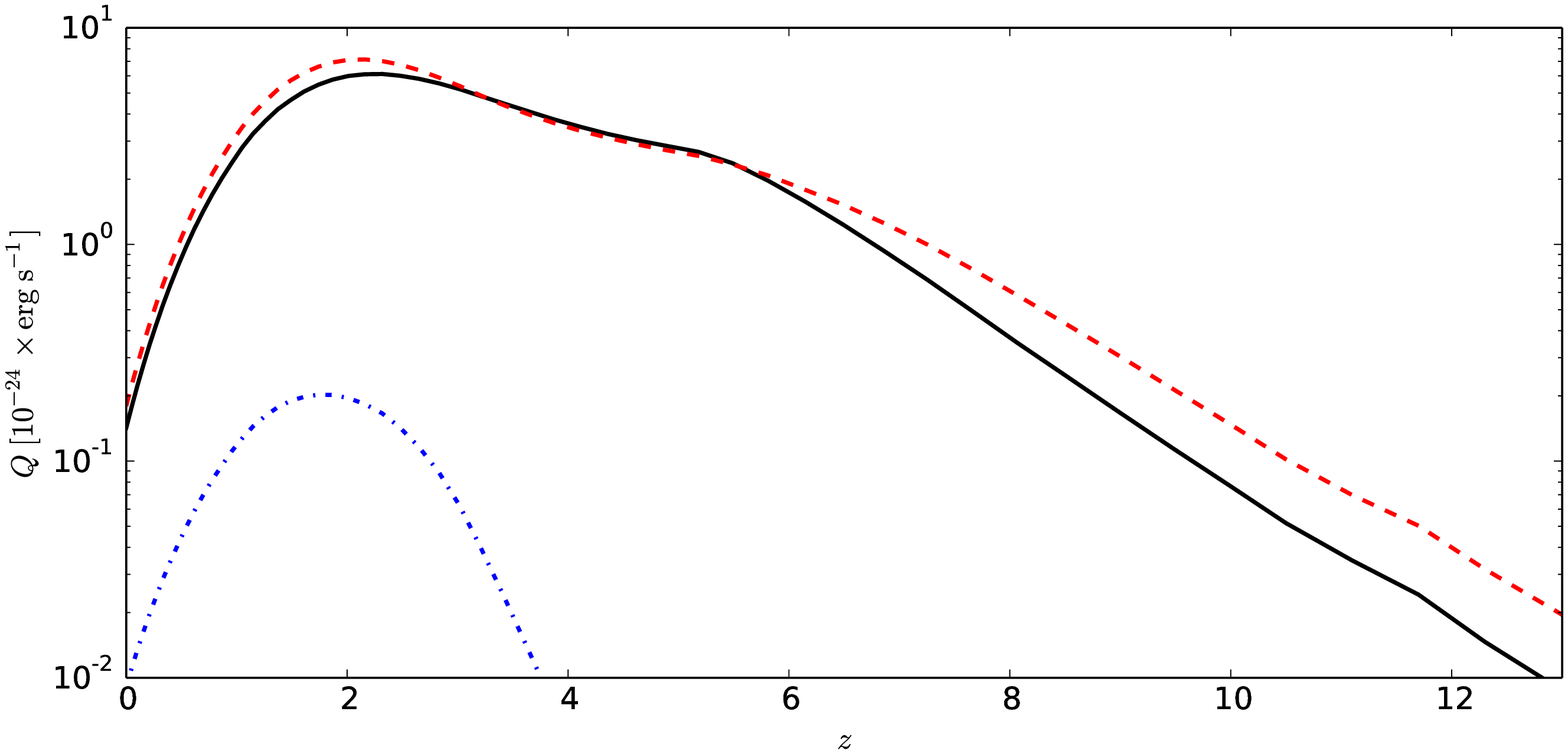,
      width=0.9\textwidth}}
  \vskip -0.1in
\caption{\label{fig:heating_rates} 
{\bf Photoheating rates per ion in our models} 
for the species \ion{H}{i} (black), 
\ion{He}{i} (red dashed) and \ion{He}{ii} (blue dotted) as a function of redshift in the HM12 model.
These rates form the basis of the range of thermal histories that 
we explore with our hydrodynamical simulations. In particular, in the 
late reionization heating model presented in the main text the 
photoheating rates are the same for $z\leq 7 $ and set to zero 
for $z>7 $, while in the increased heating model they are  
multiplied by three at all redshifts.} 
\end{figure}

As described in \cite{Nyx,Lukic15}, our simulations account for the
radiative heating and cooling via source terms in the energy
equations.  The gas in the simulations is assumed to be optically thin
and in ionization equilibrium with a spatially uniform ultraviolet
background (UVB), which results from the collective emission of
galaxies and quasars.  In this work, we use photoionization and
photoheating rates from the \cite[{\rm HM12, see Fig. 
 \ref{fig:heating_rates}}]{Haardt12} synthesis model which results in
an IGM which is already heated up to $T\sim 10^4~{\rm K}$ at $z = 15$
\cite{Puchwein15}, which is also the  de facto redshift of
reionization in simulations where the HM12 photoionization and
photoheating rates are employed. To study the impact of thermal and
reionization history on the pressure smoothing scale, we also consider
 models where the photoheating and photoionization rates are set to
zero beyond some given $z_{\rm reion}$ . At $z_{\rm reion}$
the HM12 uniform UVB and photoheating rates are then switched on, and used
for the rest of the time evolution. In the
main text we compare our results with the standard HM12 model, a
model where the reionization redshift is $z_{\rm reion}$ = 7
(”late reionization heating”) and one 
where the heating rates are increased by a factor $A = 3$, to
test whether a
increased-heating scenario might be consistent with the temperature 
measurements from \cite{Lidz09}. Rescaling heating rates is a 
standard phenomenological 
approach for modelling the thermal history, which was first adopted in 
\cite{BryanMach00}.
Further details of the three simulations we run are specified in
Table~\ref{tab:sim}.	

We analyzed a set of 13
hydrodynamical simulations, where
a wider range of heating parameters and reionization redshifts
are explored, resulting in a larger 
variation of the pressure smoothing scale. 
Beside the rescaling factor of the heating rates $A$
and the reionization redshift $z_{\rm reion}$,
we also consider models where the heating depends on density
according to  $\Delta^{B}$ \cite{BeckerBolton2011}, with $B$ being
a free parameter.  
The parameters of this set of simulations are 
reported in Table~\ref{tab:sim2}.
This allows us to better characterize the relation between the 
parameters of the thermal history in hydrodynamical models and the 
smoothing scale of the IGM measured with  phase differences and defined
in our semi-numerical dark-matter models.	
\subsection{Phase Analysis of the Hydrodynamical Simulations}\label{phase_analysis_hydro}
In this section we describe how we apply the phase angle PDF method to
a set of mock quasar pairs drawn from the hydrodynamical simulations. The
goal is to characterize the pressure smoothing of the IGM in the hydrodynamical
models using the transverse coherence of phases, consistent with what
we have done with the observational data.  This establishes the
foundation for a meaningful comparison between our measurement and the
predictions of the hydrodynamical simulations we run.

As explained in \S~\ref{dm_models}, the phase angle PDF analysis is
calibrated on a set of semi-numerical models where the smoothing is
imposed by convolving the dark matter density distribution with an
approximately-Gaussian kernel.  This approximation allows us to relate an
observable, namely the distribution of phase differences in quasar
pairs, to a definition of smoothness in the density field
($\lambda_P$).  Since this relation is used to characterize the
observational results described in the main text, we apply it to the
Ly$\alpha$ forest skewers drawn from the the hydrodynamical simulations in
order to define the pressure smoothing scale of the different hydro
simulations that we consider.

Schematically, the procedure we adopt can be described as follows:
\begin{itemize}
\item We draw 10,000 synthetic pairs of skewers from the hydrodynamical
simulation, choosing separations in multiples of the grid spacing 27.9 kpc, for a total 
path length of 140 Gpc. This constitutes our mock (noiseless) dataset.

\item  We calculate phase differences for all the pairs in the mock sample,
imposing the same limits on the dynamical range as in the data 
($k<0.1$ s/km). 

\item We evaluate the likelihood of the obtained set of phases using
the probability distributions predicted by our grid of DM-based models
at the same transverse separations.

\item The pressure smoothing scale of the simulation is defined as the pressure
smoothing scale $\lambda_P$ of the maximum-likelihood DM model. To achieve better
precision we use Gaussian processes to interpolate the  
likelihood in parameter space. 
\end{itemize}
We repeated this method for all three simulations listed in Table 
~\ref{tab:sim} and at all four redshift bins. Simulations snapshots 
are approximately drawn from the central redshift of the four 
measured bins, more precisely at $z=2,2.4,3.0,3.6$.
These values are used to determine the simulation predictions between $z=2$ 
and $z=3.6$ in Fig.~4. 
Below we explain how we extrapolate these prediction to higher and 
lower redshift. 

\subsection{Definition of the Pressure Smoothing Scale in Hydrodynamical Simulations}\label{sec:jeans_in_hydro}

Whereas $T_0$ and $\gamma$ can be easily estimated by fitting the temperature-density
distribution with a power law, defining the pressure smoothing scale in hydrodynamical
simulations is not trivial. The gas density power spectrum does not exhibit a sharp
pressure smoothing cutoff, because dense gas in collapsed halos
dominates the small-scale power masking pressure smoothing effects  \cite{Kulkarni2015}.
A possible solution has been proposed in \cite{Kulkarni2015}, where they 
define a new quantity, the real-space \mlya\ flux $F_{\rm real}$, which
naturally suppresses the impact of this dense gas, and is thus robust against the poorly
understood physics of galaxy formation, revealing pressure smoothing in the diffuse IGM.
The $F_{\rm real}$ field is calculated as the transmitted flux of \mlya\ photons 
in the fluctuating Gunn-Petersson approximation, defined by
\begin{equation}\label{eqn:freal}
F_{\rm real}(x)= \exp \left[-\frac{3\lambda^3_{\alpha}\Lambda_{\alpha}}{8\pi H(z)} n_{\rm \ion{H}{i}} \right]
\end{equation}
where $\lambda_{\alpha}= 1216$\AA\ is the rest-frame \mlya\ wavelength,
$\Lambda_{\alpha}$ is the Einstein $A$ coefficient of the transition, $H(z)$
is the Hubble parameter and $n_{\rm \ion{H}{i}}(x)$ is the neutral hydrogen number
density at the point $x$.
In the optically-thin regime, where $n_{\rm \ion{H}{i}} \propto \rho^2$
this is a non linear transformation of the  density field
that suppresses high densities, but preserves isotropy.

In \cite{Kulkarni2015} it is shown that $F_{\rm real}$ power spectrum 
is accurately described by a simple fitting function with cutoff at 
$\lambda_{P,{\rm sim}}$:
\begin{equation}\label{equation:freal_fit}
\Delta_F(k)= A k^n \exp[-(k\lambda_{P,{\rm sim}})^2] ,
\end{equation}
where $A$, $n$ and $\lambda_{P,{\rm sim}}$ are the free parameters.
This characterization allows a 
rigorous quantification of the pressure smoothing in hydrodynamical 
simulations. They have also shown that the value of $\lambda_{P,{\rm sim}}$
is strictly dependent on the thermal history, confirming its 
physical origin from the pressure smoothing of the baryons.

It is natural to consider the relation between this definition
of pressure smoothing and the one coming from the phase angle PDF analysis
described in \S~\ref{phase_analysis_hydro}, where the pressure
smoothing scale is determined by applying our phase-angle PDF method,
calibrated with semi-numerical dark matter models, to pairs of Ly$\alpha$
forest skewers drawn from hydrodynamical simulations. 
A strong correlation
between the two would be a convincing argument that our dark-matter
based modeling approach is actually measuring the small-scale coherence
of the baryon density field, which is determined by the thermal history
of the gas. To verify this, we apply our phase angle PDF analysis
to the extended set of 13 hydrodynamical simulations
described in Table~\ref{tab:sim2}.
We use the phase angle PDF analysis method described in \S~\ref{phase_analysis_hydro}
to 'measure' $\lambda_P$ for each simulation model. 
We then compute the 3D power spectrum of $F_{\rm real}$ in each of these
hydrodynamical models. The expression in eqn.~\ref{equation:freal_fit} is 
fitted to these power spectra, providing an estimate of  $\lambda_{P,{\rm sim}}$ 
for each model. 

\begin{figure}   
\centering \centerline{\epsfig{file=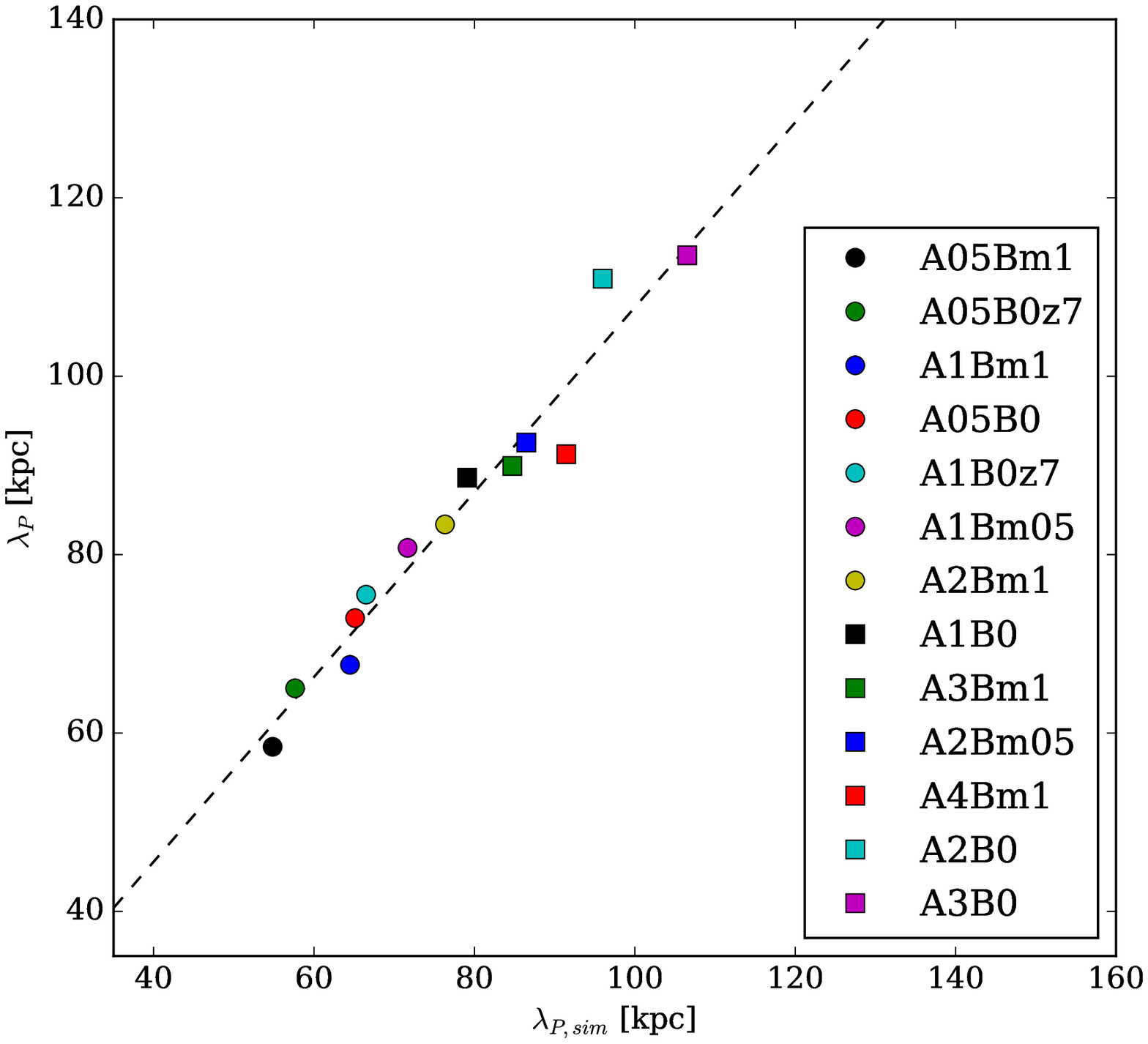,
      width=0.7\textwidth}}
  \vskip -0.1in
\caption{\label{fig:lj_lf} {\bf Pressure smoothing in hydrodynamic
simulations.}
The points show the relation between the pressure smoothing scale 
$\lambda_{P,{\rm sim}}$ 
of the hydrodynamical simulation
defined by the cutoff in the power spectrum of $F_{\rm real}$ ($x$ axis), and 
the scale $\lambda_P$ derived by the phase analysis of a mock 
sample of pairs extracted from the same simulations ($y$ axis), calculated
at $z=3$. The dashed
line represents a polynomial fit to the distribution. The legend 
lists simulations from  Table~\ref{tab:sim2},
sorted in increasing  order of $\lambda_{P,sim}$, such that 
the points appear left to right in the plot.} 
\end{figure}

At the end of this procedure we have a set of values of $\lambda_P$
(from phases and DM modeling) and of $\lambda_{P,sim}$ (from the fit
of the $F_{\rm real}$ power spectrum in hydro simulations) for all of
the thermal models. We show a comparison of these two quantities at
$z=3$ in Fig.~\ref{fig:lj_lf}. There is a clear monotonic trend between
the two measures of pressure smoothing, and an almost one-to-one
relation with a scatter of less than 5\% (see Table~\ref{tab:sim2} and
Fig.~\ref{fig:lj_lf}), indicating that they are a proxy
for the same physical quantity. Hence we argue that the
properties of $\lambda_{P,{\rm sim}} $ analysed in \cite{Kulkarni2015} are
shared by $\lambda_P$. In particular this demonstrates the sensitivity of
our measured value of $\lambda_P$ to the thermal history of the IGM.

We perform a linear fit of the relation between $\lambda_P$ and 
$\lambda_{P,{\rm sim}}$ at the four redshifts  used
for our measurement. We then assume that the fitted functions at $z=2$ and $z=3.6$
 are a 
good approximation for $\lambda_P(\lambda_{P,{\rm sim}})$ at redshift lower
and higher than the four bins, respectively. In this way we can use the values of
$\lambda_{P,{sim}}$ determined at each snapshot of the simulations in order to extrapolate
the model predictions for $\lambda_P$. This is done purely for illustrative
purposes, as we do not have a measurement to compare with outside of the
redshift  interval probed by our data $z\in[1.8,3.9]$.
These extrapolations are shown
in Fig~4 for redshifts $z<2$ and $z>3.6$. 

\subsection{Convergence and Sensitivity to Cosmological Paramaters}\label{sec:convergence_test}
In this section we quantify the level of convergence of the pressure smoothing
scale in hydrodynamical simulations with respect to resolution and box size,
and also investigate its dependence on the cosmological
parameters.  We will focus here on the hydrodynamical runs because the
dark matter simulation (described in \S~\ref{dm_models}) has both a
higher resolution and a larger size, and its resolution convergence
criteria were already tested in \cite{Rorai13}.

Our approach is to apply exactly the same procedure described in
\S~\ref{phase_analysis_hydro} in order to estimate $\lambda_P$ for a
suite of simulations where resolution, box size, or cosmological
parameters are varied, and then compare the results with the value of
the default hydrodynamical simulations that we compared to our
data. All tests are performed at $z=3$.

\subsubsection{Resolution Convergence}
The resolution convergence test is performed on three simulations 
in a 10 Mpc$/h$ box. The default setting is represented by a
run with $512^3$ cells, which has the same cell size of our standard
boxes ($l_c=$19.5 kpc$/h$). We then have two boxes with $256^3 (l_c=39.1$
kpc$/h$) and $1024^3 (l_c=9.8$ kpc$/h$) cells. The initial 
conditions and thermal histories are the
same in the three cases. The cosmology is different than the one 
employed in our thermal grid (\S~\ref{sec:nyx_sims}), as these models
were run with cosmological parameters consistent with the Wilkinson
Microwave Anisotropy Probe (WMAP) 7-year data \cite{Komatsu11}: $\Omega_m=0.275,\Omega_b=0.046,
\Omega_{\Lambda}=0.725,h=0.702$ and $\sigma_8=0.816$.

The relative discrepancy of $\lambda_P$ (with respect to the
default run) as a function
of  the cell size $c_l$ is shown in the left panel of 
Fig.~\ref{fig:convergence_tests}. While the lowest-resolution run is 
clearly not converged, the discrepancy between the default and the 
high-resolution simulations is approximately $3\%$, which is an order of 
magnitude lower than our statistical errors. 

\subsubsection{Box Size Convergence}
We examine  four hydrodynamical models with box size
$L=10,20$ (reference value) $,40,80$ Mpc$/h$ and number of cells $n=512^3,1024^3,2048^3,4096^3$,
respectively, such that the resolution is kept constant. 
All four runs have the same thermal history and 
cosmology.
The differences of $\lambda_P$ relative to the fiducial run are 
are reported in the central panel of Fig.~\ref{fig:convergence_tests}.
The simulations employed in our analysis are run in a 20 Mpc$h/$
box, which according to this test is converged at 5\% level 
with respect to the 80 Mpc$/h$ box run.

\subsubsection{Sensitivity to Cosmological Parameters}
\begin{figure}   
\centering \centerline{\epsfig{file=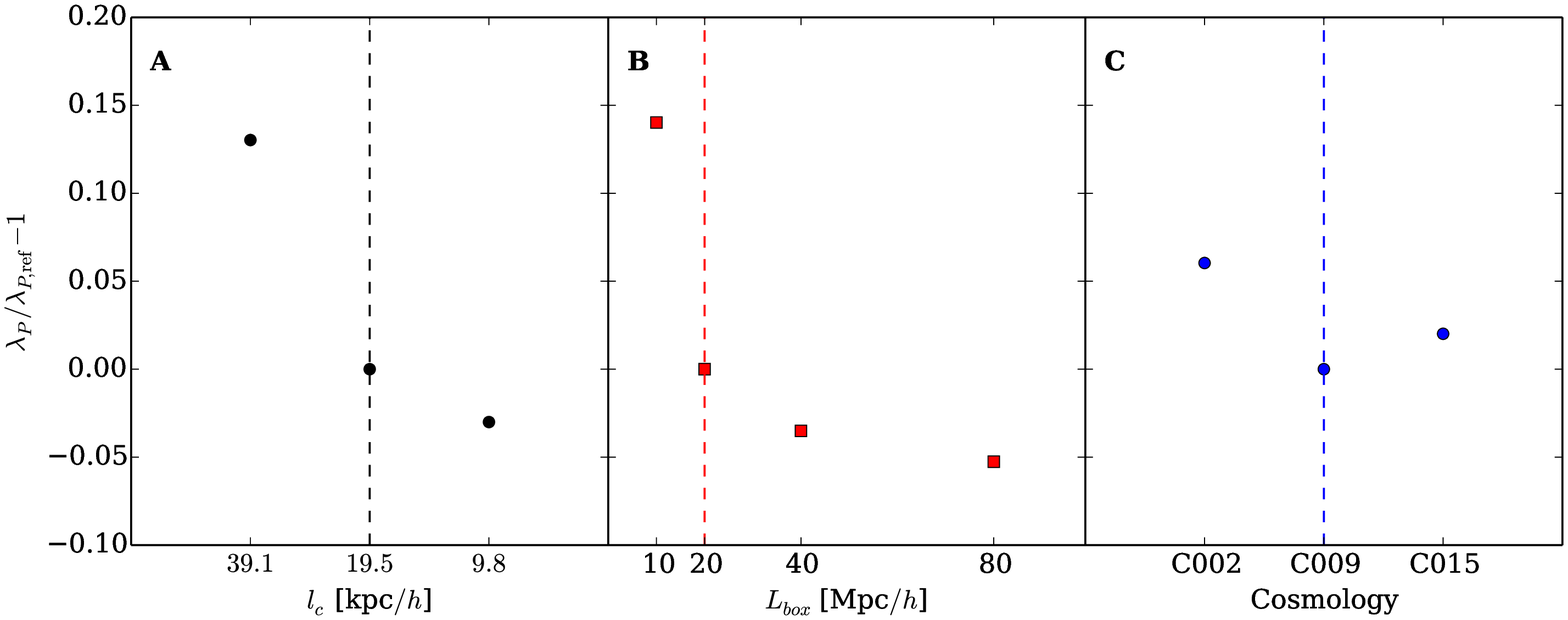,
      width=\textwidth}}
  \vskip -0.1in
\caption{\label{fig:convergence_tests} {\bf Results of the convergence
tests.} All the plotted values of $\lambda_P$ are calculated at 
$z=3$ via a phase analysis of synthetic pair spectra, as described 
in \S~\ref{phase_analysis_hydro}. Panel A, resolution test:
Relative difference in $\lambda_P$ with respect to the default run (marked by
a vertical dashed line) as 
a function of the cell size $l_c$ for three 
simulations in a 10-Mpc$/h$ box. 
Panel B, box size test: relative difference in $\lambda_P$ as 
a function of  the box size $L_{box}$. The default run, marked by the vertical
dashed line, has a box size of 20 Mpc. The resolution of the four simulation is
kept constant by choosing the appropriate number of cells. 
Panel C, cosmology dependence: sensitivity of the 
estimated $\lambda_P$ to the cosmological parameters relative to the 
fiducial model (C009, blue dashed line). The three points
represent three models in which cosmological parameters are varied consistently
with the current observational constraints (see text for details).}
\end{figure}

To investigate sensitivity to cosmological parameters, we
study three simulations in boxes of 20 Mpc$/h$ and
$1024^3$ cells. The reference run (labelled
as C009 in Fig.~\ref{fig:convergence_tests}) is the A1B0 model of the thermal
grid presented in Table~\ref{tab:sim2}, 
with fiducial values for the cosmological parameters (see \S~\ref{sec:nyx_sims}).
The other two models have the exact same thermal histories and their 
initial conditions were created using the
same random seeds, but were run using the following cosmologies:
$\Omega_m=0.298,\Omega_b=0.048,\Omega_{\Lambda}=0.702,h=0.686,
\sigma_8=0.873,n_s=0.974$ (cosmology 'C002'); 
$\Omega_m=0.333,\Omega_b=0.052,\Omega_{\Lambda}=0.667,h=0.658,
\sigma_8=0.757,n_s=0.971$ (cosmology 'C015'). These two models
were selected from the Planck posterior distribution in order
to produce conservatively large differences 
(around 5-$\sigma$) in the linear power spectrum 
compared with the mean value, used in the default case (C009).

The main effect of the cosmological parameters is modifying the 
relation  between real-space and velocity-space separations 
via $H(z)$. This means that the same smoothing scale $\lambda_P$
affects different Fourier modes of the  velocity space in  
different cosmologies, hence the uncertainty on the cosmological
parameters propagates to an uncertainty on $\lambda_P$.
This uncertainty is illustrated by the differences in $\lambda_P$
for the three model plotted in the right panel of
Fig.~\ref{fig:convergence_tests}. Large variations (5-$\sigma$)
within the Planck posterior distribution \cite{Planck2015} 
allow for shifts of $\sim 6\%$ in the
values of $\lambda_P$.

A second cosmological effect is the change in angular distance,
which sets the relation between angles in the sky and physical 
separations between the sight lines. However we have checked that 
the change induced on transverse separations are significantly 
below $1\%$, for the considered set of cosmologies.

In conclusion, the achieved level of convergence of $\lambda_P$ with respect to 
resolution and box size is around 5\%, which is also comparable to the 
change induced by varying the cosmological parameters within the current 
observational constraints. When added in quadrature to the statistical
error estimated in \S~\ref{consistency_tests}, these effects does not 
substantially contribute to the overall uncertainty.

\section{Data and Code Release}
The quasar pair data analyzed in this work can be retrieved online from public archives. All files relative to XSHOOTER observations are stored at the European Southern Observatory archive (www.archive.eso.org) and 
exposure frames for  ESI and LRIS pairs can be found at the Keck repository  (www2.keck.hawaii.edu/koa/public/koa.php). We have made the MagE data  available at the University of Cambridge data repository  (www.repository.cam.ac.uk/handle/1810/262888).

Simulated spectra from the hydrodynamical simulations and velocity/density
sight lines from the dark-matter simulations (which can be used to calculate the Lyman alpha flux for any
thermal model) have been published in a dedicated
Zenodo repository (http://doi.org/10.5281/zenodo.290181). 
There we also provide the configuration file to obtain the same initial conditions used for the hydrodynamical models described in \S~\ref{sec:nyx_sims}. 
The hydrodynamical code used for such models is available at https://github.com/BoxLib-Codes/Nyx.
Useful codes for the post-processing of simulations and for the statistical analysis employed in this work are also available at 
https://github.com/arorai/QSO-pairs-IGM-smoothing.

\begin{landscape}
\newcounter{tempfootnote}
\setcounter{tempfootnote}{\value{footnote}}
\setcounter{footnote}{0}
\renewcommand{\thefootnote}{\alph{footnote}}
 
{\footnotesize
\begin{longtabu}{X[4,c] X[c] X[c] X[c] X[c] X [c] X[c] X[2,c] X[2,c] X[2,c]}
\caption{ {\bf List of the analyzed segments of overlapping \mlya\ forest in the pair sample.} The table columns are: \\
$z_{\rm bg}$: redshift of background quasar; $z_{\rm fg}$: redshift of the foreground quasar; $z_{\rm min}$: minimum redshift of the segment; $z_{\rm max}$: maximum redshift of the segment; $\psi$ angular separation between foreground and background quasar (arcsec); $r_{\perp}$: impact parameter at the mean redshift of the segment (comoving kpc); R: mean FWHM resolution in the segment (km s$^{-1}$); ${\rm S}\slash {\rm N}_{\rm bg/fg}$: mean signal-to-noise ratio per Angstrom in the segments (background and foreground). \label{tab:sample}}\\
\hline\hline \\
Name & $z_{\rm bg}$ & $z_{\rm fg}$ & $z_{\rm min}$ & $z_{\rm max}$ & $\psi$ [\arcsec] & $r_\perp$[kpc] & Instrument &  R[km s$^{-1}$] & ${\rm S}\slash {\rm N}_{\rm bg/fg}$\\ [0.5ex]
\hline
\endfirsthead
Name & $z_{\rm bg}$ & $z_{\rm fg}$ & $z_{\rm min}$ & $z_{\rm max}$ & $\psi$ [\arcsec] & $r_\perp$[kpc] & Instrument &  R[km s$^{-1}$] & ${\rm S}\slash {\rm N}_{\rm bg/fg}$\\ [0.5ex]
\hline
\endhead
\begin{tabular}{@{}c@{}}SDSS~J000450.90-084452.0 \\ SDSS~J000450.66-084449.6\end{tabular}&3.00&3.00&2.45&2.69& 4.3&127&XSHOOTER&30&25.1/11.4\\
\begin{tabular}{@{}c@{}}SDSS~J000450.90-084452.0 \\ SDSS~J000450.66-084449.6\end{tabular}&3.00&3.00&2.82&2.91& 4.3&133&XSHOOTER&30&34.4/12.9\\
\begin{tabular}{@{}c@{}}SDSS~J005408.47-094638.3  \\ SDSS~J005408.04-094625.7 \end{tabular}&2.12&2.12&1.80&2.02&14.1&352&LRIS&171&142.7/30.0\\
\begin{tabular}{@{}c@{}}SDSS~J011707.52+315341.2 \\ SDSS~J011708.39+315338.7\end{tabular}&2.64&2.62&2.33&2.55&11.3&322&ESI&64&31.5/20.2\\
\begin{tabular}{@{}c@{}}SDSS~J025049.09-025631.7 \\ SDSS~J025048.86-025640.7\end{tabular}&2.84&2.82&2.32&2.47& 9.6&271&ESI&48&21.2/36.9\\
\begin{tabular}{@{}c@{}}SDSS~J033237.19-072219.6  \\ SDSS~J033238.38-072215.9 \end{tabular}&2.11&2.10&1.80&2.01&18.1&450&LRIS&171&19.5/19.0\\
\begin{tabular}{@{}c@{}}SDSS~J073522.43+295710.1  \\ SDSS~J073522.55+295705.0 \end{tabular}&2.08&2.06&1.80&1.99& 5.4&135&LRIS&172&61.5/39.9\\
\begin{tabular}{@{}c@{}}SDSS~J081329.49+101405.2  \\ SDSS~J081329.71+101411.6 \end{tabular}&2.08&2.06&1.80&1.99& 7.1&177&LRIS&173&20.9/17.2\\
\begin{tabular}{@{}c@{}}SDSS~J091338.97-010704.6 \\ SDSS~J091338.30-010708.7\end{tabular}&2.92&2.75&2.35&2.65&10.8&311&XSHOOTER&69&45.7/24.6\\
\begin{tabular}{@{}c@{}}SDSS~J092056.24+131057.4 \\ SDSS~J092056.01+131102.7\end{tabular}&2.43&2.42&2.06&2.35& 6.2&168&MAGE&62&25.7/37.6\\
\begin{tabular}{@{}c@{}}SDSS~J093747.24+150928.0 \\ SDSS~J093747.40+150939.5\end{tabular}&2.55&2.54&2.18&2.47&11.7&326&MAGE&51&10.9/14.4\\
\begin{tabular}{@{}c@{}}SDSS~J093959.41+184757.1 \\ SDSS~J093959.02+184801.8\end{tabular}&2.82&2.73&2.46&2.65& 7.2&209&ESI&64&17.4/21.4\\
\begin{tabular}{@{}c@{}}SDSS~J101201.68+311348.6 \\ SDSS~J101201.77+311353.9\end{tabular}&2.71&2.70&2.36&2.62& 5.4&156&ESI&64&31.9/12.0\\
\begin{tabular}{@{}c@{}}SDSS~J102116.98+111227.6 \\ SDSS~J102116.47+111227.8\end{tabular}&3.85&3.83&3.15&3.32& 7.4&240&ESI&64&58.4/25.6\\
\begin{tabular}{@{}c@{}}SDSS~J102116.98+111227.6 \\ SDSS~J102116.47+111227.8\end{tabular}&3.85&3.83&3.36&3.47& 7.4&246&ESI&64&53.3/23.0\\
\begin{tabular}{@{}c@{}}SDSS~J102116.98+111227.6 \\ SDSS~J102116.47+111227.8\end{tabular}&3.85&3.83&3.49&3.67& 7.4&251&ESI&64&60.3/29.8\\
\begin{tabular}{@{}c@{}}SDSS~J111610.69+411814.4 \\ SDSS~J111611.74+411821.5\end{tabular}&3.00&2.94&2.49&2.63&13.8&402&ESI&64&14.6/33.5\\
\begin{tabular}{@{}c@{}}SDSS~J111610.69+411814.4 \\ SDSS~J111611.74+411821.5\end{tabular}&3.00&2.94&2.72&2.86&13.8&419&ESI&64&17.5/44.2\\
\begin{tabular}{@{}c@{}}SDSS~J114958.26+430041.3 \\ SDSS~J114958.49+430048.4\end{tabular}&3.27&3.27&2.85&3.18& 7.5&237&ESI&48&94.2/17.3\\
\begin{tabular}{@{}c@{}}SDSS~J120416.69+022111.0 \\ SDSS~J120417.47+022104.7\end{tabular}&2.53&2.44&2.02&2.36&13.3&357&MAGE&62&19.2/28.5\\
\begin{tabular}{@{}c@{}}SDSS~J122545.24+564445.1  \\ SDSS~J122545.73+564440.7 \end{tabular}&2.39&2.39&1.90&2.31& 6.1&159&LRIS&163&13.3/34.1\\
\begin{tabular}{@{}c@{}}SDSS~J140502.41+444754.4  \\ SDSS~J140501.94+444759.9 \end{tabular}&2.22&2.20&1.80&2.03& 7.4&186&LRIS&173&14.0/50.9\\
\begin{tabular}{@{}c@{}}SDSS~J142023.77+283106.6 \\ SDSS~J142023.80+283055.7\end{tabular}&4.31&4.29&3.54&3.90&10.9&375&ESI&64&20.8/14.9\\
\begin{tabular}{@{}c@{}}SDSS~J142758.74-012136.2 \\ SDSS~J142758.89-012130.4\end{tabular}&2.35&2.27&1.87&2.02& 6.2&157&MAGE&62&21.8/16.2\\
\begin{tabular}{@{}c@{}}SDSS~J142758.74-012136.2 \\ SDSS~J142758.89-012130.4\end{tabular}&2.35&2.27&2.04&2.20& 6.2&165&MAGE&62&28.6/23.8\\
\begin{tabular}{@{}c@{}}SDSS~J154110.40+270231.2 \\ SDSS~J154110.37+270224.8\end{tabular}&3.63&3.62&3.36&3.52& 6.4&213&ESI&64&10.2/13.7\\
\begin{tabular}{@{}c@{}}SDSS~J161302.03+080814.3 \\ SDSS~J161301.69+080806.1\end{tabular}&2.39&2.38&1.90&2.31& 9.6&253&MAGE&62&31.9/18.3\\
\begin{tabular}{@{}c@{}}SDSS~J162210.11+070215.3 \\ SDSS~J162209.81+070211.5\end{tabular}&3.26&3.23&2.77&3.05& 5.8&180&ESI&64&116.0/18.0\\
\begin{tabular}{@{}c@{}}SDSS~J221427.03+132657.0  \\ SDSS~J221426.79+132652.3 \end{tabular}&2.01&2.00&1.80&1.93& 5.8&144&LRIS&173&40.2/42.5\\
\begin{tabular}{@{}c@{}}SDSS~J230044.52+015552.1 \\ SDSS~J230044.36+015541.7\end{tabular}&2.95&2.91&2.38&2.68&10.7&310&MAGE&51&11.9/21.7\\
\end{longtabu}}

\end{landscape}
\begin{center}
\begin{table}[h]
\caption{ {\bf Potential sources of errors we have
considered in our  measurement.} 
From left to right,  the columns are the
central redshift of each bin, the median value of the posterior distribution
of $\lambda_P$, the 16\% and 84\% percentiles of the posterior distribution 
of $\lambda_P$, the error on the median associated to noise, resolution, metal contamination and 
Lyman-limit systems, and the 16\% and 84\% percentiles corrected for 
systematics (see text for details).  All columns except the first are in comoving kpc units.\label{tab:systematics}}
\centering
\begin{tabular}{cccccccccc}
\hline\hline\\
$z$  & $\lambda_P$ & $\lambda_{P,16}$ & $\lambda_{P,84}$ & $\Delta\lambda_{P,{\rm Noise}}$ & $\Delta\lambda_{P,{\rm Res}}$ & $\Delta\lambda_{P,{\rm Met}}$ & $\Delta\lambda_{P,{\rm LLS}}$ & $\lambda_{P,16}^{\rm syst}$ & $\lambda_{P,84}^{\rm syst}$ \\
\hline\\
2.0 &   82.1   & 65.2 & 101.1 & 9.9 & 0.6  & 5.9 & 8.6 & 59.9 & 105.6 \\
2.4 &   72.5   & 55.6 & 95.8 & 0.4 & 0.3   & 5.7 & 4.5 & 54.0 & 96.9 \\
3.0 &   113.8  & 92.4 & 137.7 & 4.7 & 0.9   & 3.5 & 3.1 & 91.4 & 138.6\\
3.6 &   90.2   & 60.1 & 123.6 & 0.5 & 0.1   & 1.0 & 1.7 & 60.0 & 123.7\\
\hline\hline
\end{tabular}

\end{table}
\end{center}
\begin{center}
\begin{table}[h]
\caption{ {\bf Basic properties of the NYX simulations considered in this work.} 
All simulations have $1024^3$ dark-matter particles and grid cells, and 
all boxes have size of 20 Mpc$/h$. 
$z_{\rm reion}$ is the redshift at which the UV background
and the photoheating are switched on, below which the IGM is assumed to be
optically thin and in ionization equilibrium; $A$ is the factor by
which the heating rates by \cite{Haardt12} are rescaled; 
$T_{0,z=3}$, $\gamma_{z=3}$ and $\lambda_{P,sim,z=3}$ are the temperature 
at mean density, the slope of the temperature-density relationship and 
the $F_{\rm real}$ cut off at $z=3$. $\lambda_{P,sim,z=3}$ is 
obtained from the fit of the real-flux power spectrum as in 
eqn.~\ref{equation:freal_fit}.\label{tab:sim}}
\centering
\begin{tabular}{ccccccccc}
\hline\hline\\
Name        &  $z_{\rm reion}$ & $A$    & $\log T_{0,z=3}$ [K] & $\gamma_{z=3}$ & $\lambda_{P,{\rm sim},z=3}$ [kpc] \\
\hline\\
Standard model   & 15 & 1   & 4.03  & 1.57 & 79.1  \\
Late reionization heating    & 7  & 1   & 4.03  & 1.58 & 66.5  \\
Increased photoheating & 15 & 3 & 4.32  & 1.57 & 106.5  \\
\hline\hline
\end{tabular}
\end{table}
\end{center}
\begin{center}
\begin{table}[h]
\centering
\caption{ {\bf Basic properties of the NYX simulations used to estimate
the mapping relation between $\lambda_P $ and $\lambda_{P,{\rm sim}}$}. 
All simulations have $1024^3$ cells and a box size of 20 Mpc/$h$.
$z_{\rm reion}$ is the redshift at which the 
photoheating is switched on; $A$ is the factor by
which the heating rates by \cite{Haardt12} are rescaled;
$B$ the index of the heating-density relationship (see text) 
$T_{0,z=3}$, $\gamma_{z=3}$ and $\lambda_{P,{\rm sim},z=3}$ are the temperature 
at mean density, the slope of the temperature-density relationship and 
the $F_{\rm real}$ cut off at $z=3$. $\lambda_{P,{\rm sim}}$ is 
obtained from the fit of the real-flux power spectrum as in 
eqn.~\ref{equation:freal_fit}.\label{tab:sim2}}
\begin{tabular}{ccccccc}
\hline\hline\\
Name    & $z_{\rm reion}$ & $A$  & $B$   & $\log T_{0,z=3}$ [K] & $\gamma_{z=3}$ & $\lambda_{P,{\rm sim},z=3}$ [kpc] \\
\hline\\
A05Bm1   & 15 & 0.5 & -1 & 3.85  & 1.00 & 54.8  \\
A05B0z7    & 7 & 0.5 & 0  & 3.84  & 1.58 & 57.6  \\
A1Bm1   & 15 & 1  & -1  & 4.03  & 1.01 & 64.5  \\
A05B0  & 15 & 0.5 & 0  & 3.84  & 1.57 & 65.1  \\
A1B0z7  &	7 & 1 & 0 & 4.03 & 1.58 & 66.5\\
A1Bm05    & 15 & 1 & -0.5   &  4.03  & 1.29 & 71.7  \\
A2Bm1    & 15 & 2 & -1  &  4.22  & 1.02 & 76.3  \\
A1B0   & 15 & 1 & 0  &  4.03  & 1.57 & 79.1  \\
A3Bm1     & 15 & 3 & -1    &  4.33  & 1.02 & 84.7  \\
A2Bm05  & 15 & 2 & -0.5 & 4.22  & 1.29 & 86.5  \\
A4Bm1       & 15 & 4 & -1    &  4.41  & 1.01 & 91.4  \\
A2B0	 & 15 & 2 & 0   & 4.22  & 1.57 & 96.0  \\
A3B0	 & 15 & 3 & 0 & 4.32  & 1.57 & 106.5  \\
\hline\hline
\end{tabular}
\end{table}
\end{center}
\clearpage

\clearpage
\begin{figure}
	\centering\centerline{\epsfig{file=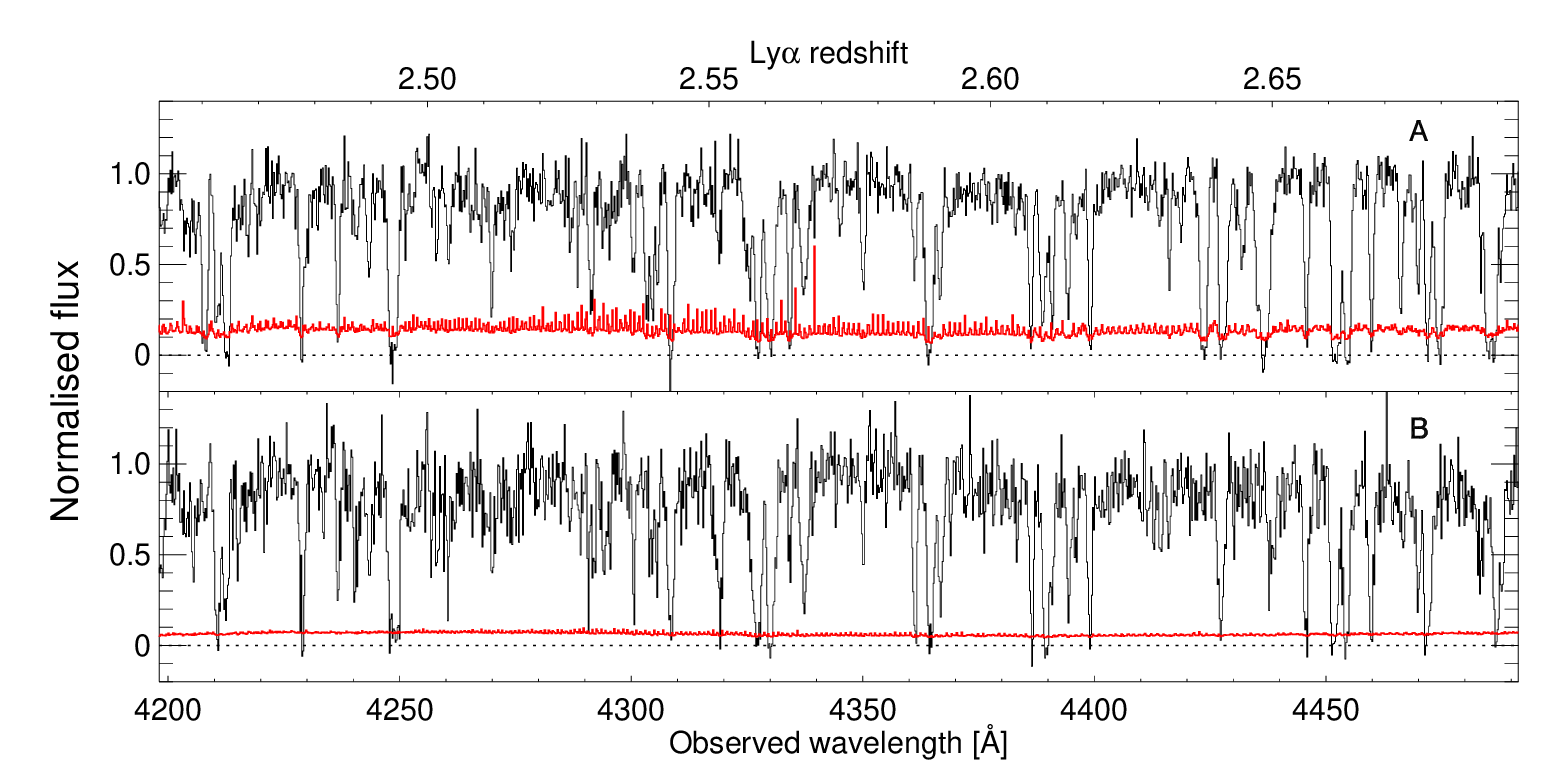,
      width=\textwidth}}
  \vskip -0.1in
\caption{\label{fig:pair_a} {\bf Spectra from the pair sample} as
in Fig.~1 of the main text, but for SDSS~J000450.90-084452.0 (panel A) and SDSS~J000450.66-084449.6 (panel B).}
\end{figure}
\begin{figure}
	\centering \centerline{\epsfig{file=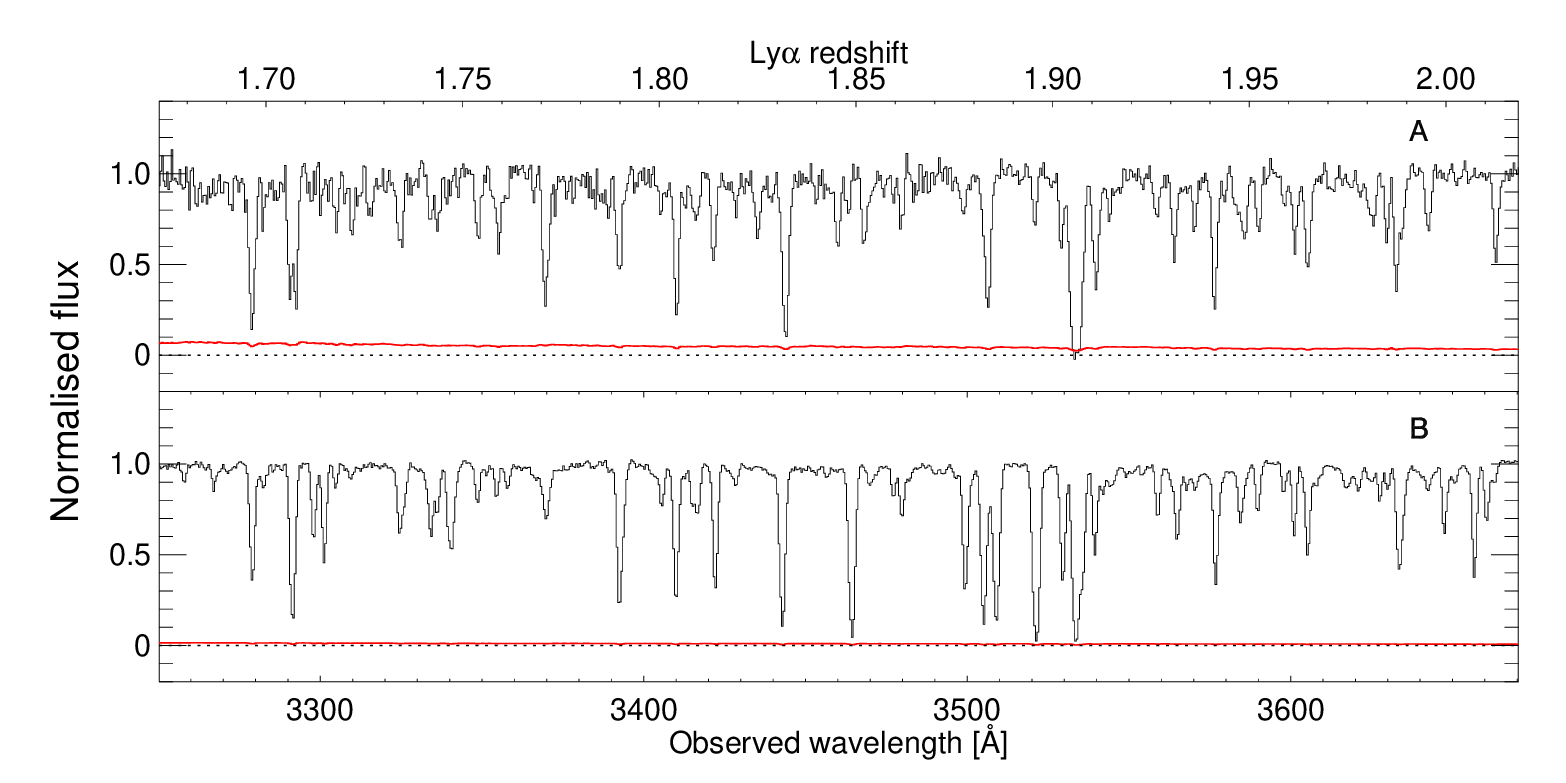,
      width=\textwidth}}
  \vskip -0.1in
\caption{\label{fig:pair_b} {\bf Spectra from the pair sample} as
in Fig.~1 of the main text, but for SDSS~J005408.47-094638.3 (panel A) and SDSS~J005408.04-094625.7 (panel B).}
\end{figure}
\begin{figure}
	\centering \centerline{\epsfig{file=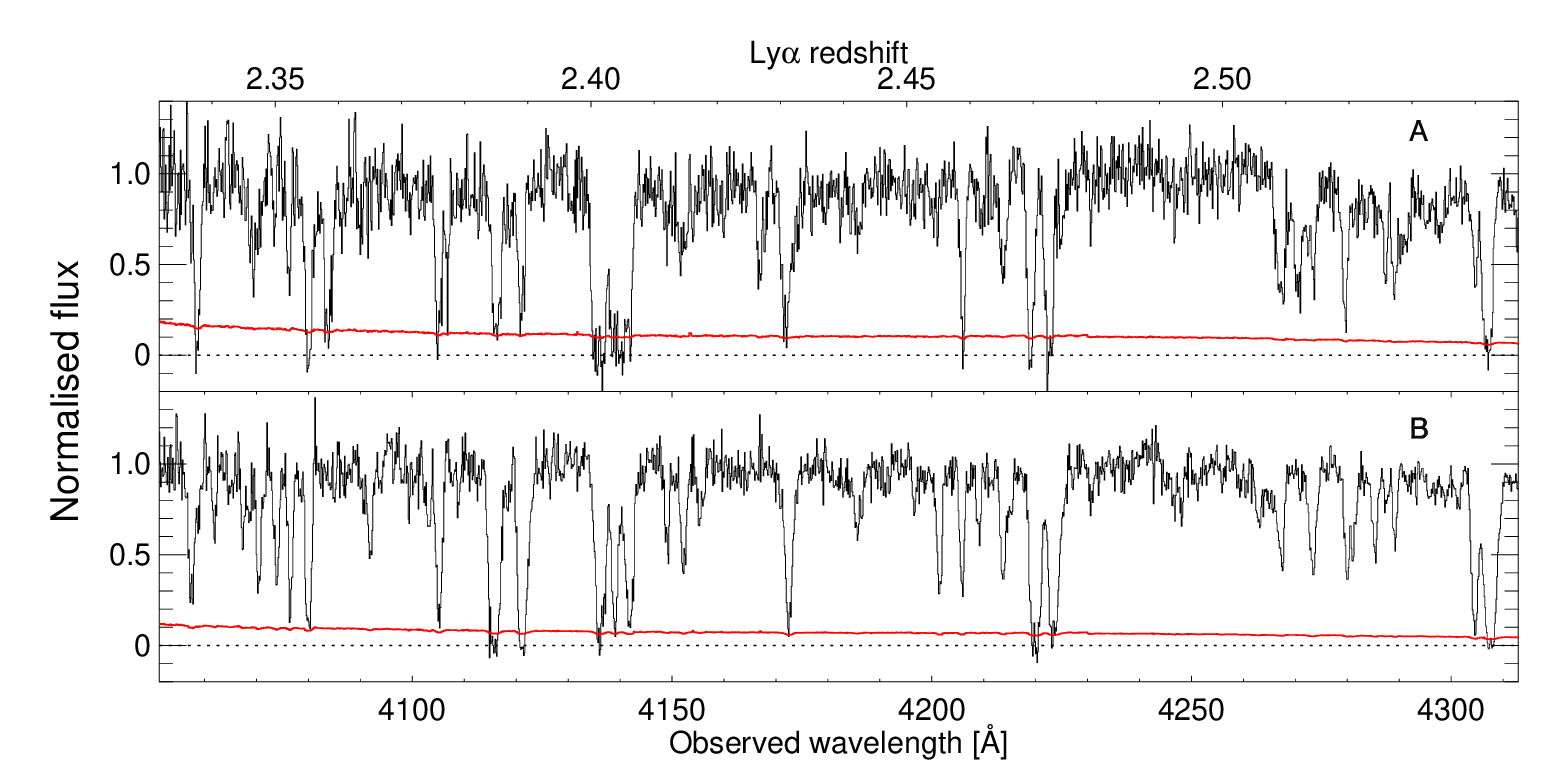,
      width=\textwidth}}
  \vskip -0.1in
\caption{\label{fig:pair_c} {\bf Spectra from the pair sample} as
in Fig.~1 of the main text, but for SDSS~J011707.52+315341.2 (panel A) and SDSS~J011708.39+315338.7 (panel B).}
\end{figure}
\begin{figure}
	\centering \centerline{\epsfig{file=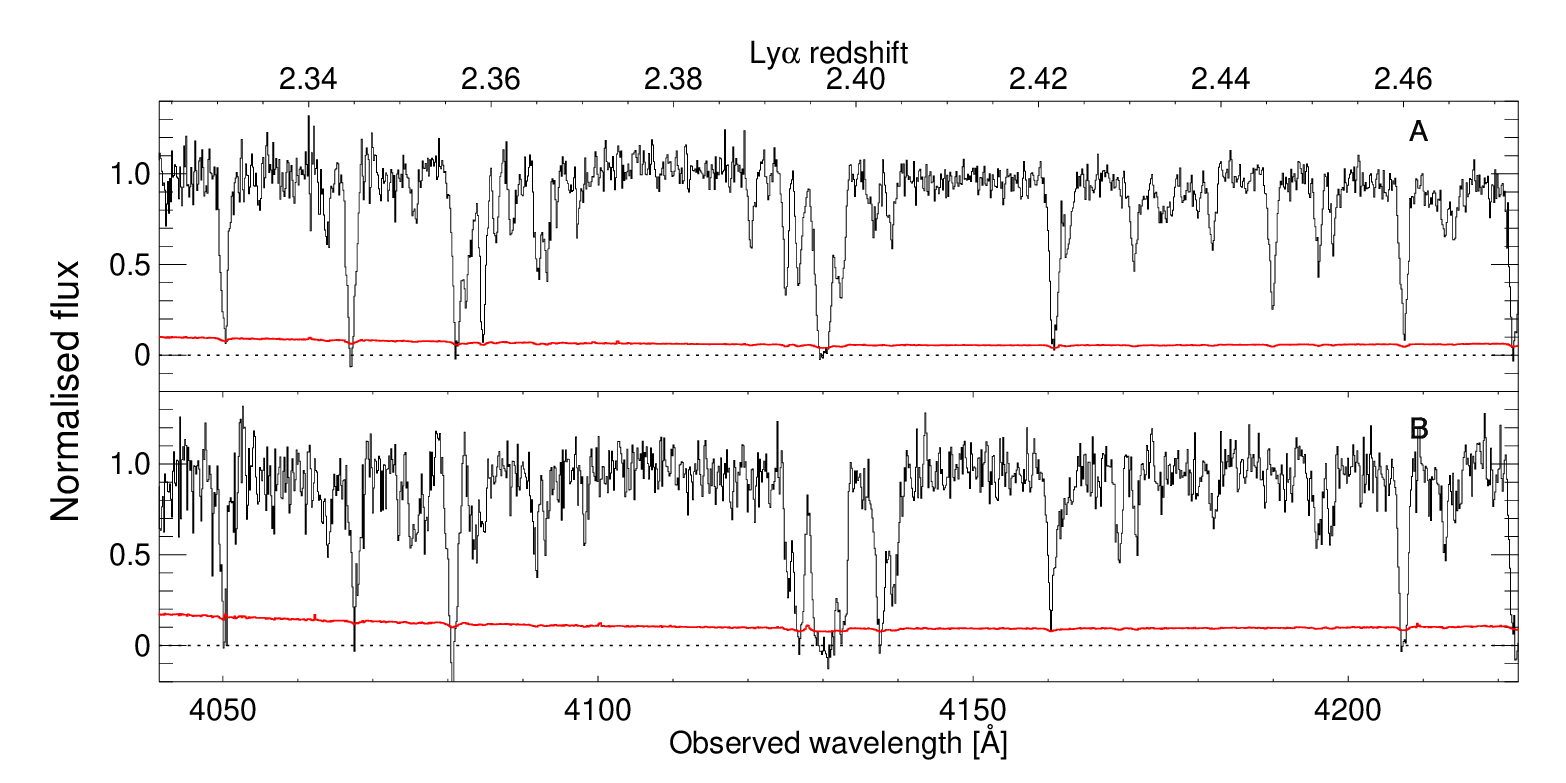,
      width=\textwidth}}
  \vskip -0.1in
\caption{\label{fig:pair_d} {\bf Spectra from the pair sample} as
in Fig.~1 of the main text, but for SDSS~J025049.09-025631.7 (panel A) and SDSS~J025048.86-025640.7 (panel B).}
\end{figure}
\begin{figure}
	\centering \centerline{\epsfig{file=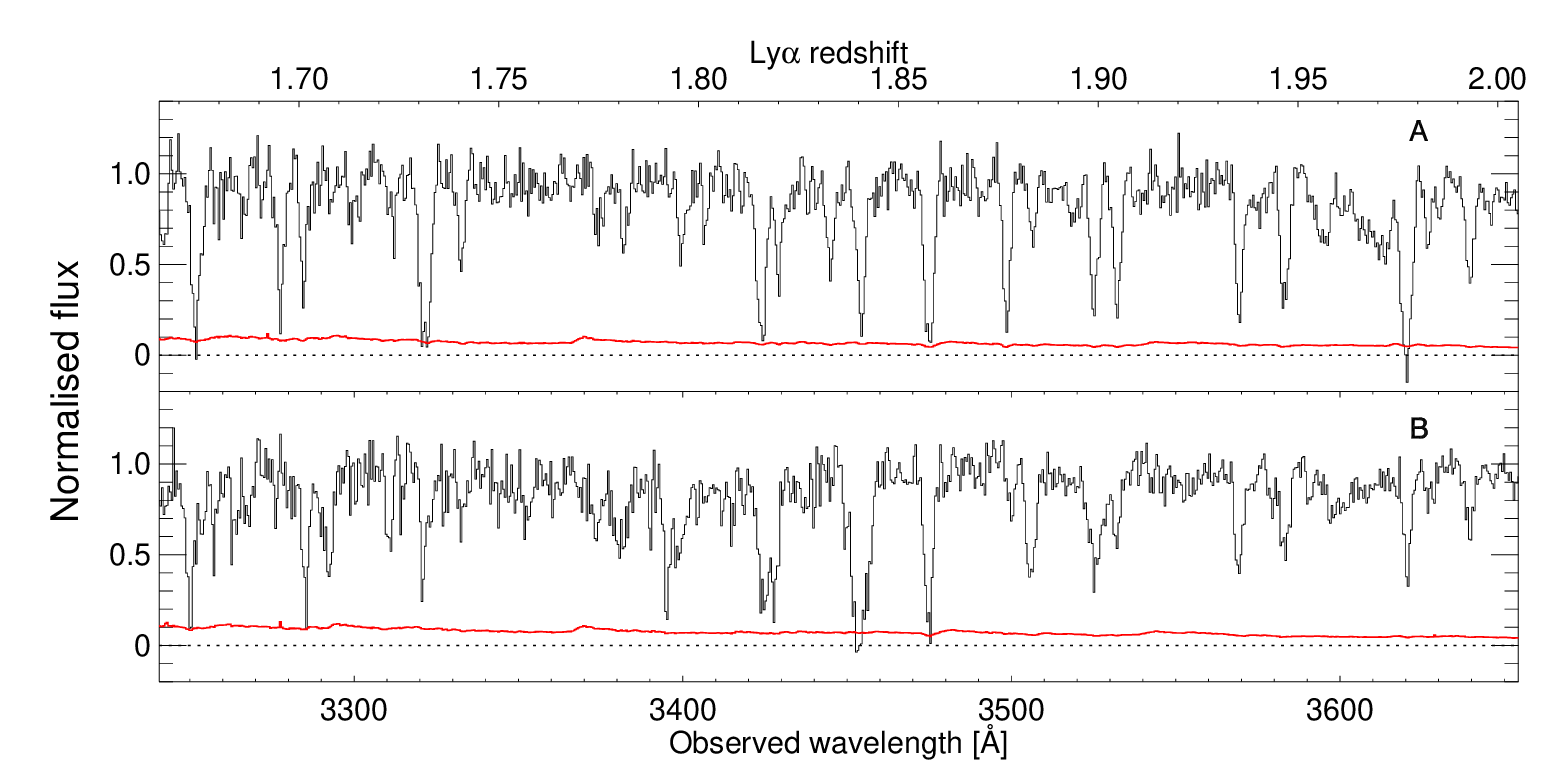,
      width=\textwidth}}
  \vskip -0.1in
\caption{\label{fig:pair_e} {\bf Spectra from the pair sample} as
in Fig.~1 of the main text, but for SDSS~J033237.19-072219.6 (panel A) and SDSS~J033238.38-072215.9 (panel B).}
\end{figure}
\begin{figure}
	\centering \centerline{\epsfig{file=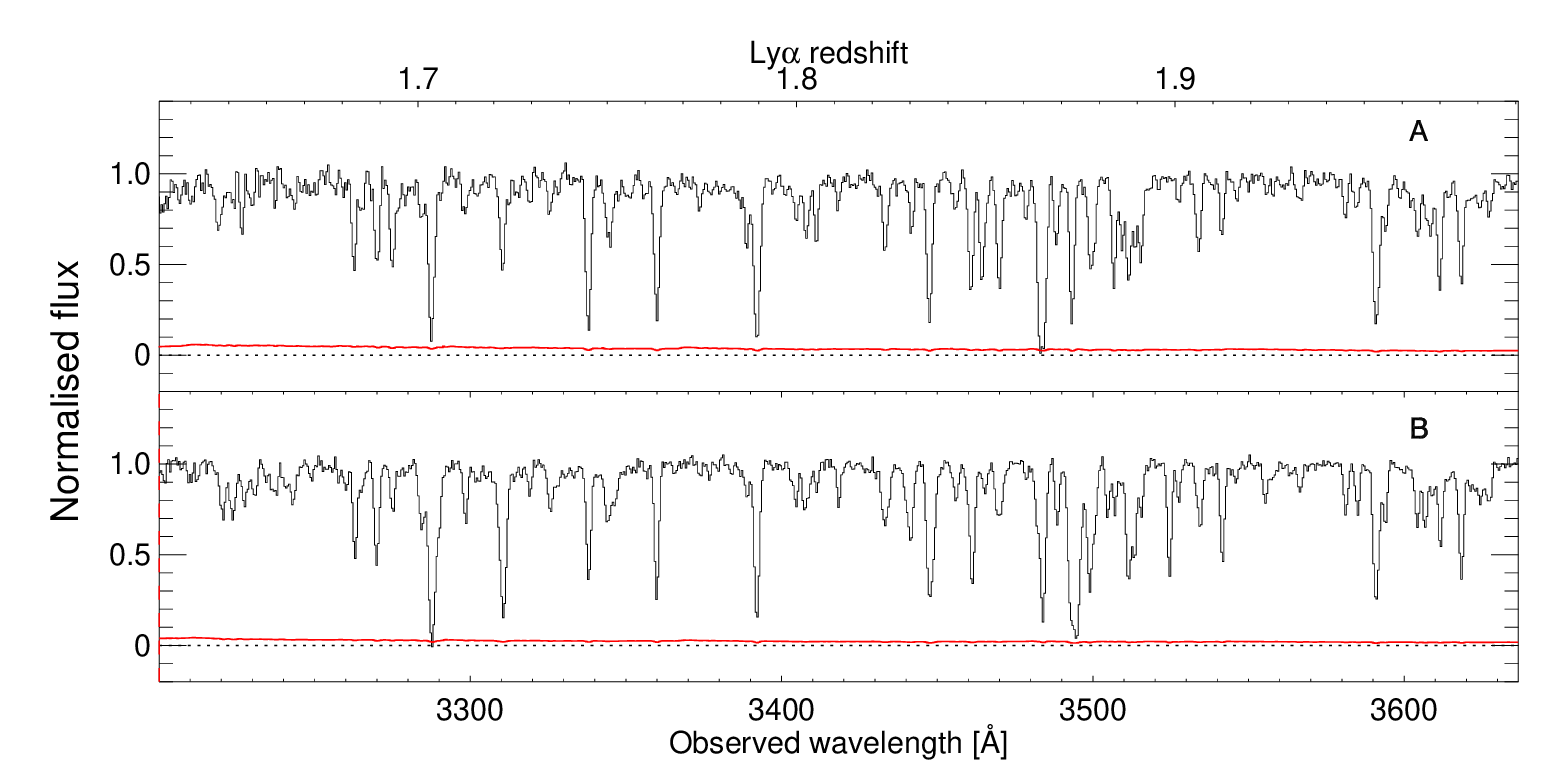,
      width=\textwidth}}
  \vskip -0.1in
\caption{\label{fig:pair_f} {\bf Spectra from the pair sample} as
in Fig.~1 of the main text, but for SDSS~J073522.43+295710.1 (panel A) and SDSS~J073522.55+295705.0 (panel B).}
\end{figure}
\begin{figure}
\centering\centerline{\epsfig{file=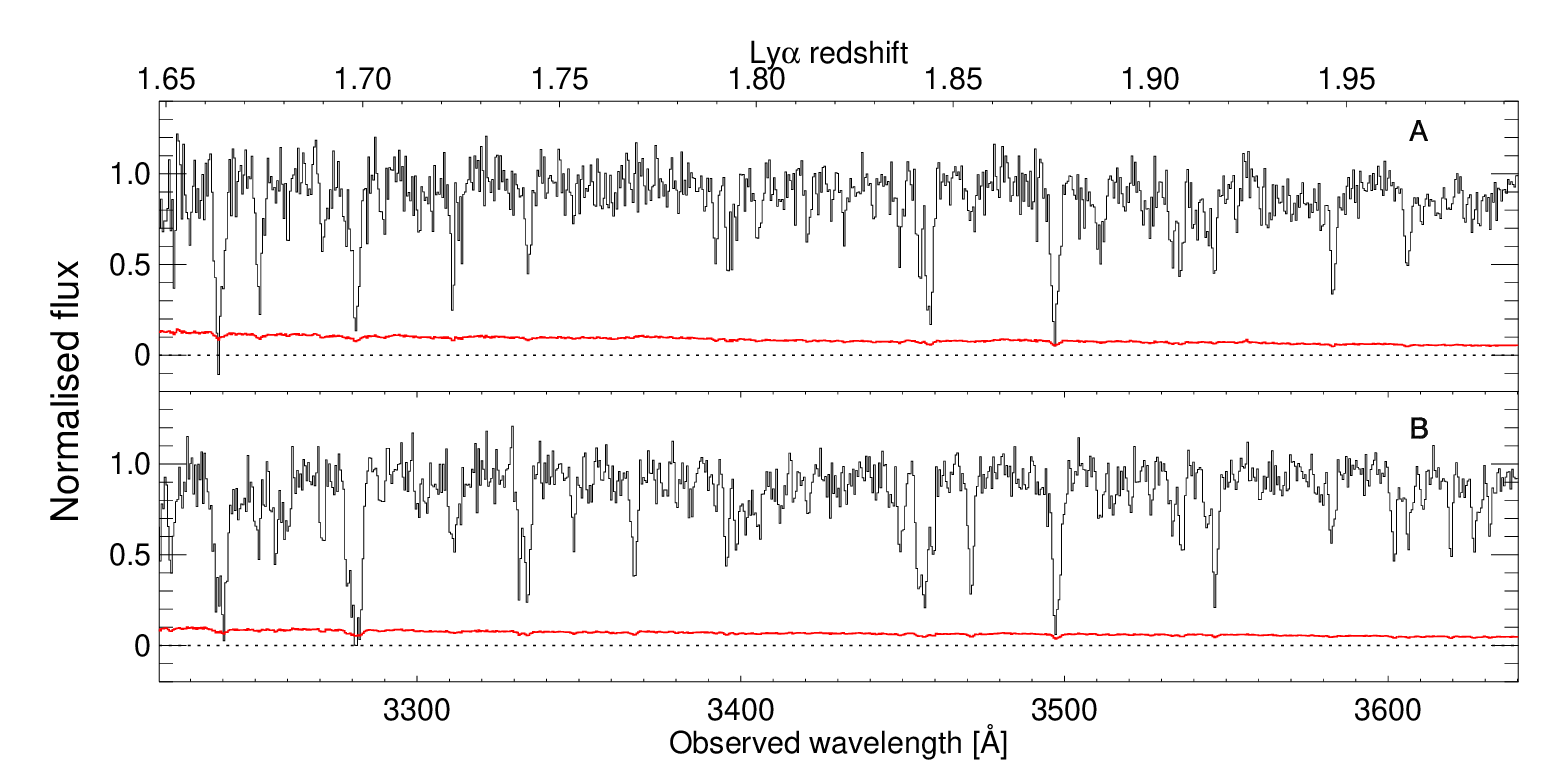,
      width=\textwidth}}
  \vskip -0.1in
\caption{\label{fig:pair_g} {\bf Spectra from the pair sample} as
in Fig.~1 of the main text, but for SDSS~J081329.49+101405.2 (panel A) and SDSS~J081329.71+101411.6 (panel B).}
\end{figure}
\begin{figure}
	\centering \centerline{\epsfig{file=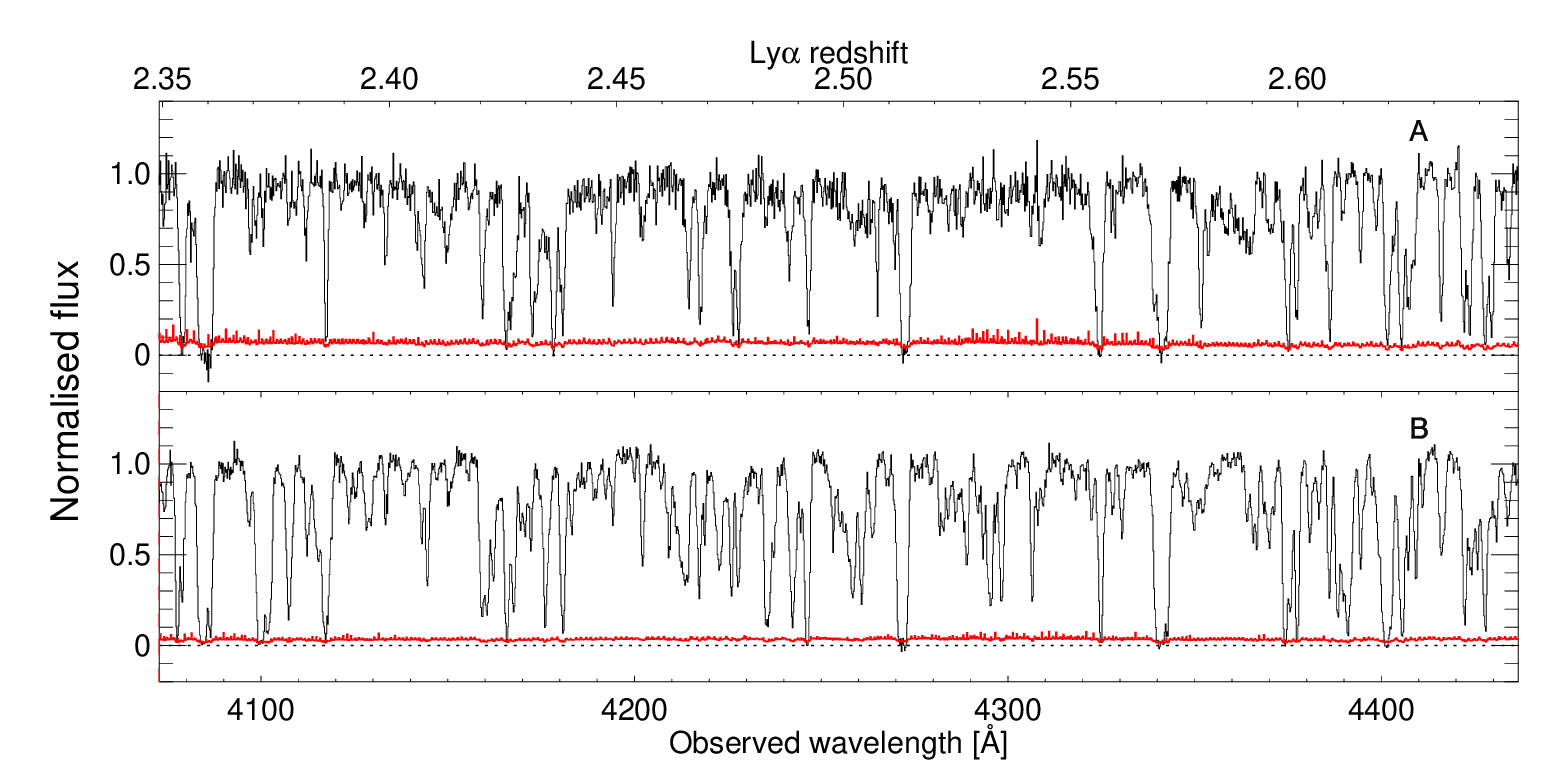,
      width=\textwidth}}
  \vskip -0.1in
\caption{\label{fig:pair_h} {\bf Spectra from the pair sample} as
in Fig.~1 of the main text, but for SDSS~J091338.97-010704.6 (panel A) and SDSS~J091338.30-010708.7 (panel B).}
\end{figure}
\begin{figure}
	\centering \centerline{\epsfig{file=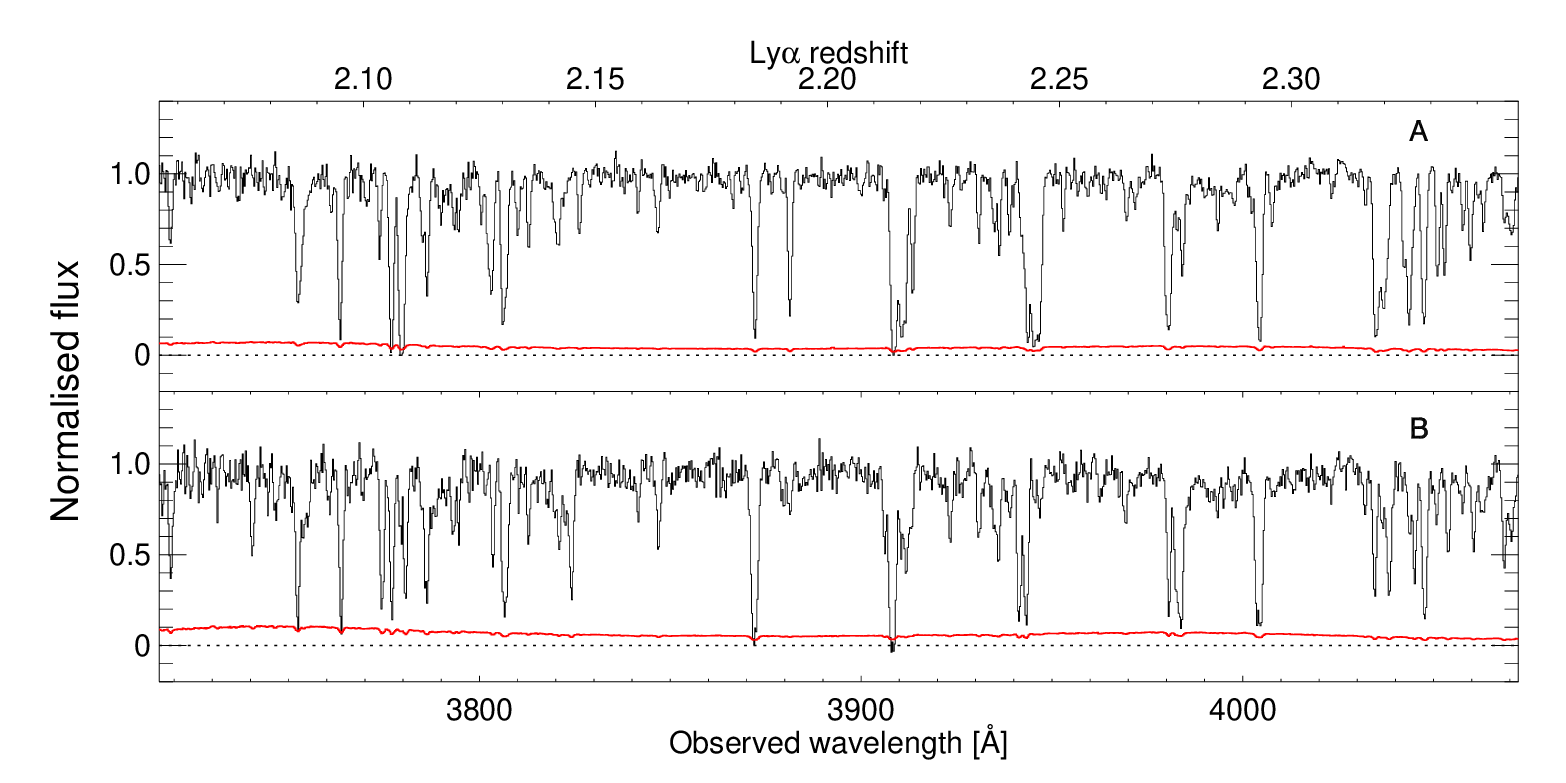,
      width=\textwidth}}
  \vskip -0.1in
\caption{\label{fig:pair_i} {\bf Spectra from the pair sample} as
in Fig.~1 of the main text, but for SDSS~J092056.24+131057.4 (panel A) and SDSS~J092056.01+131102.7 (panel B).}
\end{figure}
\begin{figure}
	\centering \centerline{\epsfig{file=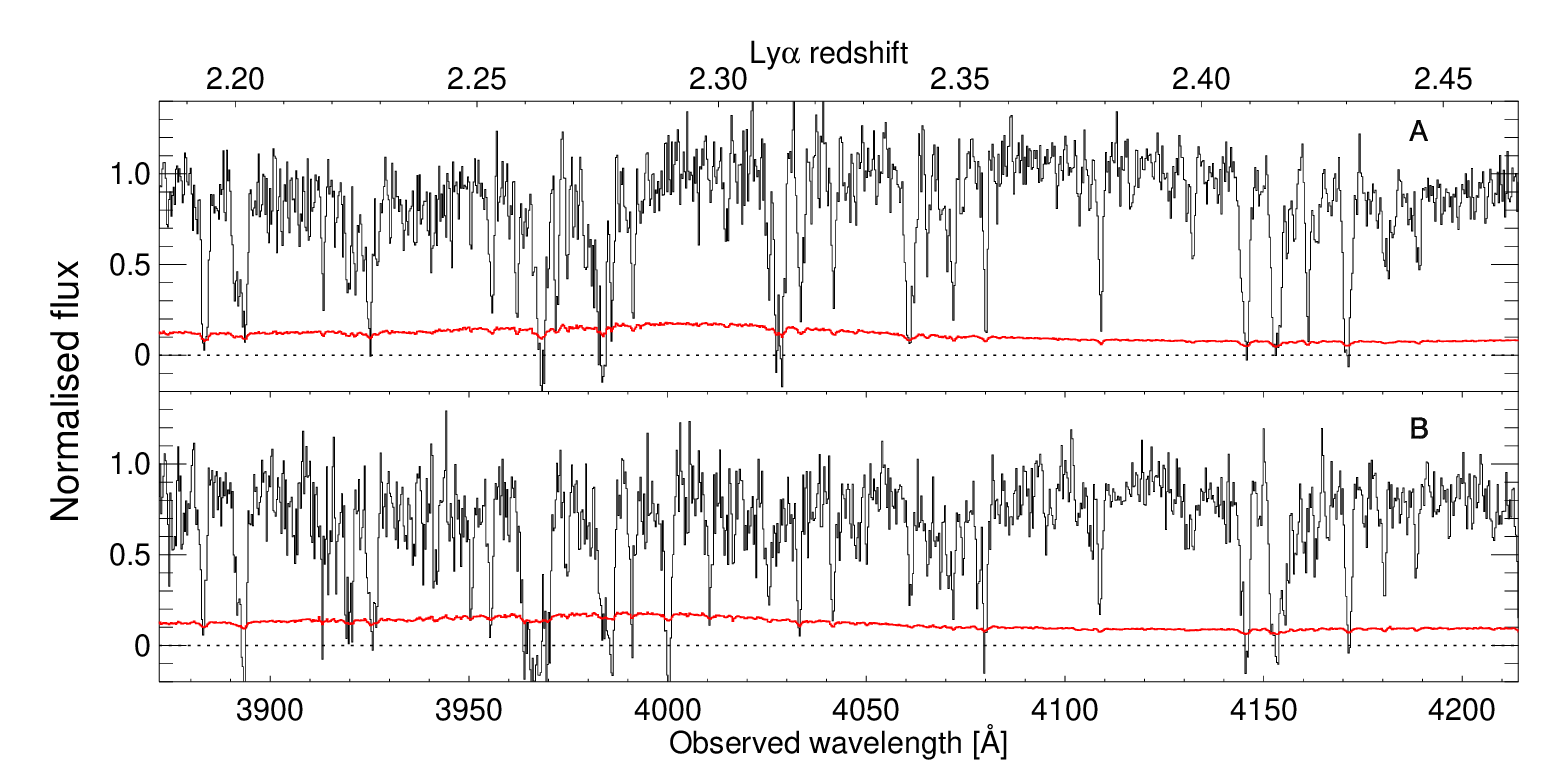,
      width=\textwidth}}
  \vskip -0.1in
\caption{\label{fig:pair_j} {\bf Spectra from the pair sample} as
in Fig.~1 of the main text, but for SDSS~J093747.24+150928.0 (panel A) and SDSS~J093747.40+150939.5 (panel B).}
\end{figure}
\begin{figure}
	\centering \centerline{\epsfig{file=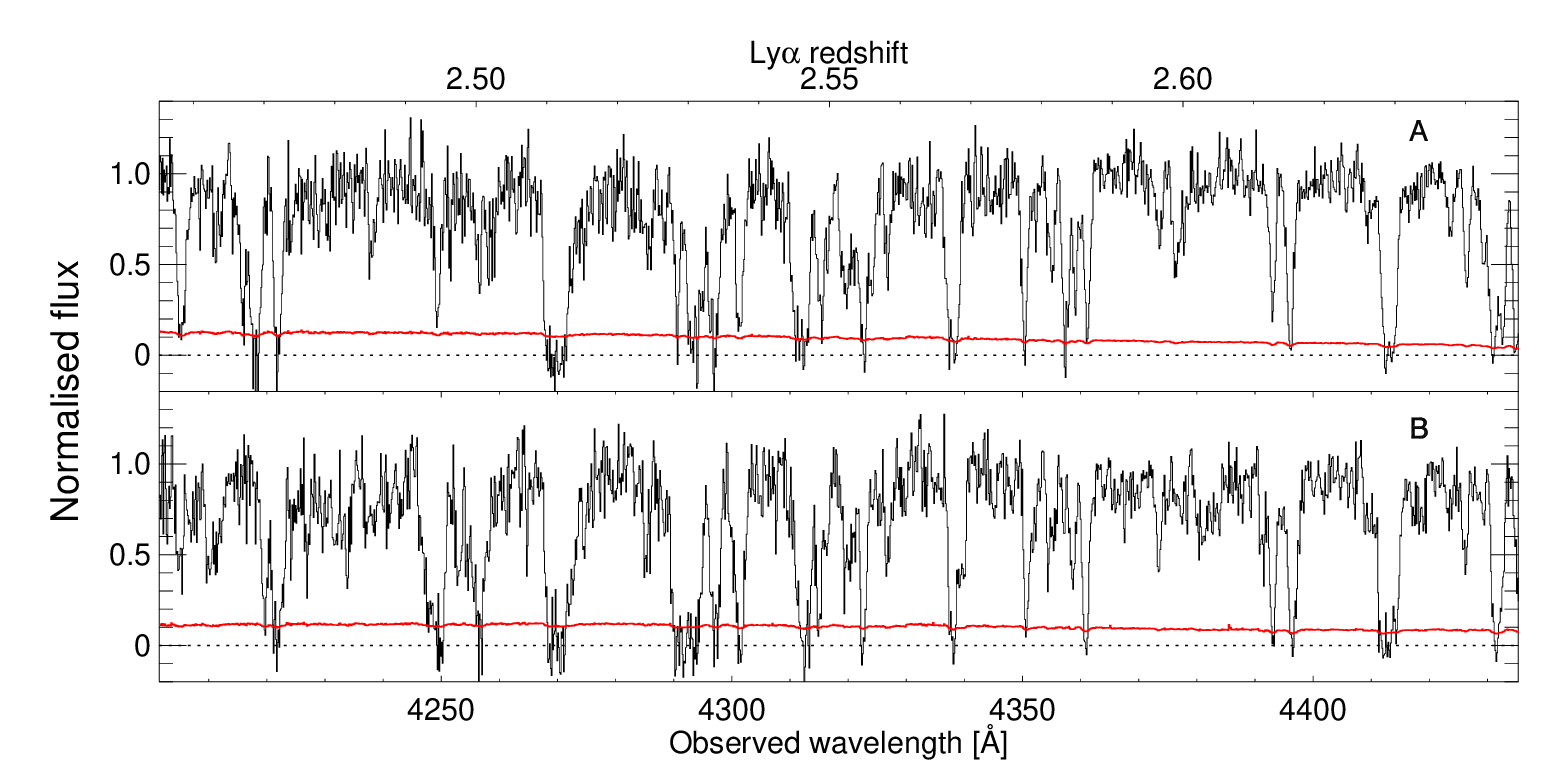,
      width=\textwidth}}
  \vskip -0.1in
\caption{\label{fig:pair_k} {\bf Spectra from the pair sample} as
in Fig.~1 of the main text, but for SDSS~J093959.41+184757.1 (panel A) and SDSS~J093959.02+184801.8 (panel B).}
\end{figure}
\begin{figure}
	\centering \centerline{\epsfig{file=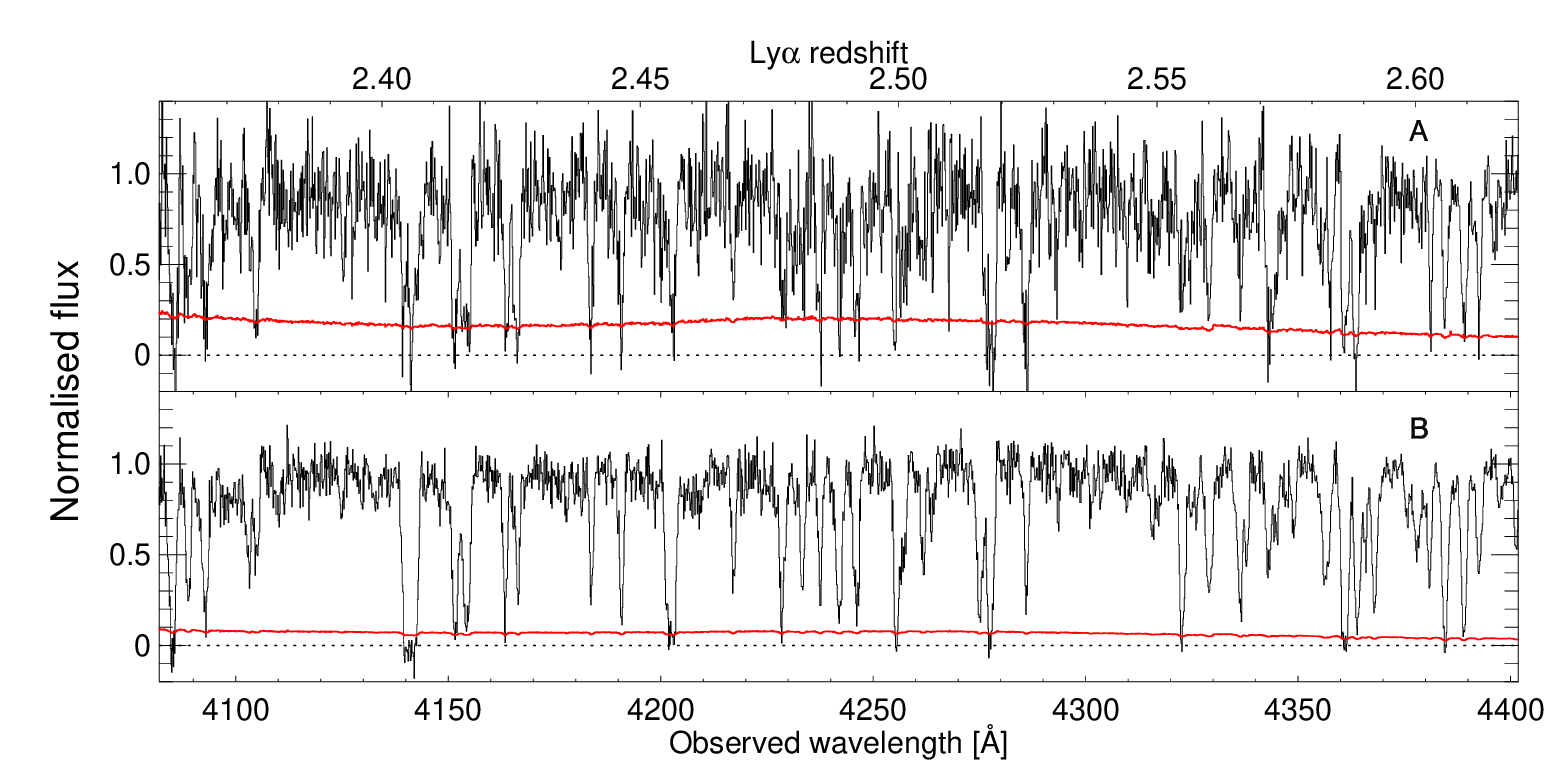,
      width=\textwidth}}
  \vskip -0.1in
\caption{\label{fig:pair_l} {\bf Spectra from the pair sample} as
in Fig.~1 of the main text, but for SDSS~J101201.68+311348.6 (panel A) and SDSS~J101201.77+311353.9 (panel B).}
\end{figure}
\clearpage
\begin{figure}
	\centering \centerline{\epsfig{file=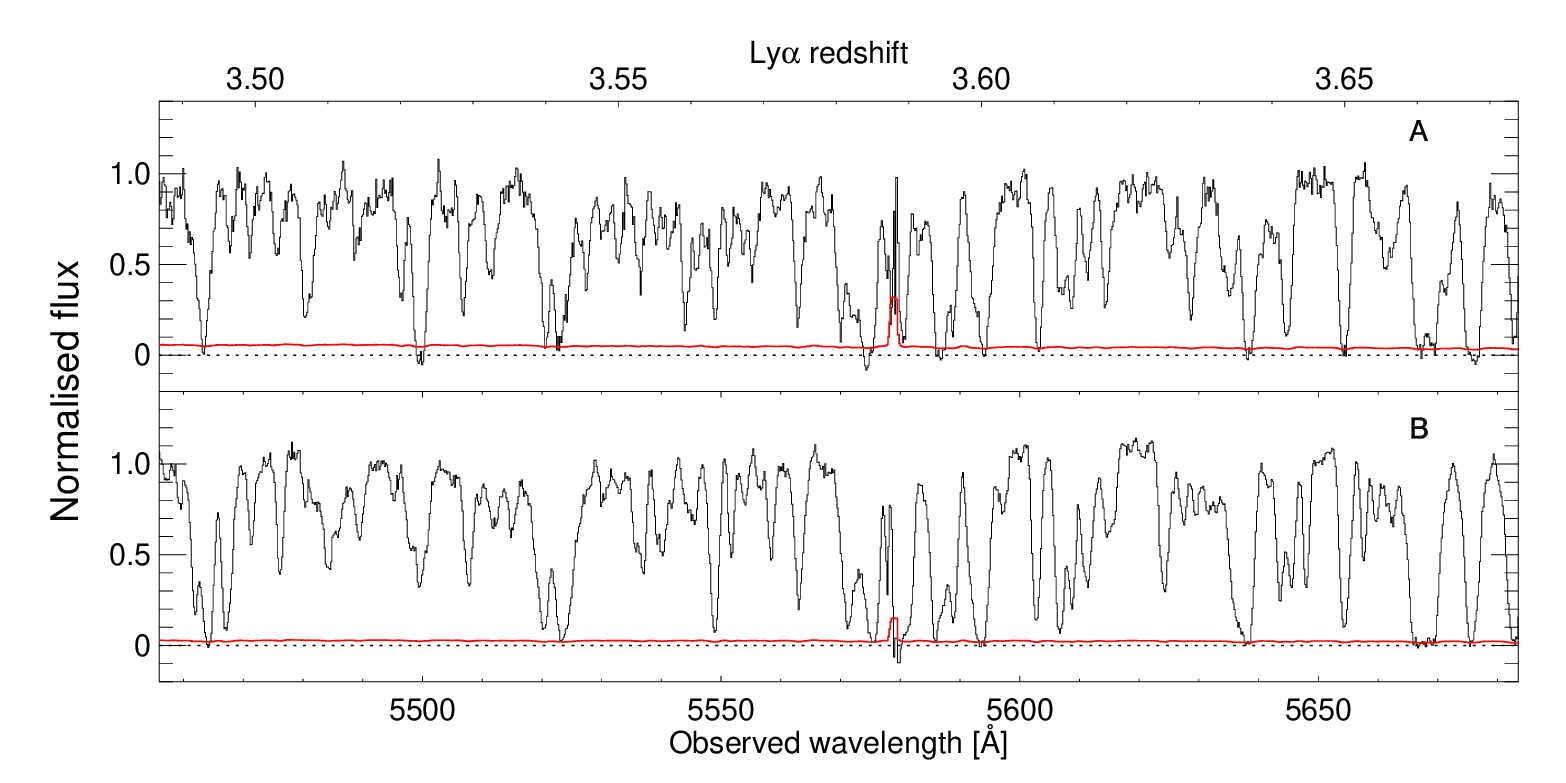,
      width=\textwidth}}
  \vskip -0.1in
\caption{\label{fig:pair_m} {\bf Spectra from the pair sample} as
in Fig.~1 of the main text, but for SDSS~J102116.98+111227.6 (panel A) and SDSS~J102116.47+111227.8 (panel B).}
\end{figure}
\begin{figure}
	\centering \centerline{\epsfig{file=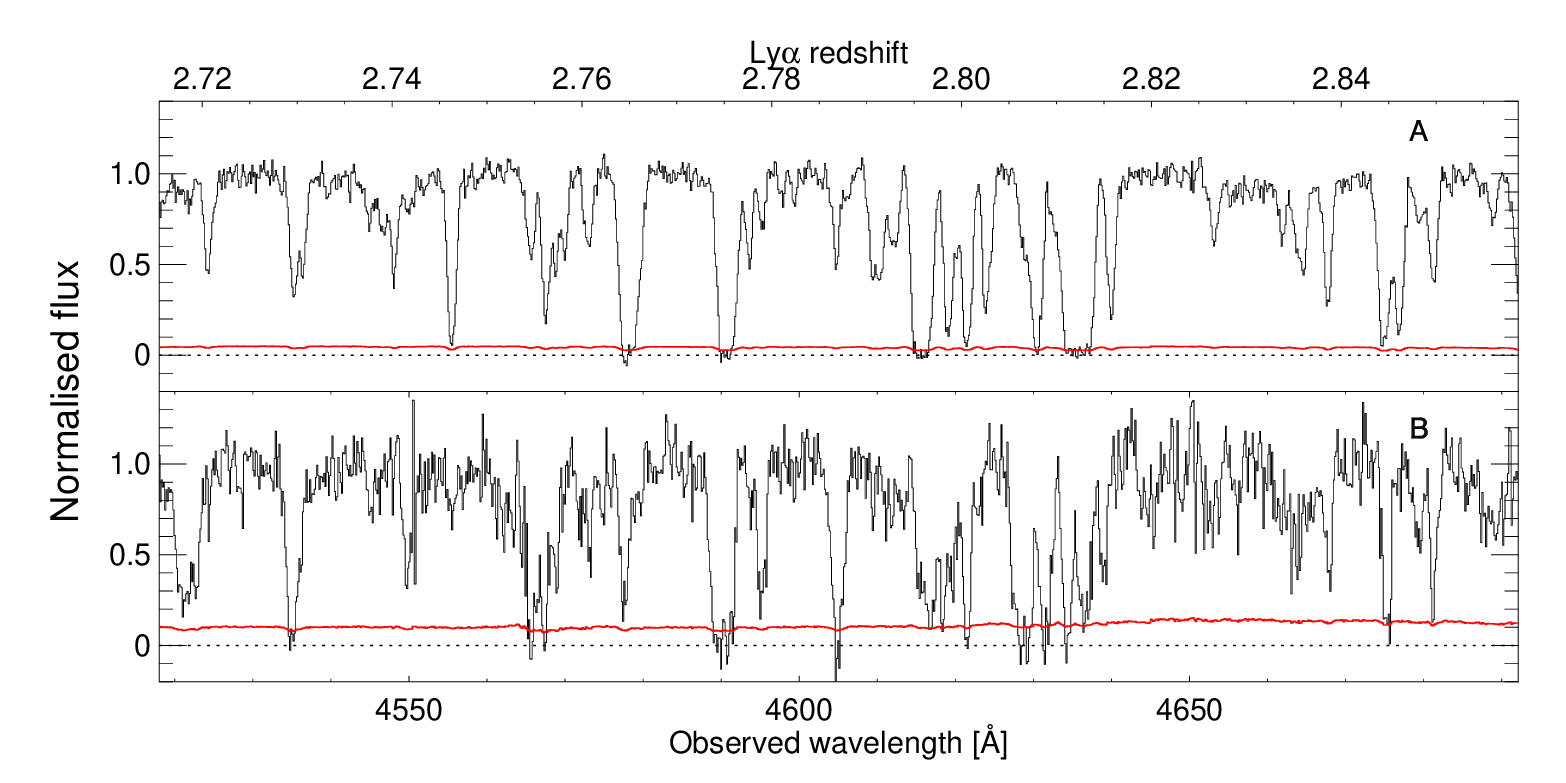,
      width=\textwidth}}
  \vskip -0.1in
\caption{\label{fig:pair_n} {\bf Spectra from the pair sample} as
in Fig.~1 of the main text, but for SDSS~J111610.69+411814.4 (panel A) and SDSS~J111611.74+411821.5 (panel B).}
\end{figure}
\begin{figure}
	\centering \centerline{\epsfig{file=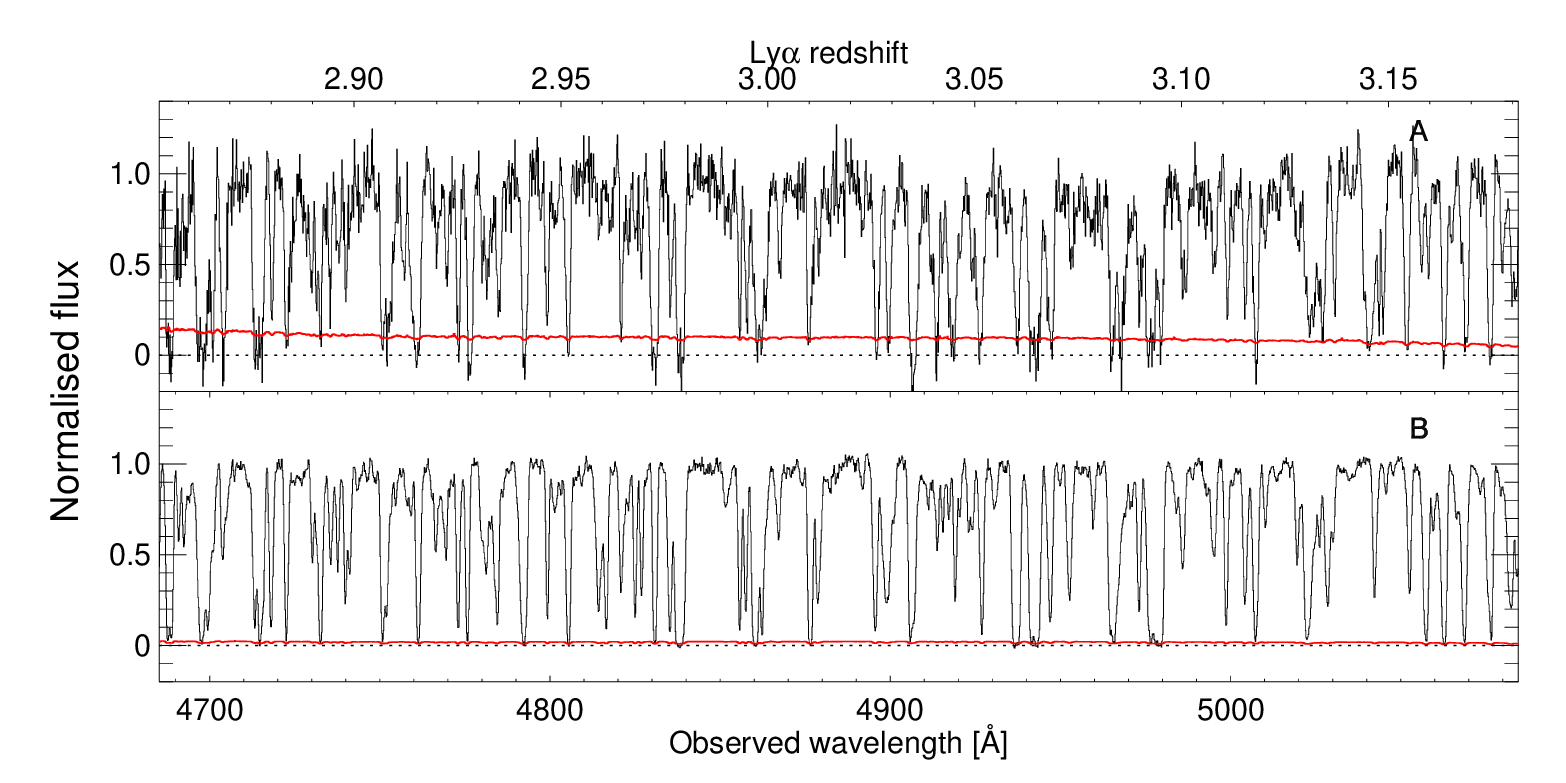,
      width=\textwidth}}
  \vskip -0.1in
\caption{\label{fig:pair_o} {\bf Spectra from the pair sample} as
in Fig.~1 of the main text, but for SDSS~J114958.26+430041.3 (panel A) and SDSS~J114958.49+430048.4 (panel B).}
\end{figure}
\begin{figure}
	\centering \centerline{\epsfig{file=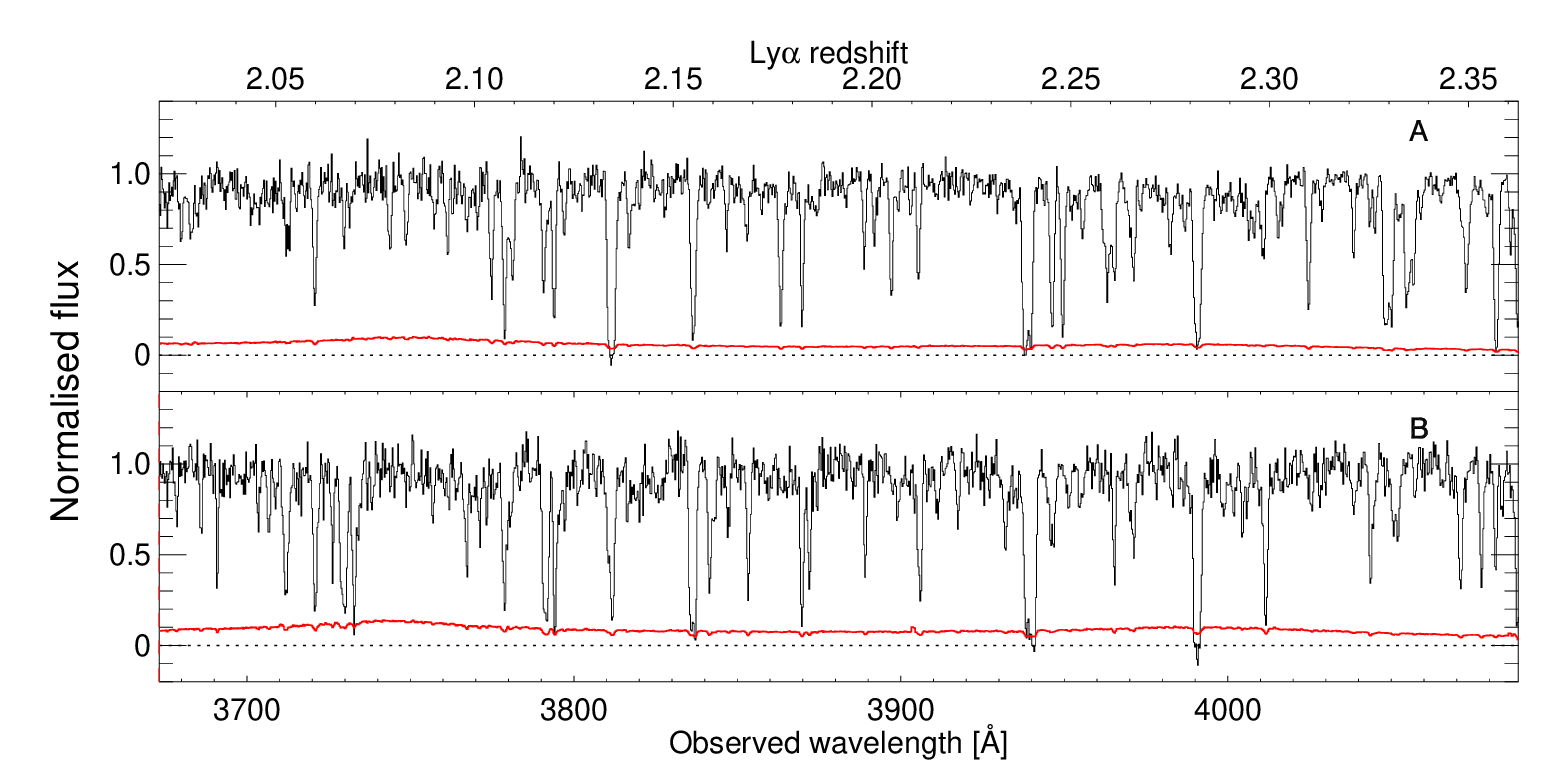,
      width=\textwidth}}
  \vskip -0.1in
\caption{\label{fig:pair_p} {\bf Spectra from the pair sample} as
in Fig.~1 of the main text, but for SDSS~J120416.69+022111.0 (panel A) and SDSS~J120417.47+022104.7 (panel B). }
\end{figure}
\begin{figure}
	\centering \centerline{\epsfig{file=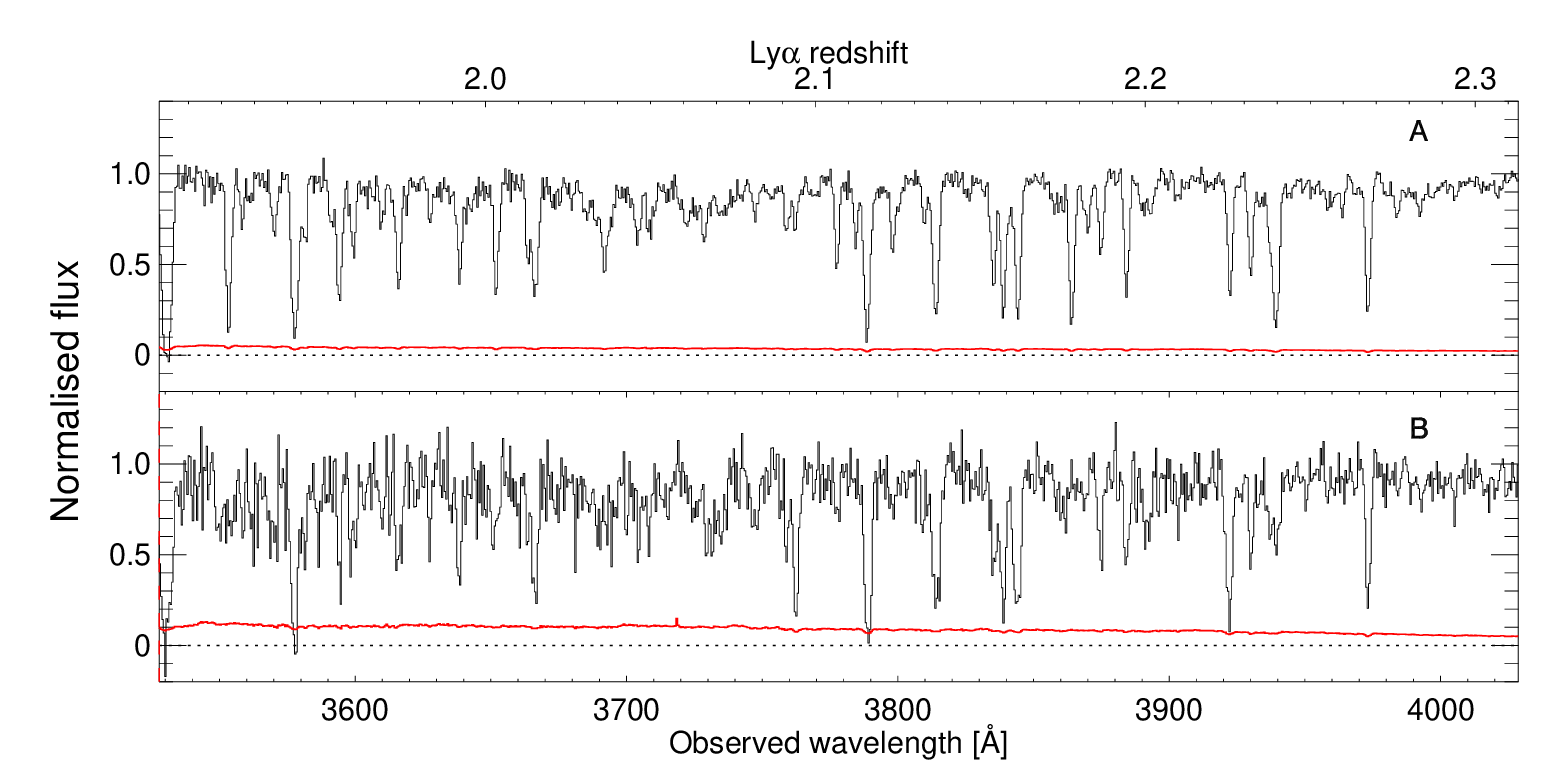,
      width=\textwidth}}
  \vskip -0.1in
\caption{\label{fig:pair_q} {\bf Spectra from the pair sample} as
in Fig.~1 of the main text, but for SDSS~J122545.24+564445.1 (panel A) and SDSS~J122545.73+564440.7 (panel B).}
\end{figure}
\begin{figure}
	\centering \centerline{\epsfig{file=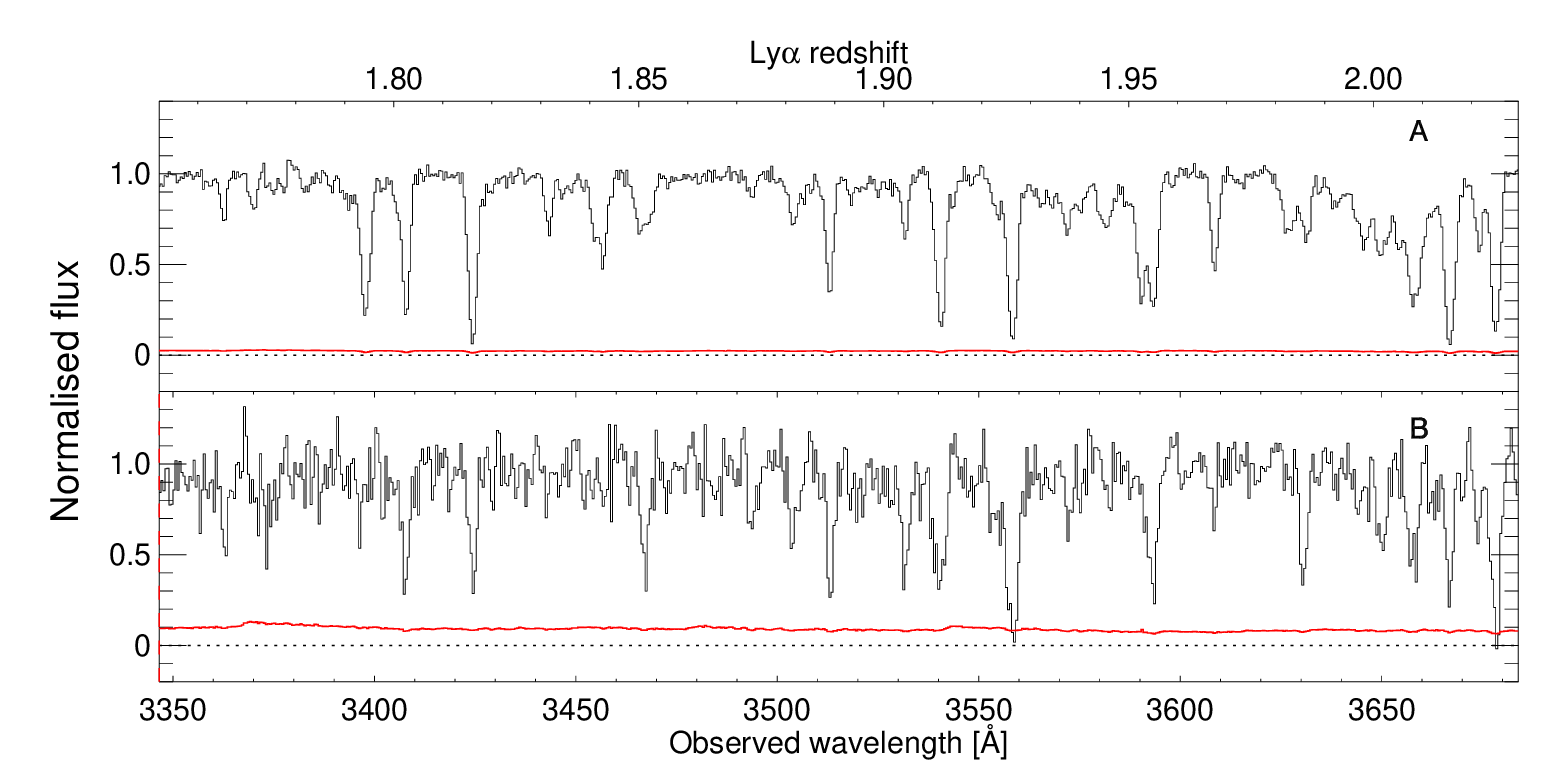,
      width=\textwidth}}
  \vskip -0.1in
\caption{\label{fig:pair_r} {\bf Spectra from the pair sample} as
in Fig.~1 of the main text, but for SDSS~J140502.41+444754.4 (panel A) and SDSS~J140501.94+444759.9 (panel B).}
\end{figure}
\begin{figure}
	\centering \centerline{\epsfig{file=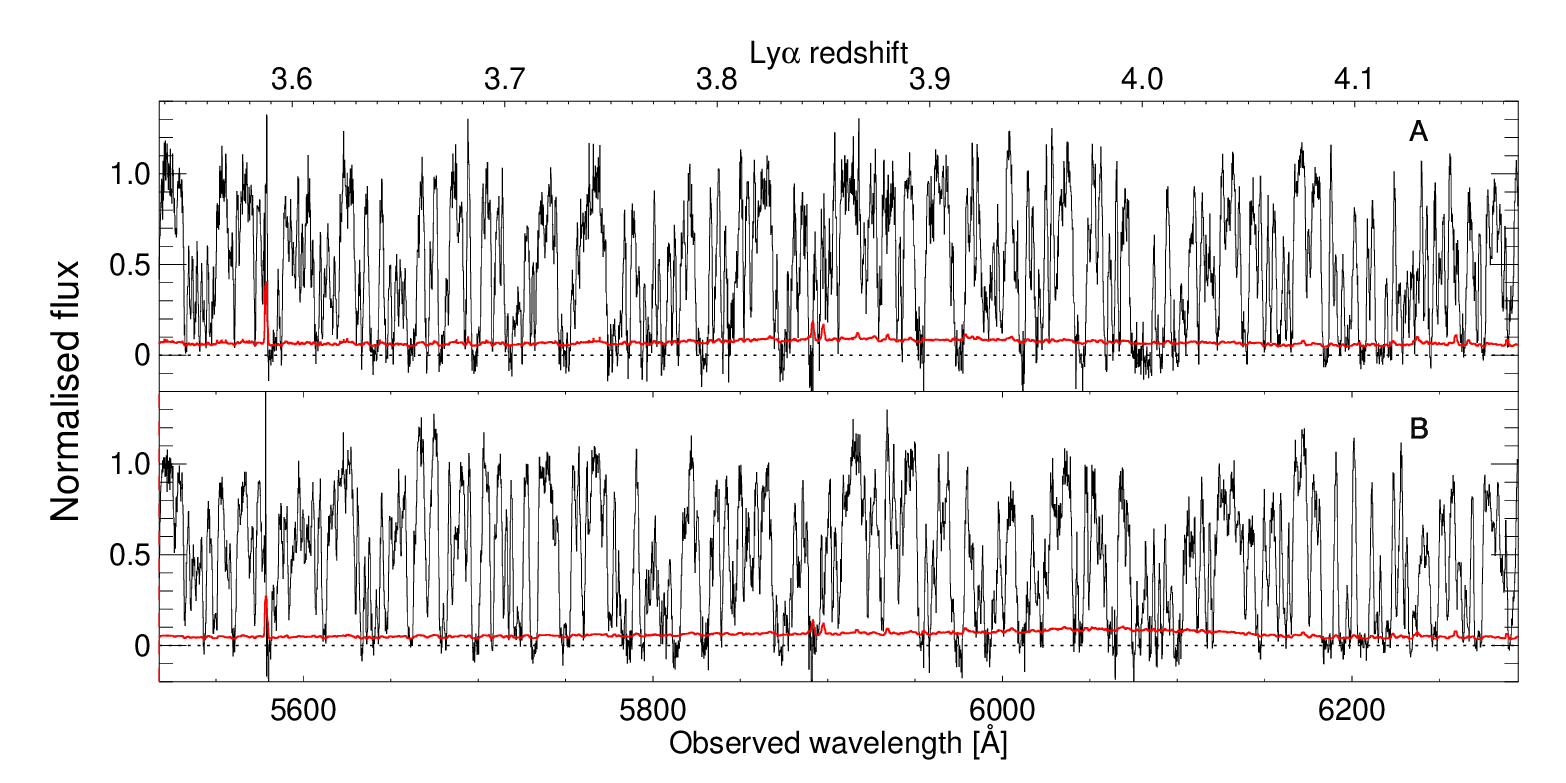,
      width=\textwidth}}
  \vskip -0.1in
\caption{\label{fig:pair_s} {\bf Spectra from the pair sample} as
in Fig.~1 of the main text, but for SDSS~J142023.77+283106.6 (panel A) and SDSS~J142023.80+283055.7 (panel B).}
\end{figure}
\begin{figure}
	\centering \centerline{\epsfig{file=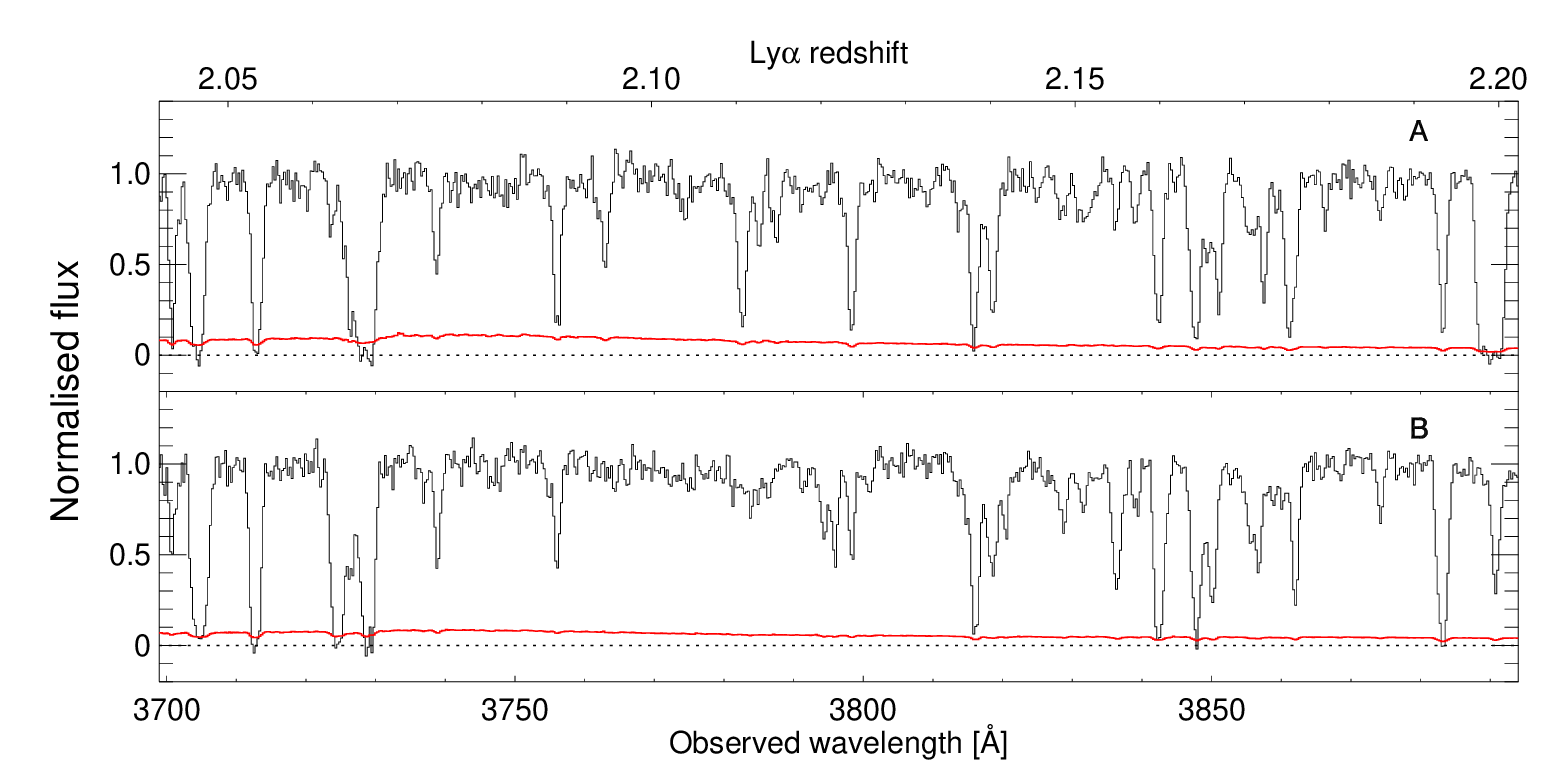,
      width=\textwidth}}
  \vskip -0.1in
\caption{\label{fig:pair_t} {\bf Spectra from the pair sample} as
in Fig.~1 of the main text, but for SDSS~J142758.74-012136.2 (panel A) and SDSS~J142758.89-012130.4 (panel B).}
\end{figure}
\begin{figure}
	\centering \centerline{\epsfig{file=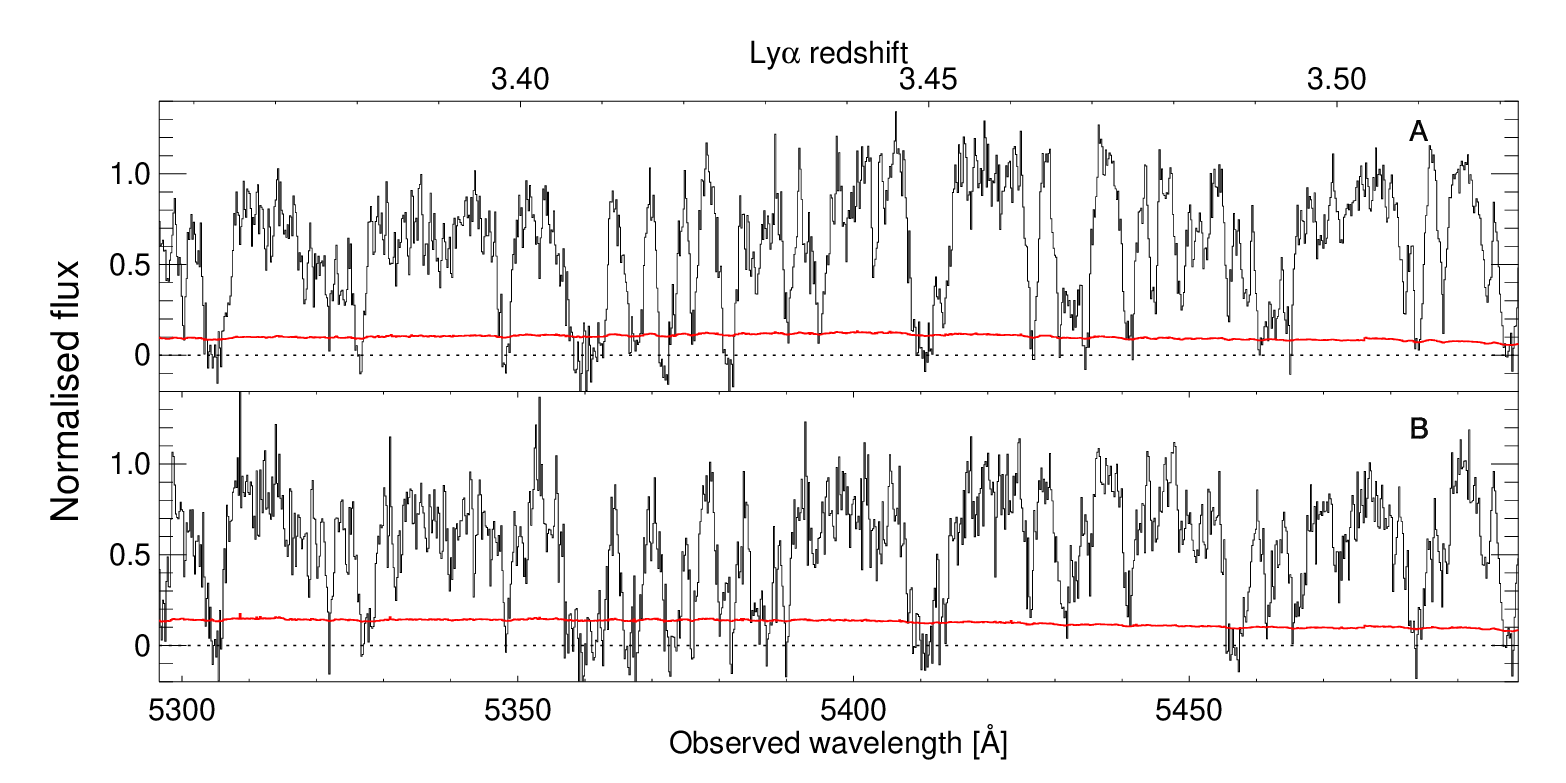,
      width=\textwidth}}
  \vskip -0.1in
\caption{\label{fig:pair_u} {\bf Spectra from the pair sample} as
in Fig.~1 of the main text, but for SDSS~J154110.40+270231.2 (panel A) and SDSS~J154110.37+270224.8 (panel B).}
\end{figure}
\begin{figure}
	\centering \centerline{\epsfig{file=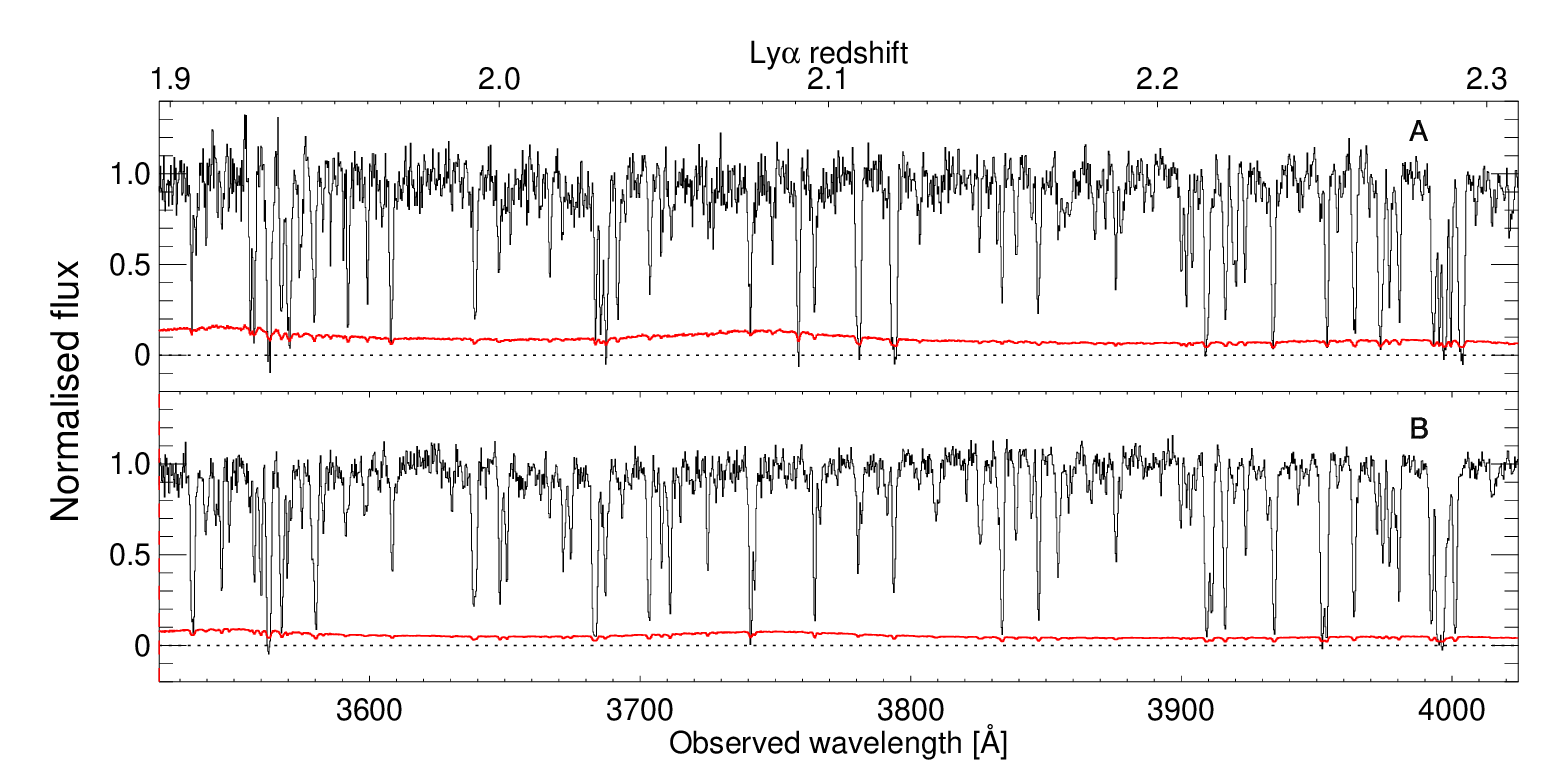,
      width=\textwidth}}
  \vskip -0.1in
\caption{\label{fig:pair_w} {\bf Spectra from the pair sample} as
in Fig.~1 of the main text, but for SDSS~J161302.03+080814.3 (panel A) and SDSS~J161301.69+080806.1 (panel B).}
\end{figure}
\begin{figure}
	\centering \centerline{\epsfig{file=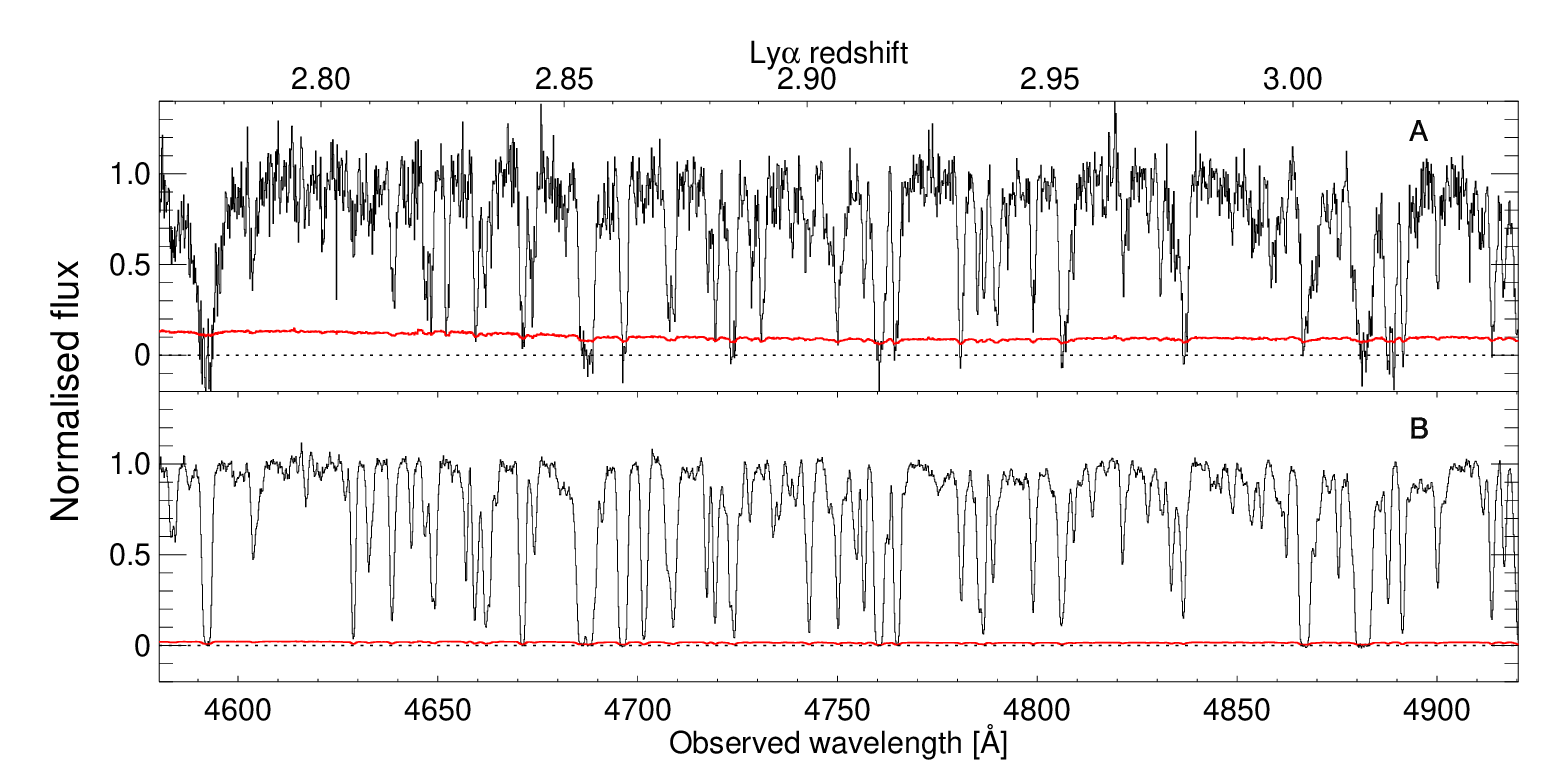,
      width=\textwidth}}
  \vskip -0.1in
\caption{\label{fig:pair_x} {\bf Spectra from the pair sample} as
in Fig.~1 of the main text, but for SDSS~J162210.11+070215.3 (panel A) and SDSS~J162209.81+070211.5 (panel B).}
\end{figure}
\begin{figure}
	\centering \centerline{\epsfig{file=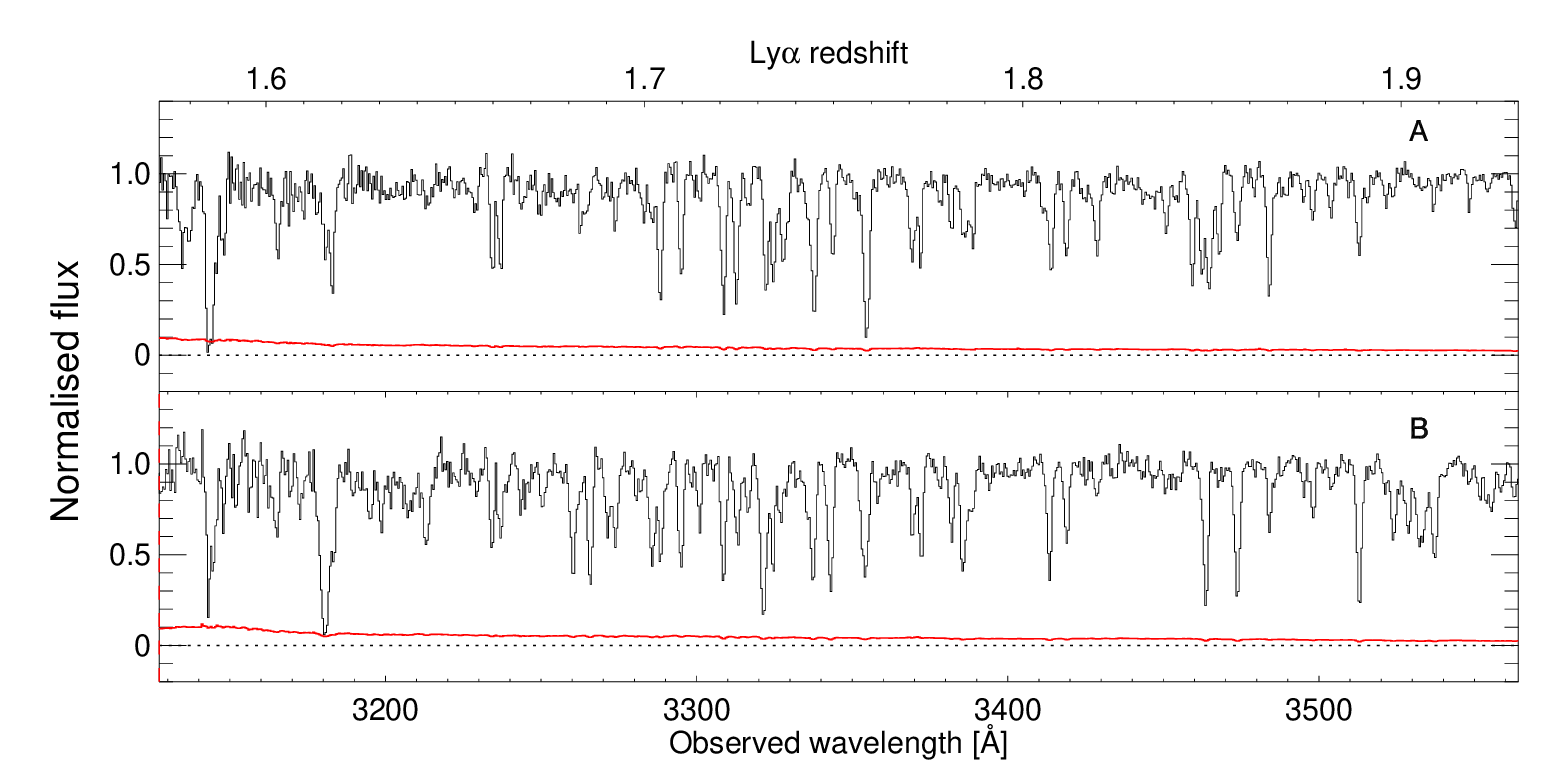,
      width=\textwidth}}
  \vskip -0.1in
\caption{\label{fig:pair_y} {\bf Spectra from the pair sample} as
in Fig.~1 of the main text, but for SDSS~J221427.03+132657.0 (panel A) and SDSS~J221426.79+132652.3 (panel B).}
\end{figure}
\begin{figure}
	\centering \centerline{\epsfig{file=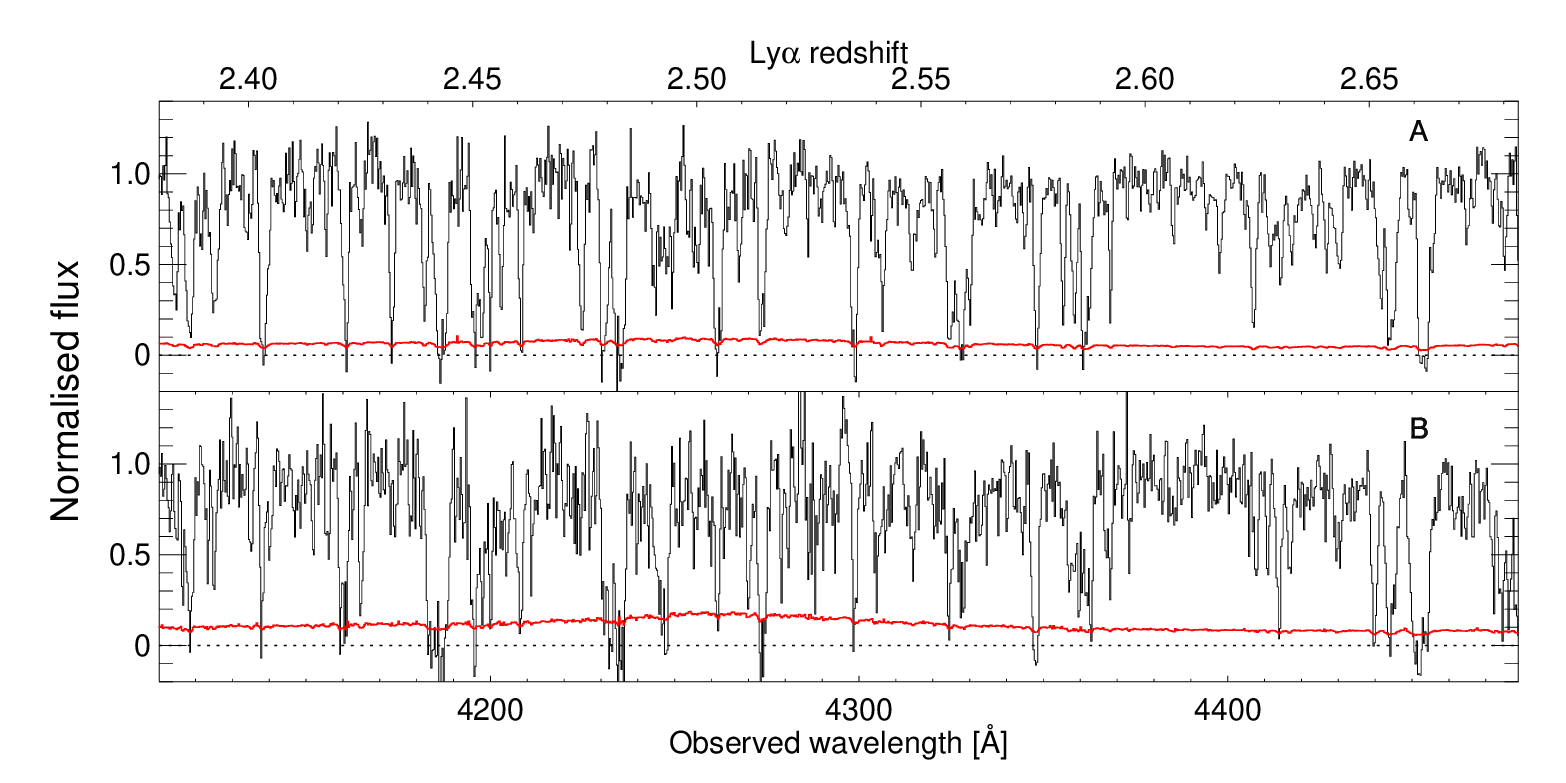,
      width=\textwidth}}
  \vskip -0.1in
\caption{\label{fig:pair_z} {\bf Spectra from the pair sample} as
in Fig.~1 of the main text, but for SDSS~J230044.52+015552.1 (panel A) and SDSS~J230044.36+015541.7 (panel B).}
\end{figure}

\clearpage

\newpage

\bibliographystyle{Science}
\bibliography{jeansbiblio}

\vskip 1cm
\noindent
{\bf Acknowledgements}\\
We thank Kate Rubin, Marie Lau, George Djorgovski, Crystal Martin,
Sara Ellison and Rob Simcoe for helping with data collection and reduction.

We thank G. Becker, M. Haehnelt, and the ENIGMA group (http://
enigma.physics.ucsb.edu) for useful discussion and comments
on early versions of this work. Some of the archival data used in
this paper were collected and/or reduced by M. Fumagalli, K. Rubin, M. Lau,
G. Djorgovski, C. Martin, S. Ellison, and R. Simcoe. Supported by
the Alexander von Humboldt Foundation and the Max Planck
Society (J.F.H.), NSF grant AST-1010004 (J.X.P.), U.S. Department
of Energy (DOE) grant DE-AC02-05CH11231 (Z.L.), and European
Research Council grant 320596 (G.K.). Calculations in this paper
used resources of the U.S. National Energy Research Scientific
Computing Center and of the Max Planck Computing and Data Facility
supported by the Max Planck Society. Some data were obtained at
the W. M. Keck Observatory, which is operated as a scientific
partnership among the California Institute of Technology, the
University of California, and NASA. The Observatory was made
possible by the financial support of the W. M. Keck Foundation. We
recognize and acknowledge the important cultural role and
reverence that the summit of Mauna Kea has always had
within the indigenous Hawaiian community; we are most
fortunate to have the opportunity to conduct observations from
this mountain. Some data were obtained with the 6.5-m
Magellan Telescopes located at Las Campanas Observatory,
Chile. Some data were obtained with the European Southern
Observatory Very Large Telescope under programs 090.A-0824
and 089.A-0855. Spectra for all quasar pairs are shown in the
supplementary materials. All the observational data analyzed
in this paper are available in public archives. Simulation
code and output files are also available in public
repositories. Full details and URLs are provided in the
supplementary materials.

\end{document}